\documentclass{article} 
\usepackage{iclr2027_conference,times}

\usepackage{amsmath,amsfonts,bm}

\def\eqref#1{equation~\ref{#1}}

\def\1{\bm{1}}

\DeclareMathAlphabet{\mathsfit}{\encodingdefault}{\sfdefault}{m}{sl}
\SetMathAlphabet{\mathsfit}{bold}{\encodingdefault}{\sfdefault}{bx}{n}

\usepackage[utf8]{inputenc} 
\usepackage[T1]{fontenc}    
\usepackage{hyperref}       
\usepackage{url}            
\usepackage{booktabs}       
\usepackage{amsfonts}       
\usepackage{nicefrac}       
\usepackage{microtype}      
\usepackage{xcolor}         
\usepackage{multirow}
\usepackage{graphicx}
\usepackage{todonotes}
\usepackage{pifont}
\usepackage{tcolorbox}
\usepackage[normalem]{ulem}
\usepackage{amsmath}
\usepackage{amssymb}
\usepackage{mathtools}
\usepackage{amsthm}
\usepackage{placeins}
\usepackage{inconsolata}
\usepackage{multirow}
\usepackage{enumitem}

\usepackage{enumitem}

\title{Brain alignment of reasoning and action representations from vision-language and action models during naturalistic gameplay}

\author{Subba Reddy Oota$^{1}$, Anant Khandelwal$^{2}$, Khushbu Pahwa$^{3}$, Satya Sai Srinath Namburi$^{4}$\\
 \textbf{Tanmoy Chakraborty$^5$, Bapi S. Raju$^6$, Manish Gupta$^{7}$}\\
\normalsize  $^1$Independent, $^2$Microsoft Research, Bangalore, India, $^3$AWS AI Labs, Amazon, $^4$GE HealthCare, USA \\
$^5$IIT Delhi, India, $^6$IIIT-Hyderabad, India, $^7$Microsoft, Hyderabad, India\\
\small \texttt{subbareddyoota@gmail.com, tanchak@iitd.ac.in, raju.bapi@iiit.ac.in,  gmanish@microsoft.com} }

\iclrfinalcopy

\usepackage{etoolbox}

\makeatletter
\patchcmd{\@maketitle}
  {\lhead{Published as a conference paper at ICLR 2027}}
  {\lhead{Preprint}}
  {}{}
\makeatother

\begin{document}

\maketitle

\begin{abstract}

Understanding how humans and artificial intelligence systems predict and plan by interacting with their environment is a fundamental challenge at the intersection of neuroscience and machine learning. Most brain-encoding studies focus on aligning artificial models with brain activity during language comprehension or passive visual processing, while interactive brain alignment studies have to date been largely limited to reinforcement-learning (RL) agents and theory-based models. 
To address this gap, we study brain alignment of representative models from two foundation-model types, namely vision-language models (VLMs) and large-action models (LAMs), using fMRI recordings from participants playing naturalistic Atari-style video games. Specifically, we examine how action-focused and reasoning-focused prompts 
shape the models' internal representations and their alignment with fMRI brain activity. First, we find that both VLMs and LAMs achieve significantly higher voxel-wise encoding performance than RL baselines, with the advantage holding even under matched feature dimensionality. Second, compared to a ``no-prompt'' baseline, prompt-driven gains are larger in higher-order frontal-parietal and motor-planning regions than in early visual cortex, roughly $2$--$2.5\times$ when averaged over region-of-interest (ROI) groups, although individual regions are heterogeneous.
Third, variance partitioning reveals a qualitatively different representational organization. VLM representations are prompt-symmetric (12.4\% unique action vs.\ 9.5\% unique reasoning), whereas LAM representations are action-dominant 
(25.6\% unique action vs.$-8.2$\% unique reasoning), with the asymmetry strongest in frontal-motor cortex. Together, these results demonstrate that action-specialized fine-tuning reorganizes multimodal game-state representations in ways that change their alignment with action-associated cortical regions, a difference that is invisible at the level of raw prediction accuracy.

\end{abstract}

\section{Introduction}

Recent work has shown that language-model representations predict human brain activity during language, visual, and multimodal processing, revealing parallels between artificial and neural representations~\citep{toneva2019interpreting,schrimpf2021neural,oota2022joint,conwell2022can,oota2024multi}.
Beyond passive tasks, brain activity during interactive gameplay has also been aligned with deep reinforcement-learning (RL) agents, linking model features to neural signals underlying action selection and goal-directed behavior~\citep{tomov2023neural,haberland2026encoding}.
However, these studies largely focus on RL or theory-based agents~\citep{van2016deep,tsividis2021human}, whose representations emphasize action learning and may not capture richer knowledge about objects, relationships, and future outcomes.


This limitation is particularly relevant given growing interest in whether modern vision-language models (VLMs) and large-action models (LAMs) acquire internal world models, structured representations of objects, goals, and environmental dynamics that support prediction and planning~\citep{johnson1983mental,ha2018recurrent,waytowich2024atari,wanglarge,xie2025play}. Interactive gameplay provides an ideal testbed because it requires perception, rule understanding, multi-step reasoning, and decision-making, allowing us to examine whether these models align with the neural representations supporting human behavior.

\begin{figure}[t]
    \centering
    \includegraphics[width=\linewidth]{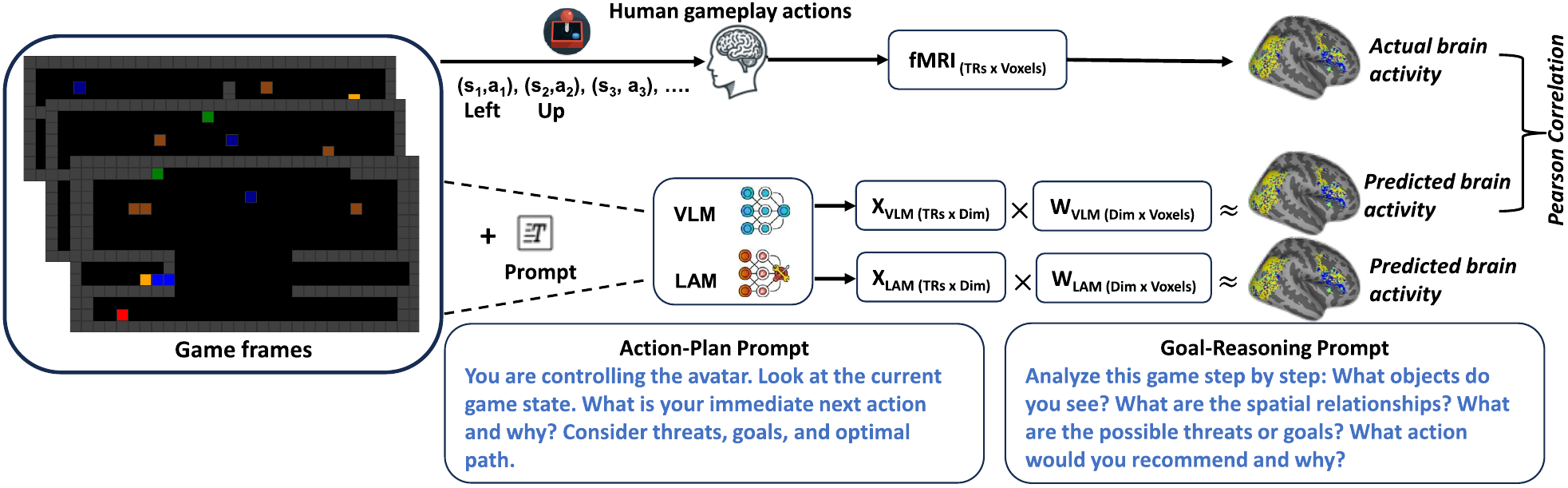}
    \caption{\textbf{Brain alignment pipeline for naturalistic Atari gameplay.} Participants played Atari-style video games during fMRI recording, producing TR-aligned brain responses (top row). The same gameplay frames were processed by two foundation-model families: vision-language models (VLMs) and large-action models (LAMs), conditioning on action-focused or reasoning-focused prompts (bottom). Voxel-wise encoding models trained on per-model per-prompt representations, $X_{\text{VLM}}$ and $X_{\text{LAM}}$, predict held-out fMRI activity, and brain alignment is measured as the Pearson correlation between predicted and actual responses. By comparing brain alignment across model families and prompting conditions, we test how action tuning and task framing shape the alignment of model game-state representations with human brain activity during interactive gameplay.
    }
    \label{fig:workflow}
\end{figure}

Successful behavior in humans is thought to rely not only on simple reactive responses, but also on richer internal representations of objects, events, goals, and causal dynamics that enable flexible behavior in new situations. Whether modern AI models acquire analogous representations is therefore an important open question, particularly for VLMs and LAMs, which are increasingly deployed as general-purpose interactive agents and may encode richer perceptual, semantic, and action-oriented structure than traditional RL policies. Aligning these models with human fMRI during gameplay, therefore, offers a principled way to assess not only what information they encode about game states, but also whether their internal representations relate to those that humans engage during interactive play.
In this paper, we investigate whether representations from VLMs and LAMs, elicited using action-focused and reasoning-focused prompts, align with human fMRI responses recorded during naturalistic Atari-style gameplay. Hence, we ask the
following research questions (RQs): 
    (1) Do VLM and LAM representations align with human brain activity during naturalistic gameplay, and does this alignment differ across action-focused and reasoning-focused prompting conditions?
    (2) How are the prompt-driven gains (compared to a ``no-prompt'' setting) in voxel-wise encoding performance distributed across cortex, from early visual to higher-order frontal-parietal regions? 
    (3) How much unique brain-relevant variance is contributed by action- versus reasoning-focused prompts, and how does this decomposition differ between VLMs and LAMs?

To address these questions, we systematically investigate both VLMs and LAMs under reasoning-focused and action-focused prompting conditions, alongside a no-prompt baseline, as shown in Fig.~\ref{fig:workflow}, and quantify brain alignment across all three settings. Using brain recordings from 32 participants playing naturalistic Atari-style video games~\citep{tomov2023neural}, we measure the brain alignment of two VLMs (Qwen2.5-VL~\citep{qwen2.5-VL} and InternVL3~\citep{chen2024internvl}) and two LAMs (UI-TARS-7B-DPO~\citep{qin2025ui,wang2025ui} and OS-Atlas-Pro-7B~\citep{wu2024atlas}). All models process visual game frames together with text prompts, enabling interactive reasoning over gameplay. We additionally probe  Qwen3.5~\citep{qwen35blog}, a thinking-mode model that produces explicit chain-of-thought traces, to test whether the model's generated reasoning aligns with brain activity differently from its final answer.
We further compare all models against two RL baselines, EMPA~\citep{tsividis2021human} and DDQN~\citep{van2016deep}. 


Our findings lead to four insights. (1) Both VLMs and LAMs significantly outperform RL baselines (EMPA, DDQN) in voxel-wise brain encoding, even at matched feature dimensionality; prompting further improves alignment 
in both. (2) Prompt-driven gains are larger in higher-order frontal-parietal and motor-planning regions (supplementary motor area (SMA), middle frontal gyrus (MFG), inferior frontal gyrus (IFG), angular gyrus (AG)) than in early visual cortex, by roughly $2$--$2.5\times$ when averaged over ROI groups, although individual regions are heterogeneous.
(3) At matched prompts, whole-brain prediction accuracy does not differ significantly between VLMs and LAMs,
but variance partitioning reveals a qualitatively different representational organization. VLMs are prompt-symmetric while LAMs are action-dominant\footnote{``prompt-symmetric'' $\implies$ action and reasoning prompts contribute similar amounts. ``action-dominant'' $\implies$ unique contribution of the action prompt substantially exceeds that of the reasoning prompt.}, with reasoning becoming redundant in frontal-motor cortex, a dissociation invisible at the level of raw accuracy. (4) In Qwen3.5, representations read out from generated reasoning traces tend to be less brain-aligned than action-output representations, a trend that holds for both last-token and mean-pooled readouts, motivating further investigation of reasoning-trace structure in thinking-mode models. 

Together, our results extend brain alignment research from passive language and visual processing to interactive gameplay, and provide the first systematic brain alignment evaluation of modern VLMs and LAMs in this setting.
By contrasting reasoning-focused and action-focused prompts, we operationalize a distinction central to world-model frameworks: the separation between understanding the environment and committing to actions within it. Our findings show that this distinction maps onto qualitatively different representational structures in VLMs and LAMs, dissociations revealed only through variance decomposition. This work opens new directions for studying how foundation models encode the structured, predictive representations that underlie adaptive behavior. Detailed related work on brain alignment, foundation models, and RL models of decision-making is in App.~\ref{app:relatedwork}. 

\section{Dataset}
\label{sec:dataset}

We use the publicly available raw BIDS-formatted fMRI data from the naturalistic Atari-style video game dataset introduced by~\citet{tomov2023neural}, comprising recordings from 32 healthy, right-handed participants (15 female and 17 male; aged 19--36 years). We preprocess these data using fMRIPrep v25.2.0~\citep{esteban2019fmriprep,esteban2020analysis}; complete preprocessing, nuisance-regression and surface-projection details are in App.~\ref{app:preprocessing}.
Each participant completed a single session of ${\sim}2.5$ hours (${\sim}1.5$ hours in scanner and ${\sim}1$ hour on BOLD acquisition during gameplay), comprising six functional runs, in which participants played six Atari-style video games.
Each run contained three blocks, and each block consisted of three levels  of increasing difficulty, of a single game; the levels were replayed for a fixed 1-minute duration (nine levels per game total). The runs included a 10-second fixation at the beginning and end, and each run lasted 566 s. For each participant, the fMRI scan was acquired at TR (Repetition Time) = 2 s, producing 283 time points per functional run. 
In this work, we aggregate across the three levels and evaluate generalization to held-out runs.


For the surface-based ROI analyses, we use the multimodal parcellation of the human cerebral cortex based on the Glasser Atlas (which consists of 180 regions of interest in each hemisphere) to report the ROI (region of interest) analysis for the brain maps~\citep{glasser2016multi}.
We perform the ROI analysis for the Atari video game dataset considering the following frontal-parietal: angular gyrus (AG), inferior frontal gyrus triangularis (IFGtriang), inferior frontal gyrus opercularis (IFGoperc), middle frontal gyrus (MFG), visual: inferior occipital gyrus (IOG), middle occipital gyrus (MOG), superior occipital gyrus (SOG), calcarine fissure (CAL), lingual gyrus (LING), cuneus (CUN), and motor processing regions: supplementary motor area (SMA), based on the prior work~\citep{tomov2023neural}.
We show the flatmap with these labeled ROIs, in Appendix Fig.~\ref{fig:language_flatmap} and list the detailed sub-ROIs of these ROIs in App.~\ref{app:detailedsubrois}.


\section{Methodology}
\label{sec:modelArch}


We use two classes of foundation models for representation extraction: 
vision-language models (VLMs) and large-action models (LAMs). Both 
process visual frames together with text prompts.

\textbf{VLMs.} (i) Qwen2.5-VL-7B-Instruct~\citep{qwen2.5-VL} 
(28 layers, hidden dim 3584) is instruction-tuned on large-scale image--text, video-text data with a dynamic-resolution vision encoder. 
(ii) InternVL3-8B~\citep{chen2024internvl} (28 layers, dim 3584) pairs a Qwen2.5-7B language model with a different vision encoder (InternViT) and training corpus.

\textbf{LAMs.} (i) UI-TARS-7B-DPO~\citep{qin2025ui,wang2025ui} is 
fine-tuned from Qwen2-VL on $\sim$50B tokens of GUI/game-interaction trajectories (SFT+DPO). (ii) OS-Atlas-Pro-7B~\citep{wu2024atlas} is separately trained from Qwen2-VL on a large GUI grounding corpus. Both LAMs share the Qwen2-VL backbone (28 layers, dim 3584). Combined with Qwen2.5-VL (matched at 28 layers, dim 3584), this enables a 
within-family VLM-vs-LAM contrast that isolates action-affordance and planning supervision while holding architecture, temporal window, and readout fixed.

\textbf{Baseline Models.} We derive TR-aligned regressors from two Atari-gameplay models. From \textbf{EMPA}~\citep{tomov2023neural}, eight per-frame regressors: four continuous: \emph{surprise} (next-state prediction error), \emph{spriteKL} (belief revision over 
sprite dynamics), \emph{R\_GG} (expected reward under the current goal-graph), \emph{num\_effects} (interaction-effect count), and four binary theory-revision flags: (\emph{replan}, \emph{theory\_change}, \emph{newEffects}, \emph{interaction\_change}). Frame values are 
averaged within each TR and trimmed by 5 TRs per run edge, yielding a $(T, 8)$ matrix per subject. From \textbf{DDQN}~\citep{van2016deep} 
(25M-step agent), we extract per-episode action sequences and compute normalized action distributions per run. More details about EMPA features reported in App.~\ref{app:baselines}.

\noindent\textbf{VLM and LAM Feature Extraction}.
\label{sec:vlm_features}
To compute model-derived predictors for each fMRI TR, we extract hidden-state representations from both VLMs and LAMs. 
Let \(\mathbf{x}_{t,j}\) denote the \(j\)-th frame sampled for fMRI TR \(t\), and let \(p\) denote the text prompt. We use \(k=4\) frames per TR. The game frame rate is approximately 20 FPS and the TR is 2 s, corresponding to approximately 40 game frames per TR. For each TR, the window ends at the last game frame in that TR, and four context frames are sampled evenly at intervals of approximately 10 frames (\(\approx 0.5\) s), spanning the final approximately 1.5 s of the TR, denoted as $\mathbf{X}_t = \big(\mathbf{x}_{t,1}, \mathbf{x}_{t,2}, \mathbf{x}_{t,3}, \mathbf{x}_{t,4}\big)$. Frames are taken from each participant's own gameplay, and windows never cross run boundaries.
The $k$ sampled frames are jointly passed as a multi-image sequence alongside $p$, exploiting the model's native multi-image input. We run \texttt{model.generate()} with greedy decoding and hidden-state outputs enabled, extracting the last hidden state at every transformer layer to yield one last-token embedding  $\mathbf{E}_t \in \mathbb{R}^{(L+1)\times d}$ over all $L+1$ layers (including the embedding layer).

\noindent\textbf{Episode-level features.} For an episode with $T$ TRs, we slide the window across the episode and stack per-TR embeddings into
$\mathbf{E} = \{\mathbf{E}_t\}_{t=1}^{T} \in \mathbb{R}^{T \times 
(L+1) \times d}$.
This tensor serves as the predictor matrix for voxel-wise encoding, placing VLM and LAM features on the same temporal footing as EMPA and DDQN. Full notation, chat-template serialization, and per-layer indexing are in App.~\ref{app:feature_extraction}.




\noindent\textbf{Prompt variants.}
To probe different forms of model reasoning, we use two prompt 
templates. The \emph{action prompt} $p_{\text{action}}$ asks the model 
to commit to a next action and briefly justify the choice (considering threats, goals, and optimal path); the \emph{reasoning prompt} $p_{\text{reasoning}}$ asks the model to analyze the frame step by step (objects, spatial relationships, threats and goals) before recommending an action. Both prompts therefore elicit an action recommendation; they differ in whether the response is organized action-first or reasoning-first. In the \emph{no-prompt} condition, the model receives the same frames with an empty text prompt. Full prompt templates are provided in App.~\ref{app:prompttemplates}. For each prompt $p_m \in \{p_{\text{action}}, 
p_{\text{reasoning}}\}$, we compute $\mathbf{E}_t^{(m)}$ separately and 
treat the resulting representations as distinct feature sets for downstream voxel-wise encoding.

\noindent\textbf{Prompt controls.}
In addition to the two primary prompt conditions, we evaluate seven control prompts to determine whether the observed effects reflect text presence, visual description, goal information, or motor commitment. These comprise \texttt{scrambled-neutral}, which contains task-irrelevant words with syntax destroyed; \texttt{scrambled}, which retains the task lexicon but destroys syntax; \texttt{lowlevel}, which requests only color statistics; \texttt{scene}, which requests object identities and spatial relationships without goals; \texttt{goal}, which requests hazards and objectives without a motor commitment; and \texttt{action-bare}, which requests only the selected movement key; and \texttt{action-full}, which requests a committed action justified against alternatives. Together with the two primary prompts, these form the nine-condition prompt battery reported in Appendix Table~\ref{tab:prompt_battery}.


\section{Experimental Setup}
\label{experimental_setup}

\noindent\textbf{Voxel-wise encoding model.} 
We train voxel-wise fMRI encoding models using bootstrap ridge regression~\citep{tikhonov1977solutions} to predict voxel responses from model representations. Feature channels and fMRI responses are z-scored separately for training and testing. To account for hemodynamic delay, we concatenate four temporal lags (2, 4, 6, and 8 s), allowing voxel-specific lag weights to be learned~\citep{toneva2019interpreting}. At time step $t$, let stimulus features be $X_{t}\in \mathbb{R}^{N \times D}$ and voxel responses $Y_{t}\in \mathbb{R}^{N \times V}$, where $N$ is the number of examples, $D$ the lagged feature dimension, and $V$ the number of voxels. Regularization parameters are selected via cross-validation. The goal is to predict each voxel's brain response from the presented stimulus.


\noindent\textbf{Layer selection.} We fit encoding models for every layer and report the best-performing layer per model. Because each layer's correlation is itself cross-validated but the layer is chosen using these same held-out scores, the reported peak values are slightly optimistic. Hence, we report full layer-wise profiles (App.~Figs.~\ref{fig:vem_vlm_rois_atari_laye_hierarchy_full}--\ref{fig:vem_vlm_rois_atari_laye_hierarchy_r005} and~\ref{fig:vem_lam_rois_atari_laye_hierarchy_full}--\ref{fig:vem_lam_rois_atari_laye_hierarchy_r005}) and verify that our conclusions do not depend on the selected layer.


\noindent\textbf{Train-test setup.}
We train subject-specific encoding models using strict leave-one-run-out (LORO) cross-validation. For each subject, in each fold, one complete scanner run is held out as the test set, while all remaining runs are used for training. This process is repeated until every run has served once as the held-out test set. Feature windows are constructed separately within each run and never cross run boundaries. Thus, no test TR shares a scanner run with the training data, preventing within-run train-test adjacency and reducing potential information leakage due to temporal autocorrelation. The detailed hyperparameter settings and statistical significance tests are provided in App.~\ref{app:hyperparameters_details} and~\ref{app:statistical_significance}.

\noindent\textbf{Evaluation metrics.}
We evaluate our models using Pearson Correlation Coefficient (PCC), which is a standard metric to evaluate brain alignment \citep{jain2018incorporating,schrimpf2021neural,goldstein2022shared}. Let TR be the number of time repetitions in the test set. Let $Y=\{Y_i\}_{i=1}^{TR}$ and $\hat{Y}=\{\hat{Y}_i\}_{i=1}^{TR}$ denote the actual and predicted value vectors for a single voxel, respectively. Thus, $Y$ and $\hat{Y}~\in \mathbb{R}^{TR}$. 
PCC is then calculated as correlation between the model's predictions $\hat{Y}$ and neural recordings $Y$. Our primary brain alignment analysis retains all cortical vertices without applying a predictivity threshold. We additionally report descriptive sensitivity analyses, and report averages over voxels with $r>0$ and $r>0.05$. We report mean $\pm$ SEM across the 32 participants. Model and prompt comparisons use paired tests across participants, with FDR correction across ROIs; details are in App.~\ref{app:statistical_significance}. Implementation details are in App.~\ref{app:hyperparameters_details}.

\noindent\textbf{Variance partitioning.}
Using a variance partitioning approach~\citep{de2017hierarchical,lebel2021voxelwise}, we test whether action- and reasoning-based prompts carry overlapping or distinct brain-relevant information, and whether LAMs, which are fine-tuned on action-specific environments reshapes this decomposition. Implementation details are in App.~\ref{app:sharedVarianceDetails}.

\noindent\textbf{Behavioral analyses.} To relate model representations to participants' behavior, we (i) compare the action recommended in each model's generated output with the participant's action at the same time point (model--human action agreement, evaluated at $2$\,Hz), and (ii) decode each participant's TR-level actions from the same model features used for encoding, using a linear classifier under leave-one-run-out cross-validation. Details are in App.~\ref{app:behavior}.



\section{Results}
\label{sec:results}


\vspace{-0.2cm}
\subsection{\textbf{[RQ1]:} VLMs and LAMs exceed RL baselines under \emph{no-prompt} conditions, and task-framed prompts further improve brain alignment.}
\vspace{-0.2cm}
To test whether VLMs and LAMs achieve higher voxel-wise encoding performance than classical RL baselines, and to isolate this from confounds of prompting and feature dimensionality, we evaluate two settings. First, a \emph{no-prompt} setting in which VLMs and LAMs receive only gameplay frames without text instruction; this isolates representational quality from prompt-induced effects and offers fair comparison with EMPA and DDQN, neither of which uses language input. Within this setting, we additionally vary feature dimensionality (8, 64, 1024) to rule out that any encoding gain simply reflects VLMs and LAMs having more features than the RL baselines.  Second, a \emph{prompted} 
setting in which the same models additionally receive text-prompt instructions (full templates in App.~\ref{app:prompttemplates}), quantifying the gain from explicit prompting.  Across both settings, VLM and LAM representations significantly outperform RL baselines, with prompting providing additional gains. We additionally test whether the prompted-alignment results generalize across two VLMs and two LAMs, quantify model--human action agreement, and examine whether participants' actions can be decoded from the model representations.

 \begin{figure}[t]
    \centering
    \includegraphics[width=0.49\linewidth]
      {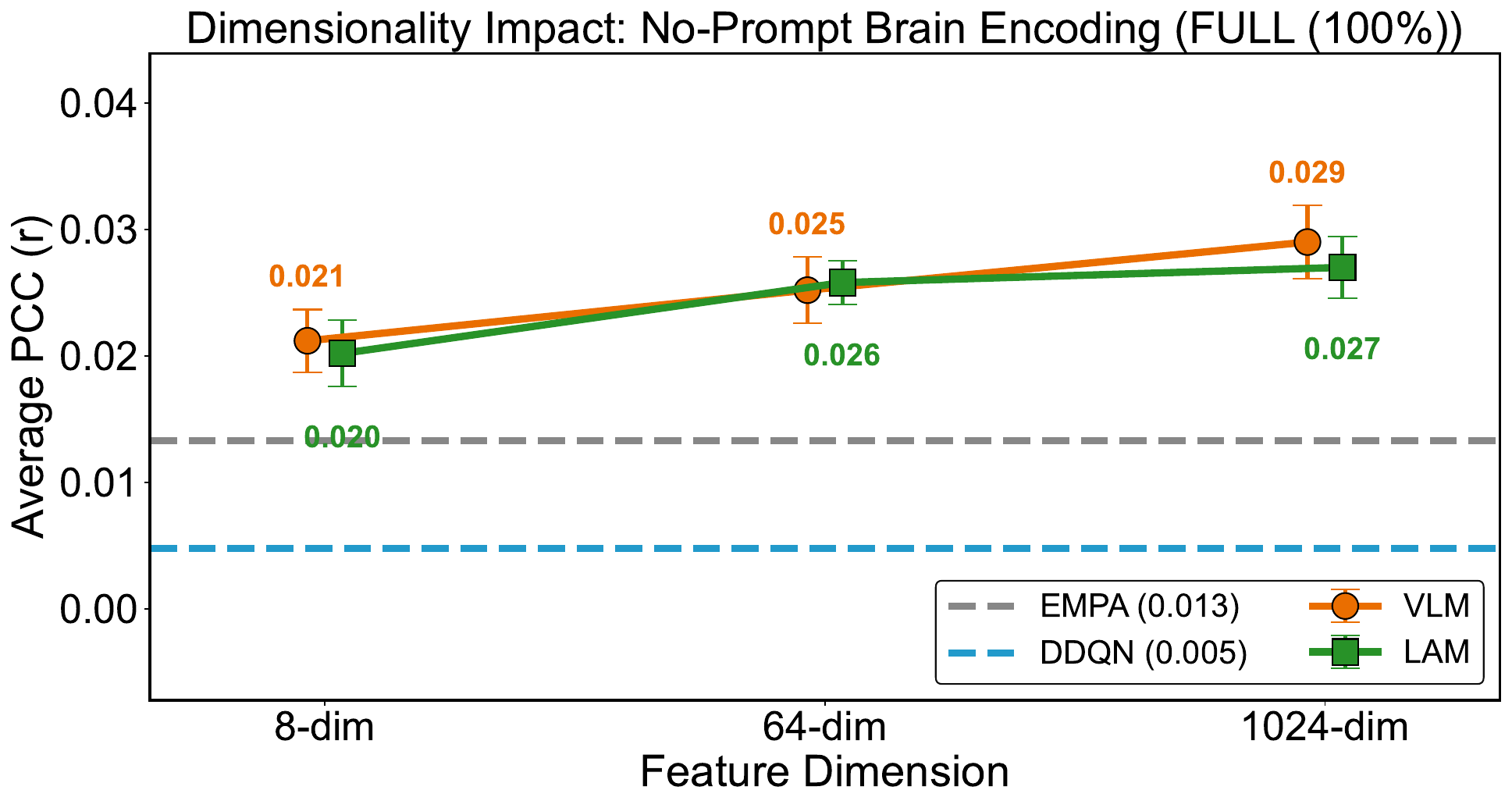}
    \includegraphics[width=0.49\linewidth]
      {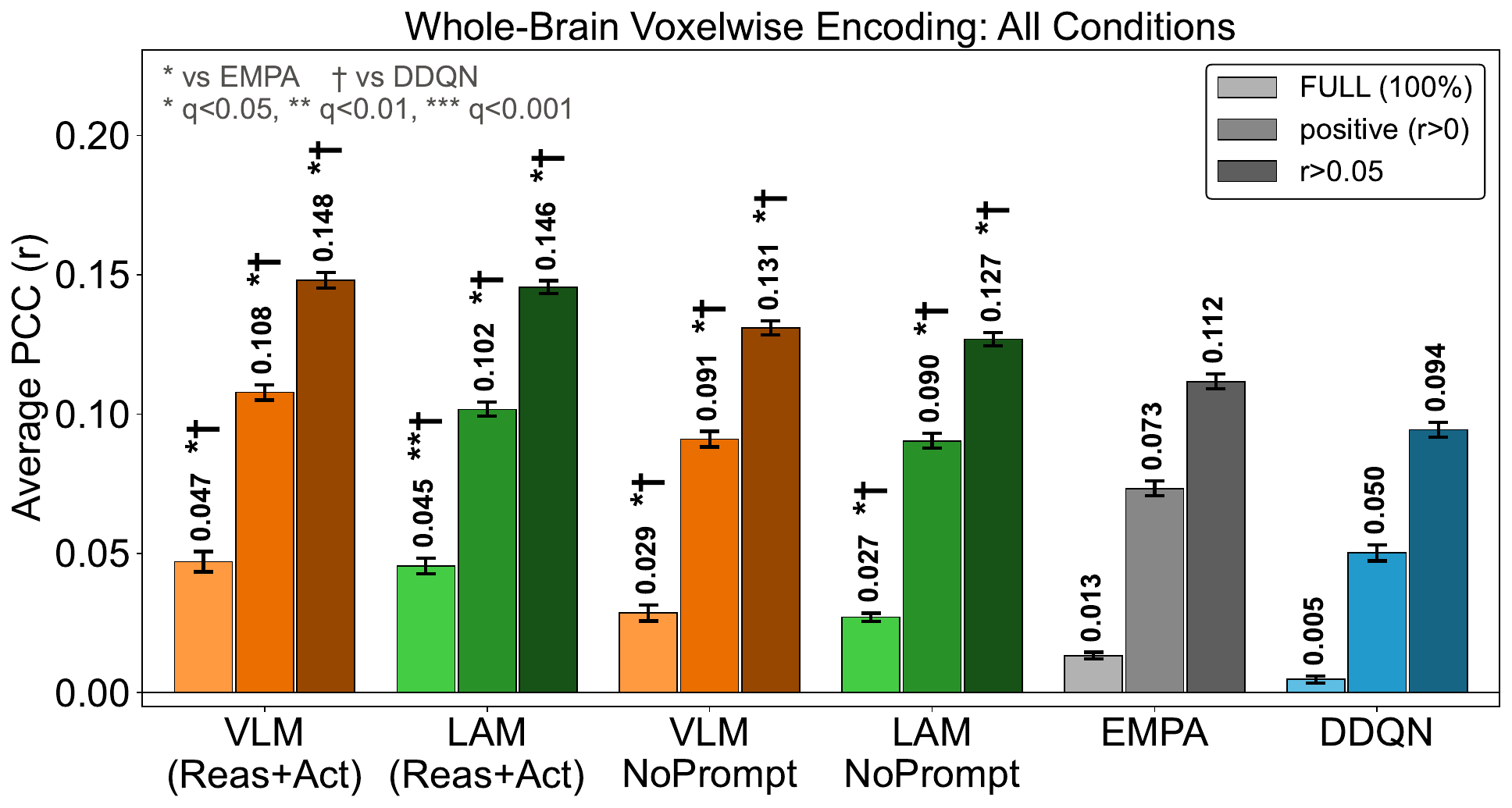}
    \caption{\textbf{Whole-brain voxel-wise encoding performance.} (left) Effect of feature dimensionality under the \emph{no-prompt} condition, shown for voxels with full voxels (unthresholded): VLM (\textcolor{orange}{orange}) and LAM (\textcolor{green}{green}) features evaluated at three dimensionalities (8, 64, 1024), compared against the RL baselines EMPA (gray dashed, r=0.013) and DDQN (light blue dashed, r=0.005). * indicates significant pairwise gain from 8 to 64 dim (paired t-test, 
$p<0.05$); the 64 to 1024 transition is non-significant, indicating saturation. At every dimensionality tested, both VLM and LAM significantly exceed both RL baselines. (right) All conditions: Whole-brain alignment for prompted (average across Reasoning and Action) and \emph{no-prompt} VLM and LAM, compared to EMPA and DDQN baselines, under three voxel-selection criteria: all voxels (Full), positively predicted (r$>$0) and (r$>$0.05). * denotes a significant difference vs.\ EMPA and $\dagger$ vs.\ DDQN (paired t-test, FDR-corrected; * $q<0.05$, ** $q<0.01$, *** $q<0.001$). Error bars: mean $\pm$ SEM across subjects.}
    \label{fig:vem_wholebrain_atari}
\end{figure}

\noindent\textbf{VLMs and LAMs outperform RL baselines under ``\emph{no-prompt}'' conditions.}
As shown in Fig.~\ref{fig:vem_wholebrain_atari} (left), both VLM and LAM representations outperform the classical baselines in whole-brain voxel-wise encoding without any prompt instruction (shown for all voxels unthresholded). Specifically, VLM (\emph{no-prompt}) achieves r=0.029 and LAM (\emph{no-prompt}) reaches r=0.027, compared to EMPA (r=0.013) and DDQN (r=0.005). The advantage is consistent across all three feature dimensionalities tested. Crucially, even when VLM and LAM features are 
reduced to 8 dimensions to match EMPA's feature count, they continue to 
significantly show higher brain alignment than both RL baselines (VLM r=0.021, LAM r=0.020), ruling out feature count as the source of the gain.
Performance saturates between 64-dim and 1024-dim, suggesting that the gain over RL baselines reflects what the features encode rather than feature dimensionality alone. Together, these results indicate that contemporary multimodal architectures learn richer latent representations of dynamic game environments than traditional task-optimized RL systems. We present similar analysis for voxels with $r>0.05$ in App.~\ref{app:lamGainsScale} Fig.~\ref{fig:vem_wholebrain_atari_r005}.

\noindent\textbf{Effect of Action and Reasoning prompts.}
We next examine whether explicit prompting further improves brain alignment by eliciting higher-level reasoning states from the same models. Fig.~\ref{fig:vem_wholebrain_atari} (right) reports all conditions on the primary unthresholded metric alongside the two sensitivity criteria. Under action and reasoning prompts, as shown in Fig.~\ref{fig:vem_wholebrain_atari} (right), VLM and LAM achieve significantly higher encoding performance than their \emph{no-prompt} counterparts (paired t-test, p$<$0.05): VLM reaches r=0.148 and LAM reaches r=0.146. At the unthresholded whole-cortex level, prompted VLM features predict brain activity significantly better than EMPA (VLM $r=0.047\pm0.005$ vs.\ EMPA $r=0.013\pm0.002$, mean $\pm$ SEM across $N=32$ participants; $\Delta r=0.034$; paired $t(31)$, $p=0.047$; Wilcoxon signed-rank $p=0.033$), and the LAM shows the same advantage ($r=0.045$; $q<0.01$ vs.\ EMPA). The advantage of both VLM and LAM features over EMPA also remains significant after AR(1) prewhitening (App.~Table~\ref{tab:ar1_control}). This shows that adding task-framed text instructions improves brain alignment beyond what gameplay frames alone provide. Prompted VLM and LAM achieve comparable whole-brain alignment, with the difference between them not reaching significance. We further investigate representational differences masked by encoding accuracy using variance partitioning in Section~\ref{sec:rq3}. A second VLM--LAM pair (InternVL3, OS-Atlas) yields comparable alignment under both prompts, with small, non-significant prompt preferences ($\leq0.006$; App.~\ref{app:vem_generalization} Fig.~\ref{fig:vem_wholebrain_atari_allmodels}); we focus subsequent analyses on Qwen2.5-VL and UI-TARS.

\noindent\textbf{Prompt controls.} To test whether prompt effects reflect text presence, visual description, goal information, or motor commitment alone, we evaluate seven control prompts (Table~\ref{tab:prompt_battery}). Scrambled text, low-level colour descriptions, and scene descriptions yield weaker alignment than the task-framed prompts ($r\approx0.006$--$0.029$), whereas action-full and goal prompts approach them. Effects are model-dependent (e.g., action-bare: LAM $\approx0.055$, VLM $\approx0.029$), so we do not interpret individual prompts as isolating distinct semantic dimensions (App.~Fig.~\ref{fig:vem_wholebrain_atari_allmodels_prompt_controls}).

\noindent\textbf{Behavior analysis.} 
Model features decode participants' TR-level actions above chance (balanced accuracy: VLM $0.342$, LAM $0.362$, EMPA $0.289$). Details in App.~\ref{app:behavior}.



\begin{figure}[t]
    \centering
    \includegraphics[width=\linewidth]
      {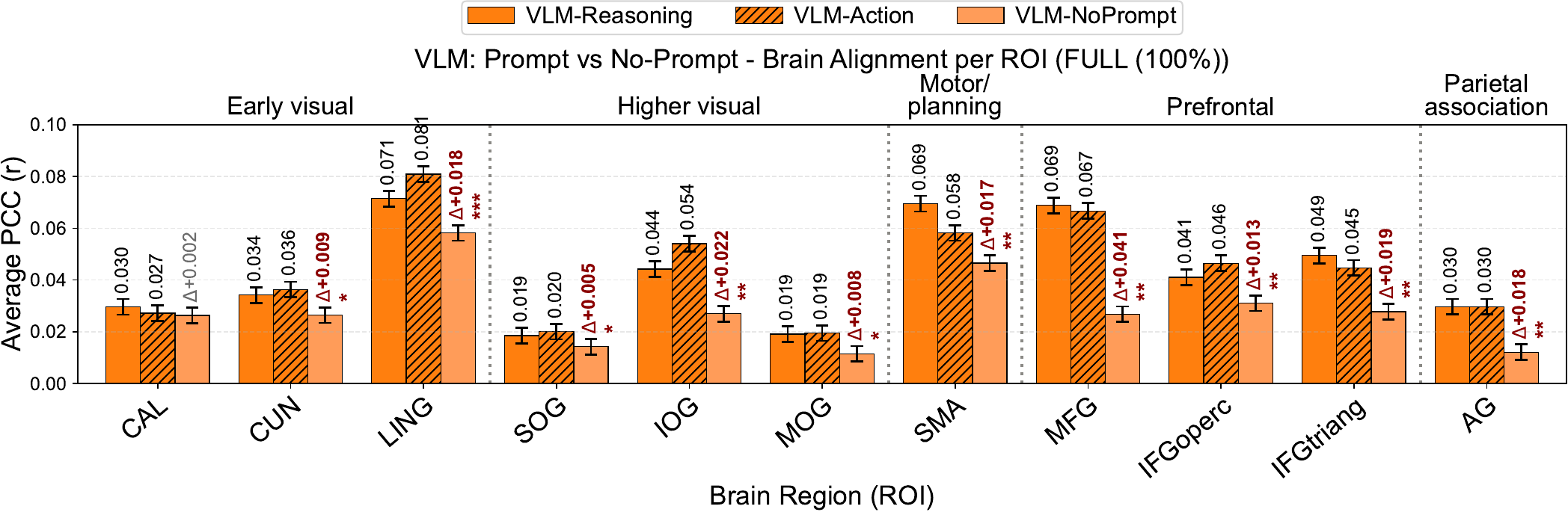}
    \caption{Brain alignment averaged across participants and voxels per ROI for prompted vs. no-prompt VLM representations. Average Pearson correlation between predicted and observed fMRI responses across 11 ROIs for VLM-Reasoning (solid), VLM-Action (hatched), and VLM-NoPrompt (light). Error bars denote mean $\pm$ SEM across subjects. Markers denote paired comparisons against no-prompt (*p $<$ 0.05, **p $<$ 0.01, **p $<$ 0.001). Both prompted conditions show higher alignment than no-prompt across all ROIs. Similar plot for LAM in Appendix Fig.~\ref{fig:vem_lam_rois_atari}. 
    }
    \label{fig:vem_vlm_rois_atari}
\end{figure}

\vspace{-0.2cm}
\subsection{\textbf{[RQ2]:} Prompt-driven gains are significantly greater in higher-order than in early visual cortex, and higher-order association regions tend to align best with deeper model layers.}
\label{sec:rq2}
\vspace{-0.2cm}

For the primary unthresholded analysis, we group the 11 ROIs into two functional groups: (1) a \emph{higher-order group} spanning frontal-parietal and motor-planning cortex (AG, IFGtriang, IFGoperc, MFG, SMA), and (2) \emph{early-visual group} (CAL, CUN, LING), and a higher-visual group (SOG, IOG, MOG). We  examine the higher-visual ROIs as an intermediate group in the descriptive analysis.

\noindent\textbf{Prompt gains differ between early-visual and higher-order ROI groups.} 
Fig.~\ref{fig:vem_vlm_rois_atari} reports unthresholded ROI means for the VLM under reasoning, action, and \emph{no-prompt} conditions, together with the composite prompt gain $\Delta r=r_{\text{prompted}}-r_{\text{no-prompt}}$, where $r_{\text{prompted}}$ is averaged across the reasoning and action conditions; the corresponding LAM results are shown in App.~\ref{app:lamGainsScale} Fig.~\ref{fig:vem_lam_rois_atari}. For the VLM, prompting significantly raises alignment in 10 of 11 ROIs (paired $t$-tests, FDR-corrected). The largest gains occur in MFG ($\Delta r=+0.041$, $q<0.01$), followed by IOG ($+0.022$, $q<0.01$), IFGtriang ($+0.019$),  LING and AG ($+0.018$), SMA ($+0.017$), IFGoperc ($+0.013$), with smaller gains in CUN ($+0.009$), MOG ($+0.008$), and SOG ($+0.005$) (all $q<0.05$). CAL ($+0.002$, not significant (ns)). For the LAM, prompting significantly raises alignment in all 11 ROIs. Gains are largest in SMA ($+0.057$, $q<0.001$), IOG ($+0.059$, $q<0.001$) and MFG ($+0.042$, $q<0.001$), intermediate in IFGtriang ($+0.030$), SOG and AG (both $+0.023$), LING ($+0.021$), and MOG ($+0.019$), and smallest in IFGoperc ($+0.016$), CUN ($+0.010$), and CAL ($+0.009$) (all $q<0.05$). The LAM \emph{no-prompt} baseline is close to zero in SOG, MOG, and AG, so prompted alignment in these regions arises almost entirely from prompting. For both families, prompted alignment is highest in LING, SMA, and MFG and lowest in SOG and MOG.

\noindent\textbf{Early-visual vs.\ higher-order cortex.}
To test the group-level regional organization of prompt effects, we grouped ROIs into an early-visual, a higher-visual, and a higher-order group. We computed each participant's composite prompt gain within each ROI group and compared the subject-wise values. Using all voxels, prompt gains are significantly larger in higher-order than in early-visual cortex for both model families (VLM: higher-order minus early-visual $\Delta r=+0.012$, $t(31)=3.83$, $q=0.019$; LAM: $\Delta r=+0.020$, $t(31)=3.31$, $q=0.039$). Higher-order gains are thus roughly $2$--$2.5\times$ early-visual gains in both families. In both families, the higher-visual group is intermediate (VLM: $+0.012$; LAM: $+0.024$), yielding an early-visual $<$ higher-visual $<$ higher-order ordering at the ROI-group level. The contrast remains significant when restricting the analysis to vertices with $r>0.05$ (VLM: $+0.029$ vs.\ $+0.012$, $\Delta r=+0.018$, $t(31)=2.85$, $q=0.041$; LAM: $+0.041$ vs.\ $+0.016$, $\Delta r=+0.025$, $t(31)=2.13$, $q=0.046$), where the higher-visual group is likewise intermediate (VLM: $+0.013$; LAM: $+0.028$).
These higher-order ROIs partially overlap with prefrontal regions whose activity~\citet{tomov2023neural} found to be predicted by EMPA's theory-based representations, providing anatomical correspondence with prior findings from the same dataset.

These results support a modest macro-scale regional effect rather than a strictly monotonic cortical hierarchy. Individual ROIs remain heterogeneous: IOG, a higher-visual region, shows gains exceeding several frontal ROIs in both families (VLM: $+0.022$; LAM: $+0.030$), whereas CAL shows the smallest gain in both families.
In contrast to prompt gains, absolute prompted alignment differs only slightly between the early-visual and higher-order groups (VLM: $r=0.047$ vs.\ $0.050$; LAM: $r=0.045$ vs.\ $0.051$). We therefore restrict our regional conclusion to the significant group-level difference in prompt-induced gains and make no claim about an absolute regional ordering or, beyond the descriptive ROI-group ordering, a monotonic progression across individual ROIs.


\noindent\textbf{Layer-wise and model comparisons.} Visual ROI groups peak at early layers (VLM L4; LAM L7), whereas motor/planning and prefrontal groups peak at intermediate-to-deep layers (L18--L21); these peaks are descriptive (App.~Table~\ref{tab:roi_peak_layers}, Figs.~\ref{fig:vem_vlm_rois_atari_laye_hierarchy_full}--\ref{fig:vem_lam_rois_atari_laye_hierarchy_full}).

\noindent\textbf{VLM--LAM regional comparison.}
Raw encoding accuracy does not differ significantly between the VLM and LAM in any ROI after FDR correction. Under the action prompt, all ROI-wise comparisons yield $q>0.90$, while under the reasoning prompt the smallest adjusted value is $q=0.08$. Within the LAM, action-prompt alignment is numerically equal to or higher than reasoning-prompt alignment in 10 of 11 ROIs, with IFGtriang as the only exception, whereas the VLM exhibits a more mixed regional pattern. However, none of the within-model action-versus-reasoning differences reaches ROI-level significance, so this pattern is descriptive. 
In Section~\ref{sec:rq3}, we test whether this numerical action-leaning pattern corresponds to differences in the shared and prompt-unique information encoded by the two model families.


\noindent\textbf{Spatial visualization.} 
Group-averaged across subjects per-voxel prompt gains, $r_{\text{prompted}}$-$r_{\text{no-prompt}}$, under both reasoning (App.~\ref{app:lamGainsScale} Fig.~\ref{fig:vlm_prompt_noprompt_flatmap_groupmean_atari}a) and action (App.~\ref{app:lamGainsScale} Fig.~\ref{fig:vlm_prompt_noprompt_flatmap_groupmean_atari}b) prompts, reveal a consistent spatial pattern: large positive gains (red) concentrated in frontal-parietal and motor-planning cortex (MFG, SMA, IFG, AG), substantially smaller gains in mid-level visual regions, and a mixture of positive and negative gains within early visual cortex (CAL) and mid-level visual (SOG). The spatial pattern recapitulates the ROI-level ordering and clarifies that the small ROI-mean gains in early visual regions reflect heterogeneous voxel-level effects rather than uniform improvement, consistent with these regions being driven primarily by direct visual-input encoding rather than prompt-conditioned representations.

\vspace{-0.2cm}
\subsection{\textbf{[RQ3]:} Variance partitioning reveals prompt-symmetric alignment in VLMs but action-dominant brain alignment in LAMs.} 
\label{sec:rq3}
\vspace{-0.2cm}

\begin{figure}[t]
    \centering
    \includegraphics[width=\linewidth]
      {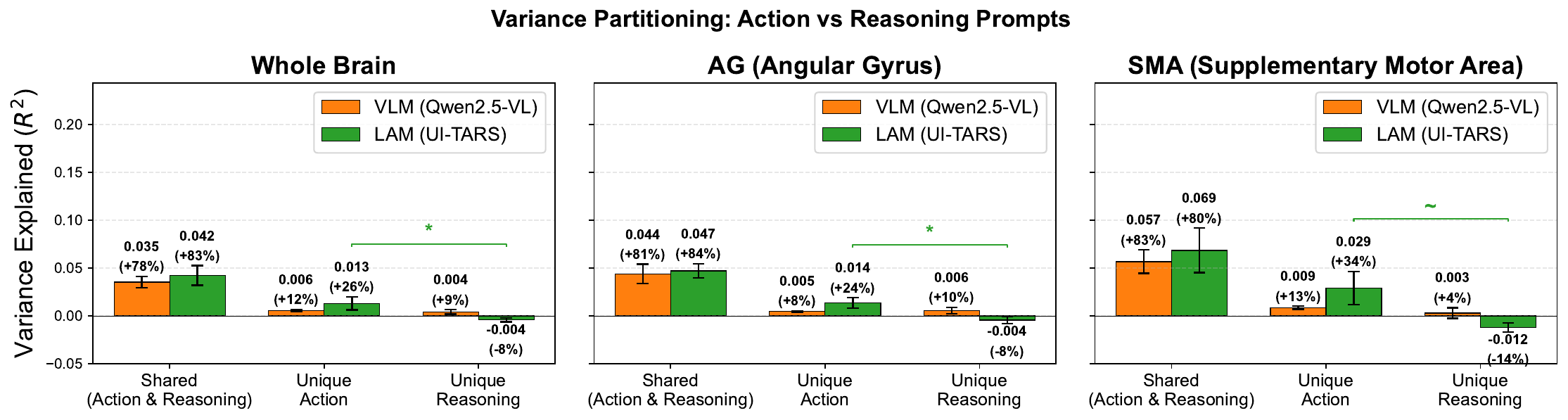}
    \caption{
    \textbf{Variance partitioning of action vs.\ reasoning prompts.} $R^2$ decomposed into shared variance and prompt-unique components for VLM (Qwen2.5-VL, orange) and LAM (UI-TARS-7B-DPO, green), averaged across participants. Numbers above bars show absolute $R^2$ and 
(in parentheses) percentage of joint explained variance. VLM unique variance is balanced across prompts; LAM is action-dominant, with negative unique-reasoning variance in whole brain, AG and SMA. Brackets: VLM-vs-LAM comparison on unique reasoning, * $p<0.05$, $\sim$ marginal 
($p<0.10$). Error bars: mean $\pm$ SEM across subjects.}
    \label{fig:variance_partition_wholebrain_atari}
\end{figure}

Although VLM and LAM achieve comparable raw prediction accuracy at matched prompts (refer [RQ2]), this similarity may mask differences in what the two models encode. To probe this, we apply variance partitioning to decompose the brain variance explained by Reasoning (R) and Action (A) prompts into shared and prompt-unique components, separately for each model. For this, we use standard variance partitioning approach
discussed in Section~\ref{experimental_setup}. Fig.~\ref{fig:variance_partition_wholebrain_atari} presents the shared variance between
action and reasoning prompts for the VLM and LAM models, averaged across all subjects.

As shown in Fig.~\ref{fig:variance_partition_wholebrain_atari}, for VLM, the two prompts contributed statistically indistinguishable unique variance (paired t-test p = 0.75). For action prompts, $u_{A}$ = 0.006 (12.4\% of explained variance). For reasoning prompts, $u_{R}$ = 0.004 (9.5\% of explained variance). This is over a large shared component (0.035, 78.1\% of explained variance), indicating prompt-symmetric representations. In contrast, the LAM exhibits a significant asymmetry. Action prompts contributed substantially more unique variance than Reasoning prompts ($u_{A}$ = 0.013 (25.6\%) vs $u_{R}$ = -0.004 (-8.2\%); p = 0.032), while the shared component remained high (0.042, 82.4\%). The LAM's unique-reasoning variance is significantly lower than the VLM's at the whole-brain level ($p<0.05$). This dissociation is regionally specific. In AG, VLM shows balanced prompt effects (8\% action vs. 10\% reasoning), whereas LAM shifts toward action-dominant representations, with $u_{A}$ = 24\% of variance and $u_{R}$ = $-8\%$. In SMA, the contrast is strongest: both VLM unique components are small (13\% action vs. 4\% reasoning), whereas the LAM devotes one-third of explainable variance to action ($u_{A}$ = 34\%, with $u_{R}$=$-14\%$).

\begin{figure}[t]
    \centering
    \includegraphics[width=\linewidth]
      {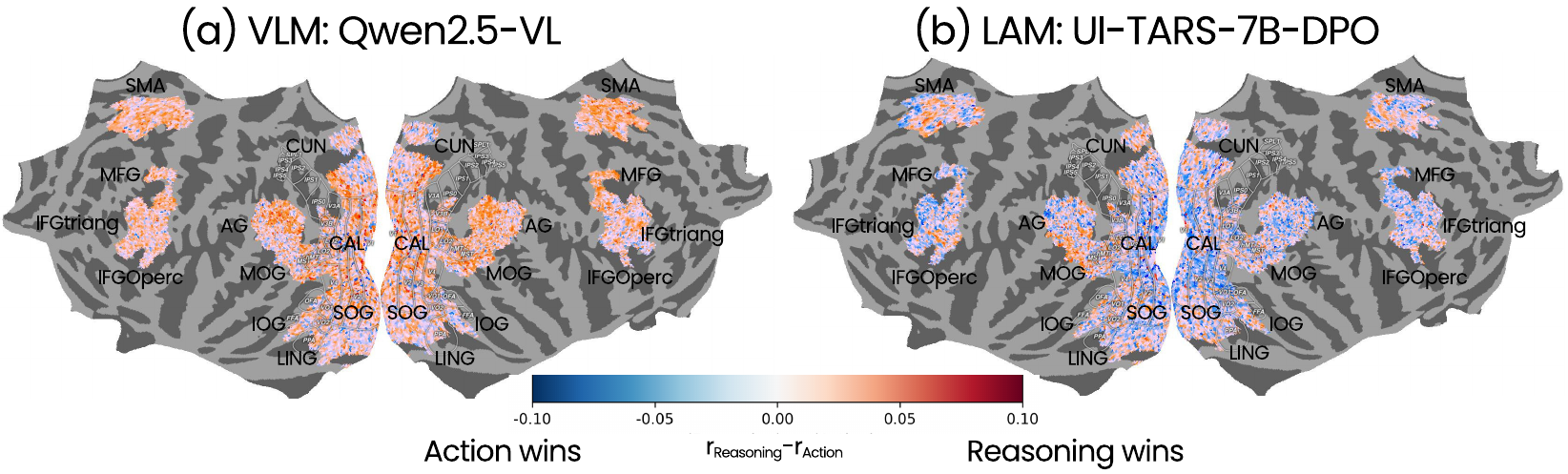}
    \caption{\textbf{Spatial visualization confirms the VLM/LAM dissociation.}
    Group-averaged across subjects per-voxel difference maps (r$_{Reasoning}$ - r$_{Action}$) for
    (a) VLM (Qwen2.5-VL): positive differences (red, r$_{Reasoning}$ $>$ r$_{Action}$) dominate dorsal stream and lateral occipital cortex, with localized negative differences in early visual cortex (V1, V2). (b) LAM (UI-TARS): the same difference map is substantially more balanced, with widespread negative differences (blue, r$_{Action}$ $>$ r$_{Reasoning}$) across early visual and ventral stream regions, alongside positive differences in dorsal regions. 
    }
    \label{fig:flatmap_groupmean_atari}
\end{figure}

Together, these results reveal a clear dissociation. VLM representations are prompt-symmetric (12.4\% vs. 9.5\% unique variance), whereas LAM representations are action-dominant (25.6\% vs. -8.2\% unique variance). This suggests that action-specialized fine-tuning reorganizes model representations toward action-associated cortical regions, despite similar overall predictive accuracy. Variance-partitioning results for all seven ROIs are reported in App.~\ref{app:variancePartition}, Fig.~\ref{fig:variance_partition_remain_rois_atari}.

\noindent\textbf{Spatial visualization confirms the VLM/LAM dissociation.} 
Group-averaged voxel-wise difference flatmaps across subjects ($r_{\text{Reasoning}}-r_{\text{Action}}$; Fig.~\ref{fig:flatmap_groupmean_atari}) show VLMs are \textcolor{red}{reasoning-dominant (red)}, with positive differences across dorsal stream and lateral occipital cortex and only localized Action-favoring regions in early visual cortex (Fig.~\ref{fig:flatmap_groupmean_atari}a). In contrast, LAMs exhibit widespread \textcolor{blue}{action-dominant (blue)} voxels across early visual and ventral regions, with only scattered Reasoning-favoring voxels dorsally (Fig.~\ref{fig:flatmap_groupmean_atari}b). These maps mirror the ROI-level prompt comparisons (Fig.~\ref{fig:vem_vlm_rois_atari}) and variance-partitioning (Fig.~\ref{fig:variance_partition_wholebrain_atari}) results, indicating prompt-symmetric VLM representations but action-leaning LAM representations despite comparable prediction accuracy.

\noindent\textbf{Generalization across VLM--LAM pairs.} To test whether the prompt-symmetric VLM vs action-dominant LAM pattern generalizes beyond the Qwen-VL family, we additionally ran variance partitioning on second pair InternVL3 (Qwen2.5-VL-based VLM) vs OS-Atlas-Pro (Qwen2-VL-based LAM). The qualitative pattern is preserved. At whole-brain, OS-Atlas shows substantially more $u_{A}$ variance than $u_{R}$ (0.0102 vs. 0.002), while InternVL3 is approximately prompt-symmetric (0.004 vs. 0.006). The shared component remained high in LAM vs. VLM (0.031, 0.037). ROI analyses (AG, SMA) show the same OS-Atlas-Pro action-dominance, with InternVL3 shifting slightly toward Reasoning-leaning representations in higher-order cortex. Detailed analysis is in App.~\ref{app:cross_family_vp}.

\noindent\textbf{Thinking-mode readouts.} In Qwen3.5, the last-token reasoning readout shows weaker brain alignment than the final-answer readout ($r=0.012$ vs.\ $0.031$; paired $t$-test $p=0.032$), but mean-pooling over the reasoning span raises alignment to $r=0.036$, numerically comparable to the final-answer readout, suggesting that the difference is sensitive to the readout method (App.~\ref{app:qwen35}).

\section{Discussion and Conclusion}
\label{sec:discussion}

We evaluate how VLMs and LAMs align with human brain activity during Atari-style gameplay using voxel-wise encoding on an fMRI dataset. Comparing two VLMs and two LAMs against RL baselines yields four key findings.
\textbf{First}, both VLMs and LAMs substantially outperform RL baselines, even after matching feature counts, showing that multimodal foundation models encode more brain-relevant information than traditional task-optimized RL agents.
\textbf{Second}, prompt-driven gains are strongest in higher-order frontal-parietal and motor-planning regions such as MFG, SMA, IFG, and AG, with smaller gains in early visual cortex. This suggests prompts improve representations related to goals, planning, and decision-making, likely reflecting multiple-demand network activity rather than language alone~\citep{duncan2010multiple,fedorenko2024language}.
\textbf{Third}, although VLMs and LAMs achieve similar whole-brain prediction accuracy, variance partitioning reveals distinct internal organization. VLMs show balanced contributions from Action and Reasoning prompts, whereas action-tuned LAMs are strongly action-dominant, especially in frontal-motor regions. This indicates that action fine-tuning already captures much of the information reasoning prompts provide.
\textbf{Fourth}, in a preliminary extension, Qwen3.5 reasoning-trace representations tend to align less with whole-cortex brain activity than its final-answer representations ($r=0.012$ vs.\ $0.031$ for last-token readouts), suggesting that explicit chain-of-thought reasoning does not necessarily improve brain alignment.
Together, these findings extend brain alignment research from passive perception to interactive gameplay and uncover representational differences between VLMs and action-tuned models that emerge only through variance decomposition. 
We discuss limitations in App.~\ref{app:limitations}.

\bibliography{neurips_2026}
\bibliographystyle{iclr2027_conference}

\appendix
\noindent{\Large\textbf{Overview of Appendix Sections}}
\begin{itemize}
\item App.~\ref{app:relatedwork}: Related Work
\item App.~\ref{app:preprocessing}: fMRI Preprocessing
\item App.~\ref{app:detailedsubrois}: Detailed sub-ROIs of frontal-parietal, visual, and motor regions
\item App.~\ref{app:baselines}: Baseline Models Features
\item App.~\ref{app:prompttemplates}: Prompt Templates for Brain Encoding Models
\item App.~\ref{app:hyperparameters_details}: Hyperparameter Details  
\item App.~\ref{app:sharedVarianceDetails}: Details of explained variance partitioning
\item App.~\ref{app:statistical_significance}: Statistical significance  
\item App.~\ref{app:feature_extraction}: Detailed Feature Extraction  
\item App.~\ref{app:behavior}: Behavioral Analysis
\item App.~\ref{app:vem_generalization}: Generalization across model families
\item App.~\ref{app:promptcontrols}: Prompt Controls
\item App.~\ref{app:lamGainsScale}: VLM and LAM: Prompt-driven alignment gains across ROIs
\item App.~\ref{app:variancePartition}: Variance Partitioning Results
\item App.~\ref{app:qwen35}: Alignment using generated reasoning traces of thinking-mode models.
\item App.~\ref{app:limitations}: Limitations
\end{itemize}

\section{Related Work}
\label{app:relatedwork}

\noindent\textbf{Foundation models and brain alignment.}
Our work also relates to a growing literature that investigates the alignment between human brains and language models. A number of studies have used text-based language models have been shown to 
predict both text- and speech-evoked brain responses with high fidelity~\citep{jain2018incorporating,deniz2019representation,
toneva2019interpreting,caucheteux2020language,antonello2021low,
oota2022neural,oota2022joint}. Advances in Transformer-based speech 
models have similarly motivated work on speech-evoked brain alignment~\citep{millet2022toward,vaidya2022self,tuckute2022many,
oota2024speech,chen2023cortical}. More recently, 
multimodal Transformers have been tested under both unimodal~\citep{
dong2023interpreting,oota2022visio,oota2025correlating} and multimodal naturalistic 
stimuli~\citep{dong2023vision,nakagi2024brain,oota2024multi}. Our work is complementary: we apply 
recent foundation models, specifically vision-language and large-action models, to study brain alignment during interactive Atari-style gameplay, with a particular focus on how action and reasoning prompt-conditioned representations differ across the cortical hierarchy.

\noindent\textbf{Reinforcement learning, world models, and brain alignment during interactive behavior.}
A complementary line of work has aligned brain activity recorded during interactive gameplay with deep reinforcement-learning (RL) agents and 
theory-based models~\citep{tomov2023neural,haberland2026encoding}. Notably,~\citet{tomov2023neural} introduced the naturalistic Atari-style fMRI dataset and find that theory-based RL representations~\citep{tsividis2021human} in prefrontal cortex, showing 
that EMPA outperforms model-free baselines such as 
DDQN~\citep{van2016deep} in predicting frontal activity during gameplay. These studies demonstrate that features from RL agents map onto neural activity associated with decision-making and goal-directed control. However, because RL agents are optimized primarily for reward-based action learning, their representations tend to encode 
action-relevant features and may miss the richer perceptual, semantic, and predictive structure humans use to represent objects, dynamics, and 
future outcomes during gameplay. 

World models, structured internal representations of the environment that support prediction and 
planning, were originally proposed in cognitive 
science~\citep{Ritchie_1944,johnson1983mental} and later formalized in deep-RL~\citep{ha2018recurrent}. Recent work has begun probing whether modern foundation models, particularly VLMs and LAMs, acquire analogous internal world models when interacting with games and GUIs~\citep{waytowich2024atari,wang2025ui,xie2025play}, but these evaluations focus on behavioral benchmarks (task success, generalization) 
rather than neural alignment. Our work bridges these two lines: we systematically evaluate the brain alignment of modern VLMs and LAMs on 
the same Atari-style fMRI dataset previously used to study theory-based RL~\citep{tomov2023neural}, comparing them directly against EMPA and 
DDQN baselines and probing how reasoning- and action-focused prompts 
shape brain-relevant representations across the cortical hierarchy.


\section{fMRI Preprocessing}
\label{app:preprocessing}

\noindent\textbf{Preprocessing pipeline.}
We preprocess the publicly available raw BIDS-formatted fMRI data using fMRIPrep v25.2.0~\citep{esteban2019fmriprep,esteban2020analysis}. This performs the same core preprocessing stages as the original SPM12-based analysis of~\citet{tomov2023neural} such as, head-motion correction, coregistration, and normalization to MNI space, using a standardized, validated pipeline. We subsequently applied spatial smoothing with an 8-mm full-width-at-half-maximum (FWHM) Gaussian kernel, and included the estimated motion parameters as nuisance regressors in the encoding analysis.

\noindent\textbf{Nuisance regression.}
Before surface projection, each MNI-space functional run is denoised using \texttt{nilearn.image.clean\_img} with 14 nuisance regressors: the six rigid-body motion parameters, their six temporal derivatives, and mean cerebrospinal-fluid and white-matter signals obtained from fMRIPrep. Confound regression is performed jointly with linear detrending and high-pass filtering at (1/128) Hz (($\mathrm{TR}=2$) s). This procedure controls for residual motion-related and physiological variance beyond the initial head-motion correction. The same denoised time series is used for all VLM, LAM, EMPA, and DDQN encoding models.

\noindent\textbf{Surface projection and parcellation.}
The denoised MNI-space functional data are projected to the \texttt{fsaverage6} cortical surface using \texttt{nilearn.surface.vol\_to\_surf}~\citep{Nilearn}. The pial and white-matter surfaces define depth-based sampling across the cortical ribbon, yielding 40,962 vertices per hemisphere. We remove the first and last five TRs from each run, leaving 273 TRs per run, and apply an \texttt{fsaverage6}-compatible Glasser parcellation for the ROI analyses.

\noindent\textbf{Additional confound controls.}
We address game identity and temporal train-test leakage using complementary cross-validation scheme. Leave-one-run-out cross-validation evaluation holds out complete scanner runs and eliminates within-run adjacency between training and test TRs. Input windows are constructed separately within each run and never cross run boundaries.

We do not include separate button-press or task-specific visual-event regressors identical to those used by~\citet{tomov2023neural}; therefore, we cannot completely exclude contributions from motor execution or task-specific visual events. As a complementary control, we evaluate low-level color-statistics and object/spatial scene-description prompts, which yield lower alignment than task-framed prompts containing goal and state-conditioned action information. However, these prompt controls do not replace explicit button-press regression. We therefore interpret frontal-motor findings as representational alignment and not as direct evidence of motor planning or execution.

\noindent\textbf{Temporal autocorrelation control (AR(1) prewhitening).}
To test whether the advantage of VLM and LAM features over EMPA could be driven by temporal autocorrelation in the fMRI time series, we repeat the unthresholded whole-cortex comparison after AR(1) prewhitening (Table~\ref{tab:ar1_control}). At all five layers tested (L4, L11, L18, L21, and L28), both VLM and LAM features retain higher alignment than EMPA, with paired gaps of $+0.0060$ to $+0.0088$ for the VLM and $+0.0084$ to $+0.0109$ for the LAM; all comparisons remain significant after Bonferroni correction (all corrected $p<0.01$).

\begin{table}[!ht]
\centering
\small
\setlength{\tabcolsep}{4pt}
\caption{\textbf{VLM and LAM vs.\ EMPA after AR(1) prewhitening.}
Unthresholded whole-cortex alignment (all vertices) at five layers.
The common EMPA mean is $r=0.0241$.
The paired gap is $\Delta r=r_{\mathrm{model}}-r_{\mathrm{EMPA}}$.
The last column reports Bonferroni-corrected $p$-values.}
\label{tab:ar1_control}
\begin{tabular}{|l|c|c|c|c|c|}
\hline
\textbf{Model} & \textbf{Layer} & \textbf{Model} $r$
& \textbf{Gap} $\Delta r$ & \textbf{Paired} $p$
& \textbf{Bonferroni} $p$ \\
\hline
VLM & L4  & $0.0329\pm0.0013$ & $+0.0088\pm0.0025$ & $0.00035$ & $0.00173$ \\
VLM & L11 & $0.0301\pm0.0012$ & $+0.0060\pm0.0020$ & $0.00073$ & $0.00366$ \\
VLM & L18 & $0.0311\pm0.0021$ & $+0.0070\pm0.0025$ & $0.00101$ & $0.00503$ \\
VLM & L21 & $0.0306\pm0.0013$ & $+0.0065\pm0.0022$ & $0.00079$ & $0.00393$ \\
VLM & L28 & $0.0316\pm0.0020$ & $+0.0075\pm0.0025$ & $0.00073$ & $0.00366$ \\
\hline
LAM & L4  & $0.0340\pm0.0013$ & $+0.0099\pm0.0030$ & $0.00047$ & $0.00470$ \\
LAM & L11 & $0.0325\pm0.0012$ & $+0.0084\pm0.0028$ & $0.00073$ & $0.00732$ \\
LAM & L18 & $0.0350\pm0.0014$ & $+0.0109\pm0.0032$ & $0.00040$ & $0.00405$ \\
LAM & L21 & $0.0335\pm0.0013$ & $+0.0094\pm0.0030$ & $0.00060$ & $0.00598$ \\
LAM & L28 & $0.0345\pm0.0020$ & $+0.0104\pm0.0032$ & $0.00050$ & $0.00504$ \\
\hline
\end{tabular}
\end{table}

\section{Detailed sub-ROIs of frontal-parietal, visual, and motor regions}
\label{app:detailedsubrois}
\begin{figure}[!ht]
    \centering
    \includegraphics[width=\linewidth]{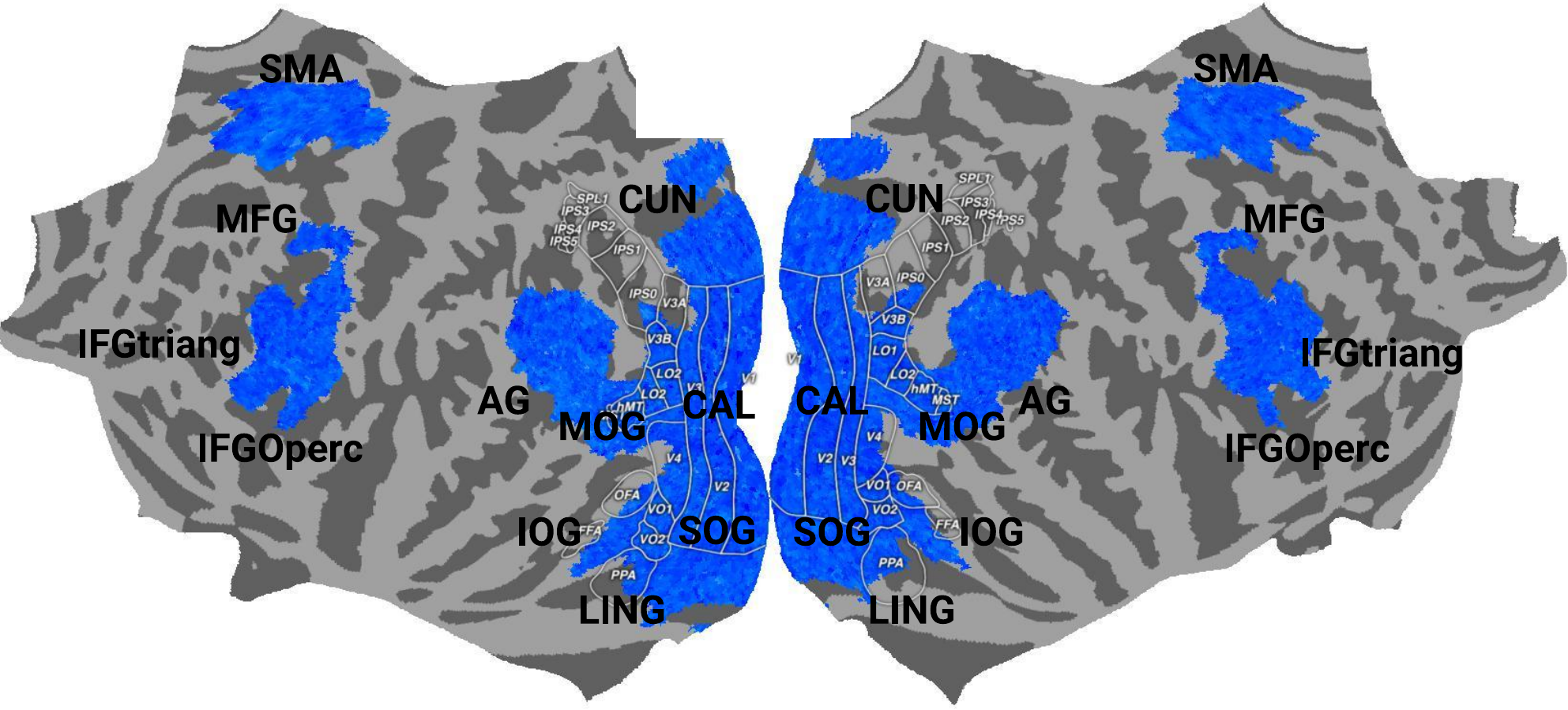}
    \caption{Flattened cortical surfaces for
frontal-parietal, visual- and motor-selective regions displayed on the `fsaverage' surface, used as the mask for all participants.}
    \label{fig:language_flatmap}
\end{figure}

The data covers eleven brain regions of interest (ROIs) in the human brain with the following sub-divisions: (i) angular gyrus (AG):  PFm, PGs, PGi, TPOJ2, TPOJ3, (ii) inferior frontal gyrus triangularis (IFGtriang): 44, 45, IFJa, IFSp, (iii) inferior frontal gyrus opercularis (IFGoperc): 6r, 6v, FOP1, FOP4, (iv) middle frontal gyrus (MFG): 55b, (v) inferior occipital gyrus (IOG): V8, VVC, VMV3, VMV2, VMV1, (vi) middle occipital gyrus (MOG): LO1, LO2, LO3, V3CD, MT, MST, FST, V4t, (vii) superior occipital gyrus (SOG): V7, V3a, V3b, V3cd, (viii) calcarine fissure (CAL): V1, V2, (ix) lingual gyrus (LING): VMV1, VMV2, VMV3, (x) cuneus (CUN): V3, POS1, V6,  and (xi) supplementary motor area (SMA): SFL, 6ma, 6mp, SCEF~\citep{baker2018connectomic,milton2021parcellation,desai2022proper}.
Fig.~\ref{fig:language_flatmap} shows flattened cortical surfaces for
frontal-parietal, visual- and motor-selective regions displayed on the `fsaverage' surface, used as the mask for all participants.

\section{Baseline Models Features}
\label{app:baselines}
We derived TR-aligned regressors from two computational models of Atari gameplay. For EMPA~\citep{tomov2023neural}, we extracted eight per-frame regressors. Four are continuous: 
\emph{surprise} (the agent's prediction error on the observed next state), \emph{spriteKL} (KL divergence between the agent's prior and updated 
beliefs about sprite dynamics, indexing belief revision over object behavior), \emph{R\_GG} (expected reward under the current goal-graph theory), and \emph{num\_effects} (count of distinct interaction effects observed at the current timestep). Four are binary indicators of theory revision: \emph{replan\_flag} (the agent re-planned its policy), \emph{theory\_change\_flag} (the agent's theory of game rules changed), \emph{newEffects\_flag} (a previously-unseen interaction effect was observed), and \emph{interaction\_change\_flag} (beliefs about an interaction between two object types changed).  
Concatenating these episodes within each level in instance order and averaging frame values within each TR over the level's expected start/end window, then trimming 5 TRs from each run edge to match fMRI preprocessing, yielding a (T, 8) feature matrix per subject. For DDQN (25M-step agent)~\citep{van2016deep}, we loaded dqn\_regressors\_25M.bson and extracted per-episode action sequences as (keycode, frame\_idx, timestamp) tuples, mapping pygame keycodes to action labels and computing normalized action distributions per run for behavioral comparison against human and EMPA policies.

\section{Prompt Templates for Brain Encoding Models}
\label{app:prompttemplates}



\subsubsection*{Action Prompt}
\begin{tcolorbox}[colback=green!5]
\textbf{Action Plan:} \\
``You are controlling the avatar in this game. \\
Look at the current game state. \\
What is your immediate next action and why? \\
Consider threats, goals, and optimal path.''
\end{tcolorbox}

\subsubsection*{Reasoning Prompt}
\begin{tcolorbox}[colback=purple!5]
\textbf{Reasoning:} \\
``Analyze this game step by step: \\
What objects do you see? \\
What are the spatial relationships? \\
What are the possible threats or goals? \\
What action would you recommend and why?''
\end{tcolorbox}

\subsubsection*{Action Prompt (full)}
\begin{tcolorbox}[colback=green!5]
\textbf{Motor commitment $+$ justification:} \\
``You are controlling the avatar in this game. \\
Choose the movement key to press right now \\
and explain why it beats the other options open to you. \\
Answer briefly.''
\end{tcolorbox}

\subsubsection*{Action Prompt (bare)}
\begin{tcolorbox}[colback=green!5]
\textbf{Motor commitment, no justification:} \\
``You are controlling the avatar in this game. \\
Name the single movement key you would press right now. \\
Give the key and nothing else at all. \\
Answer briefly.''
\end{tcolorbox}

\subsubsection*{Goal Prompt}
\begin{tcolorbox}[colback=blue!5]
\textbf{Hazards/objectives, no motor commitment:} \\
``You are looking at a frame from a game. \\
Identify which visible things are hazards that threaten the avatar \\
and which are objectives it must reach. \\
Answer briefly.''
\end{tcolorbox}

\subsubsection*{Scene Prompt (baseline)}
\begin{tcolorbox}[colback=gray!5]
\textbf{Object semantics $+$ spatial relations, no goals:} \\
``You are looking at a frame from a game. \\
Describe the objects that are visible in it \\
and where each one sits relative to the others around it. \\
Answer briefly.''
\end{tcolorbox}

\subsubsection*{Low-level Prompt}
\begin{tcolorbox}[colback=gray!5]
\textbf{Image-referring task, no game semantics or goals:} \\
``You are looking at a still image. \\
Report which colours cover the largest areas of the picture \\
and roughly what fraction of it each one covers. \\
Answer briefly.''
\end{tcolorbox}

\subsubsection*{Scrambled Prompt (task-lexicon control)}
\begin{tcolorbox}[colback=gray!5]
\textbf{Text present, syntax destroyed, task lexicon retained:} \\
``and this are press controlling to key options game the beats movement \\
why now the you in open right the explain You avatar it Choose to other. \\
Answer briefly.''
\end{tcolorbox}

\subsubsection*{Scrambled Prompt (neutral control)}
\begin{tcolorbox}[colback=gray!5]
\textbf{Text present, syntax destroyed, task-irrelevant lexicon:} \\
``a volumes room quiet library the and where bound shelves a cloth \\
spring windows has narrow fountain old since reading faded face The \\
hold courtyard in stood stone. \\
Answer briefly.''
\end{tcolorbox}



\subsubsection*{Qwen3.5 Prompt}
\label{qwen3.5_prompt}
\begin{tcolorbox}[colback=purple!5, sharp corners]
\textbf{Reasoning prompt (with thinking enabled):} \\
"Analyze this game frame step by step: \\
What objects/entities do you see? \\
What are the spatial relationships? \\
What are the possible threats or goals? \\
What action would you recommend and why? \\
\
IMPORTANT: Be brief. Keep your private reasoning under 800 tokens, then end the reasoning block and give a one-line final answer."
\end{tcolorbox}

\begin{table}[!ht]
\centering
\small
\caption{Prompt battery used to isolate the effects of text presence, visual and goal information, and motor commitment.}
\label{tab:prompt_battery}
\setlength{\tabcolsep}{4pt}
\begin{tabular}{|@{}p{0.23\linewidth}|p{0.21\linewidth}|p{0.46\linewidth}@{}|}
\hline
\textbf{Condition} & \textbf{Axis} & \textbf{Isolates} \\
\hline
\texttt{action}
& Primary prompt
& Original action prompt \\

\texttt{reasoning}
& Primary prompt
& Original reasoning prompt \\

\texttt{scrambled-neutral}
& Text presence
& Text present, syntax destroyed, and task-irrelevant lexicon \\

\texttt{scrambled}
& Text presence
& Text present, syntax destroyed, and task lexicon retained \\

\texttt{lowlevel}
& Visual vs.\ goal
& Colour statistics only, without game semantics \\

\texttt{scene}
& Visual vs.\ goal
& Object semantics and spatial relationships, without goals \\

\texttt{goal}
& Action vs.\ reasoning
& Hazards and objectives, without motor commitment \\

\texttt{action-bare}
& Action vs.\ reasoning
& Motor commitment only: name the movement key without justification \\

\texttt{action-full}
& Action vs.\ reasoning
& Motor commitment with justification relative to alternative actions \\
\hline
\end{tabular}
\end{table}

\noindent\textbf{Prompt-dependent changes in model representations.}
We next examined whether the prompt conditions altered the model representations independently of their relationship with brain activity. For each condition, we constructed a representation matrix $\mathbf{X}\in\mathbb{R}^{n\times d}$ from the same $n=126$ matched game levels. We decomposed its total energy into a condition-mean component,
$n\lVert\boldsymbol{\mu}\rVert_2^2/\lVert\mathbf{X}\rVert_F^2$, and a stimulus-varying component,
$\lVert\mathbf{X}-\mathbf{1}\boldsymbol{\mu}^{\top}\rVert_F^2/\lVert\mathbf{X}\rVert_F^2$.
We then used linear centered kernel alignment (CKA) to compare representational geometry across prompt conditions. Uncentered CKA retains the condition means, whereas centered CKA compares only the stimulus-varying components.

For the VLM, the condition mean accounted for $48.2\%$--$97.3\%$ of the total representational energy, depending on the prompt. Consequently, uncentered cross-condition CKA was uniformly high ($0.857$--$0.995$). After removing each condition mean, however, cross-condition CKA decreased to $0.028$--$0.173$. A similar pattern was observed for the LAM, for which centered cross-condition CKA ranged from $0.050$ to $0.251$. Scene-based prompts retained the largest stimulus-varying component in the VLM ($51.8\%$), whereas action-oriented representations were generally more mean-dominated; this was most pronounced for the LAM action-bare condition, for which the condition mean accounted for $98.4\%$ of the total energy. These results show descriptively that prompts affect not only global representational offsets but also the geometry of stimulus-dependent variation. This analysis characterizes the model representations themselves and does not establish their correspondence with neural activity.

\begin{table}[!ht]
\centering
\caption{\textbf{Energy decomposition of VLM representations across prompt
conditions.} Values indicate the percentage of total embedding energy
attributable to the condition-specific mean and to the remaining
stimulus-varying component across 126 matched game levels.}
\label{tab:vlm_embedding_energy}
\scriptsize
\setlength{\tabcolsep}{5pt}
\begin{tabular}{|l|c|c|}
\hline
\textbf{Condition} &
\textbf{Condition mean (\%)} &
\textbf{Stimulus-varying (\%)} \\
\hline
Action             & 97.3 &  2.7 \\
Reasoning          & 72.2 & 27.8 \\
Action-bare        & 89.4 & 10.6 \\
Action-full        & 60.2 & 39.8 \\
Goal               & 89.5 & 10.5 \\
Scene              & 48.2 & 51.8 \\
Low-level          & 60.9 & 39.1 \\
Scrambled          & 60.8 & 39.2 \\
Scrambled-neutral  & 79.1 & 20.9 \\
\hline
\end{tabular}
\end{table}

\begin{table*}[!ht]
\centering
\caption{\textbf{Uncentered linear CKA between VLM representations.}
Condition means are retained in this analysis. Uncentered CKA is uniformly high across conditions, consistent with the strongly mean-dominated representations reported in Table~\ref{tab:vlm_embedding_energy}. Abbreviations: Act., Action; Reas., Reasoning; A-bare, Action-bare; A-full, Action-full; Low, Low-level; Scr., Scrambled; and Scr.-N, Scrambled-neutral.}
\label{tab:vlm_cka_uncentered}
\scriptsize
\setlength{\tabcolsep}{3.2pt}
\begin{tabular}{|l|c|c|c|c|c|c|c|c|c|}
\hline
 & \textbf{Act.} & \textbf{Reas.} & \textbf{A-bare} &
 \textbf{A-full} & \textbf{Goal} & \textbf{Scene} &
 \textbf{Low} & \textbf{Scr.} & \textbf{Scr.-N} \\
\hline
Action
 & 1.000 & 0.924 & 0.995 & 0.934 & 0.988 & 0.965 & 0.966 & 0.963 & 0.973 \\
Reasoning
 & 0.924 & 1.000 & 0.925 & 0.857 & 0.913 & 0.899 & 0.895 & 0.891 & 0.909 \\
Action-bare
 & 0.995 & 0.925 & 1.000 & 0.931 & 0.984 & 0.963 & 0.964 & 0.960 & 0.971 \\
Action-full
 & 0.934 & 0.857 & 0.931 & 1.000 & 0.921 & 0.908 & 0.910 & 0.913 & 0.913 \\
Goal
 & 0.988 & 0.913 & 0.984 & 0.921 & 1.000 & 0.953 & 0.955 & 0.951 & 0.961 \\
Scene
 & 0.965 & 0.899 & 0.963 & 0.908 & 0.953 & 1.000 & 0.936 & 0.936 & 0.943 \\
Low-level
 & 0.966 & 0.895 & 0.964 & 0.910 & 0.955 & 0.936 & 1.000 & 0.939 & 0.943 \\
Scrambled
 & 0.963 & 0.891 & 0.960 & 0.913 & 0.951 & 0.936 & 0.939 & 1.000 & 0.938 \\
Scrambled-neutral
 & 0.973 & 0.909 & 0.971 & 0.913 & 0.961 & 0.943 & 0.943 & 0.938 & 1.000 \\
\hline
\end{tabular}%
\end{table*}

\begin{table*}[!ht]
\centering
\caption{\textbf{Centered linear CKA between the stimulus-varying components of VLM representations.} Each condition-specific mean was removed before computing CKA. The low cross-condition similarities indicate that different prompts yield distinct stimulus-varying representational geometries. Abbreviations follow Table~\ref{tab:vlm_cka_uncentered}.}
\label{tab:vlm_cka_centered}
\scriptsize
\begin{tabular}{|l|c|c|c|c|c|c|c|c|c|}
\hline
 & \textbf{Act.} & \textbf{Reas.} & \textbf{A-bare} &
 \textbf{A-full} & \textbf{Goal} & \textbf{Scene} &
 \textbf{Low} & \textbf{Scr.} & \textbf{Scr.-N} \\
\hline
Action
 & 1.000 & 0.053 & 0.071 & 0.058 & 0.087 & 0.108 & 0.077 & 0.076 & 0.042 \\
Reasoning
 & 0.053 & 1.000 & 0.046 & 0.029 & 0.041 & 0.099 & 0.061 & 0.047 & 0.028 \\
Action-bare
 & 0.071 & 0.046 & 1.000 & 0.045 & 0.028 & 0.074 & 0.051 & 0.054 & 0.036 \\
Action-full
 & 0.058 & 0.029 & 0.045 & 1.000 & 0.045 & 0.124 & 0.074 & 0.143 & 0.066 \\
Goal
 & 0.087 & 0.041 & 0.028 & 0.045 & 1.000 & 0.141 & 0.095 & 0.073 & 0.034 \\
Scene
 & 0.108 & 0.099 & 0.074 & 0.124 & 0.141 & 1.000 & 0.170 & 0.173 & 0.085 \\
Low-level
 & 0.077 & 0.061 & 0.051 & 0.074 & 0.095 & 0.170 & 1.000 & 0.129 & 0.036 \\
Scrambled
 & 0.076 & 0.047 & 0.054 & 0.143 & 0.073 & 0.173 & 0.129 & 1.000 & 0.070 \\
Scrambled-neutral
 & 0.042 & 0.028 & 0.036 & 0.066 & 0.034 & 0.085 & 0.036 & 0.070 & 1.000 \\
\hline
\end{tabular}%
\end{table*}



\section{Hyperparameter Details}
\label{app:hyperparameters_details}

We use the HuggingFace \texttt{transformers} interface to load the pretrained VLM and LAM models and their corresponding processors, and use 40-core GPU with 128GB-RAM with half precision (FP16) as compute.

We used bootstrap ridge-regression with the following parameters: MSE loss function, and L2-decay ($\lambda$) varied from  10$^{1}$ to 10$^{3}$; best $\lambda$ was chosen by tuning on validation data that comprised a randomly chosen 10\% subset from train set used only for hyper-parameter tuning.

\section{Details of explained variance partitioning}
\label{app:sharedVarianceDetails}

Using a variance partitioning approach~\citep{de2017hierarchical,lebel2021voxelwise}, we test whether action- and reasoning-based prompts carry overlapping or distinct brain-relevant information, and whether LAMs, which are fine-tuned on action-specific environments reshapes this decomposition. In particular, we decomposed each model's joint $R^{2}$ into three components: variance shared between action and reasoning prompts, variance uniquely explained by action prompt, and variance uniquely explained by reasoning prompt~\citep{vaidya2022self,deniz2019representation}. 
For each subject, we fit three voxel-wise encoding models: one using the concatenated action and reasoning representations as joint predictors (yielding $R^{2}_{joint}$), one using only action 
representations ($R^{2}_{action}$), and one using only reasoning representations ($R^{2}_{reasoning}$). We then computed each component's contribution following standard variance-partitioning formulations: $R^{2}_{joint} - R^{2}_{reasoning}$ for unique action, $R^{2}_{joint} - R^{2}_{action}$ for unique reasoning, and $R^{2}_{action} + R^{2}_{reasoning} - R^{2}_{joint}$ for shared 
variance. 
We performed this analysis separately for two VLMs (Qwen2.5-VL, and InternVL3-8B) and  LAMs (UI-TARS-7B-DPO and OS-Atlas-Pro-7B) 
on all 32 subjects, and evaluated each component at both whole-brain and ROI levels.

\section{Statistical significance}
\label{app:statistical_significance}
To determine whether prediction accuracy is significantly higher than chance, we use block permutation tests. We employ the standard implementation of a block permutation test for fMRI data, which is to split the fMRI data into blocks of 10 contiguous TRs and permute the order of these blocks, while maintaining the original order of the TRs within each block. 
By permuting predictions 5000 times, we create an empirical distribution for chance performance, from which we estimate the p-value of the actual performance.
To estimate the statistical significance of performance differences, such as between the model's predictions and chance, we utilized the Wilcoxon signed-rank test, applying it to the mean correlations for the participants.
To compare models or prompt conditions, we use paired $t$-tests across the 32 participants and additionally report Wilcoxon signed-rank tests where indicated. For ROI-wise comparisons, $p$-values are corrected using the Benjamini--Hochberg false discovery rate (FDR) procedure, and $q$ denotes the adjusted $p$-value. In figures, significance is marked as * ($<0.05$), ** ($<0.01$), and *** ($<0.001$), with $\sim$ denoting marginal effects ($p<0.10$).

\section{Detailed Feature Extraction}
\label{app:feature_extraction}

Let $\mathbf{x}_{t,j}$ denote the $j$-th RGB game frame sampled for fMRI TR $t$ (converted to a PIL image), and let $p$ denote the text prompt (e.g., action or reasoning; empty in the \emph{no-prompt} condition). Rather than conditioning on a single frame, we give the model a short multi-frame window for each TR. The game runs at ${\sim}20$ FPS, so each 2-s TR spans ${\sim}40$ frames; we sample $k=4$ frames evenly ${\sim}10$ frames (${\approx}0.5$\,s) apart, ending at the last frame of the TR:
\begin{equation}
\mathbf{X}_t = \big(\mathbf{x}_{t,1}, \mathbf{x}_{t,2}, \mathbf{x}_{t,3}, \mathbf{x}_{t,4}\big).
\end{equation}
Frames are taken from each participant's own gameplay, and windows are constructed within each run, never crossing run boundaries. We build a chat-style input in which the $k$ frames are passed as an ordered sequence of image items followed by the prompt:
\begin{equation}
\mathrm{input}_t = \big[\texttt{<image>}{=}\mathbf{x}_{t,1};\,\dots;\,\texttt{<image>}{=}\mathbf{x}_{t,4};\,\texttt{<text>}{=}p\big],
\end{equation}
which is serialized with each model's chat template (with the generation marker added) and tokenized by the model-specific processor. This uses each model's native multi-image input and gives the model access to short-term motion cues rather than a single static observation.

\paragraph{Generation and hidden states.}
We run \texttt{model.generate()} with greedy decoding, and hidden-state outputs enabled. Let $L$ denote the number of transformer blocks and $d$ the hidden dimension. At each decoding step, the model returns hidden states for all $L+1$ layers (embedding layer and $L$ transformer blocks). We take the hidden states at the final decoding step, i.e., the representation from which the last output token is predicted, excluding the end-of-sequence token. Because generation length varies across TRs, this readout is taken at a variable position. For layer $\ell\in\{0,\dots,L\}$, let $\mathbf{e}_t^{(\ell)}\in\mathbb{R}^{d}$ denote this final-step hidden state. Because of causal self-attention over the full context, $\mathbf{e}_t^{(\ell)}$ aggregates information from all $k$ frames, the prompt, and the generated response. We concatenate the per-layer vectors into
\begin{equation}
\mathbf{E}_t = \big[\mathbf{e}_t^{(0)};\,\mathbf{e}_t^{(1)};\,\dots;\,\mathbf{e}_t^{(L)}\big] \in \mathbb{R}^{(L+1)\times d}.
\end{equation}

\paragraph{Run-level features.}
For a run with $T$ TRs, we compute one feature tensor per TR and stack them into
\begin{equation}
\mathbf{E} = \{\mathbf{E}_t\}_{t=1}^{T} \in \mathbb{R}^{T\times(L+1)\times d}.
\end{equation}
These tensors serve as predictors for the voxel-wise encoding models, enabling layer-wise analyses and placing VLM and LAM features on the same TR-level footing as the EMPA regressors.

\section{Behavioral Analysis}
\label{app:behavior}

\noindent\textbf{Model--human action agreement.} Gameplay logs are resampled at $2$\,Hz (every tenth frame of the ${\approx}20$\,Hz logs), so agreement is evaluated at a finer temporal resolution than the TR-level encoding features. At each sampled frame, we give the model the action prompt (App.~\ref{app:prompttemplates}), which asks for its immediate next action, together with a four-frame window constructed as in the encoding analysis but ending at that frame. We extract one action from the generated text with a parser fixed before inspecting any model output: a closed alias table maps 30 surface forms onto the six canonical actions (\texttt{LEFT}, \texttt{RIGHT}, \texttt{UP}, \texttt{DOWN}, \texttt{SPACEBAR}, \texttt{NOOP}), including semantic variants (``fire''/``shoot''$\rightarrow$\texttt{SPACEBAR}; ``do nothing''/``wait''$\rightarrow$\texttt{NOOP}). A bare action word in first position takes precedence; otherwise, the earliest whole-word alias in the response is used. Responses containing no alias are treated as parse failures and excluded from the agreement computation rather than scored as disagreement; parse-failure rates ($0$--$2.3\%$) are reported in Table~\ref{tab:model_human_action_agreement}.

The participant's action at a sampled frame is the key held at that timestamp (each key-down remains active until the next, and the final one for $1$\,s); if none is held, we use the nearest key-down within $0.5$\,s, and otherwise drop the frame. Agreement is the fraction of matched frames with identical actions, computed per participant and game, and compared with two baselines: uniform chance over the actions the participant used ($1/k$, $k=4$--$5$) and the majority-class rate (always predicting the participant's most frequent action). Because action distributions are skewed toward no-ops and repeated moves, we report Cohen's $\kappa$, macro-$F_1$, and balanced accuracy as primary statistics (Table~\ref{tab:model_human_action_agreement}); raw accuracy against uniform chance overstates agreement and is included only for comparability.

\begin{table}[!ht]
\centering
\small
\caption{Model--human action agreement across games (mean $\pm$ SD across 32 participants). Chance denotes the uniform-chance level for the available actions in each game.}
\label{tab:model_human_action_agreement}
\begin{tabular}{|l|c|c|c|}
\hline
\textbf{Game} & \textbf{Chance} &
\textbf{VLM agreement} & \textbf{LAM agreement} \\
\hline
bait          & $25.0\%$ & $24.5\pm9.1\%$  & $27.2\pm8.3\%$  \\
chase         & $25.0\%$ & $21.9\pm14.3\%$ & $30.5\pm13.9\%$ \\
helper        & $25.0\%$ & $28.1\pm12.4\%$ & $31.3\pm13.0\%$ \\
lemmings      & $20.0\%$ & $29.7\pm6.3\%$  & $26.9\pm5.7\%$  \\
plaqueAttack  & $20.0\%$ & $20.4\pm1.0\%$  & $17.7\pm3.1\%$  \\
zelda         & $20.0\%$ & $35.2\pm12.3\%$ & $36.1\pm13.5\%$ \\
\hline
\end{tabular}
\end{table}

\noindent\textbf{Model--human action agreement.}
To contextualize the representational findings with participants' behavior, we compare the actions recommended by the VLM and LAM with the actions performed by each participant. As shown in Table~\ref{tab:model_human_action_agreement}, agreement is modest and varies across games. Both model families are numerically above chance for \emph{helper}, \emph{lemmings}, and \emph{zelda}, with the highest agreement observed for \emph{zelda}. However, several model--game combinations remain near or below chance, including the VLM on \emph{bait} and \emph{chase} and the LAM on \emph{plaqueAttack}. Overall, neither model family shows a consistent advantage across games. Moreover, because the models process observations offline without executing actions or receiving subsequent states and rewards, closed-loop measures such as reward, win rate, and task success are not defined in the present setting. These results therefore provide behavioral context for the neural alignment findings but do not establish that either model reproduces the participants' action policy.

\noindent\textbf{Action decoding.} We test whether participants' actions can be decoded from model representations during gameplay. For each TR, the target is the participant's action in that TR, defined as the most frequent key within the TR (a TR-level analogue of the frame-level definition used for agreement). The predictors are the same TR-level features used in the encoding analysis: the final-step hidden states of the action-prompt generation at the layer selected for encoding (VLM L18, LAM L21); for EMPA, we use its eight TR-level regressors. We train an L2-regularized logistic regression, with regularization strength selected by inner cross-validation, per participant with leave-one-run-out cross-validation, so that no test TR shares a run with the training data. Because the action distribution is imbalanced, we report balanced accuracy as the primary metric, with chance defined per participant as $1/k$ for the $k$ actions present in that participant's data (mean chance = 0.167) (Table~\ref{tab:action_decoding}).

\noindent\textbf{Decoding participants' actions from model representations.}
We next test whether the model representations contain information predictive of participants' actions. Because the model representations are constant within each game level, the prediction target is each participant's dominant action for that level; this is therefore a level-wise rather than TR-wise decoding analysis. Using the layers selected in the brain-encoding analysis, both model families exceed the mean chance level of $0.167$ under leave-one-run-out cross-validation (Table~\ref{tab:action_decoding}). Balanced accuracy reaches $0.342\pm0.083$ for the VLM and $0.362\pm0.088$ for the LAM, compared with $0.289\pm0.073$ for EMPA. However, raw accuracy is only slightly higher than the majority-action baseline for the VLM ($0.503$ vs.\ $0.475$) and LAM ($0.512$ vs.\ $0.475$). Moreover, the LAM exceeds the VLM by only $0.020$ in balanced accuracy, indicating a small numerical difference rather than a reliable action-tuning advantage. Thus, the model representations contain modest level-wise information about participants' actions, but these results do not establish strong behavioral prediction or human-like policy use.

\begin{table}[!ht]
\centering
\scriptsize
\caption{Decoding participants' dominant level-wise actions from model features using leave-one-run-out cross-validation. Values are mean $\pm$ SD across 32 participants. Balanced accuracy is compared with the six-action chance floor of $0.167$.}
\label{tab:action_decoding}
\begin{tabular}{|l|c|c|c|c|c|}
\hline
\textbf{Model} &
\textbf{Layer} &
\textbf{Balanced acc.} &
\textbf{$\Delta$ chance} &
\textbf{Raw acc.} &
\textbf{Majority baseline} \\
\hline
VLM  & L18 & $0.342\pm0.083$ & $+0.175$ & $0.503$ & $0.475$ \\
LAM  & L21 & $0.362\pm0.088$ & $+0.195$ & $0.512$ & $0.475$ \\
EMPA & N/A & $0.289\pm0.073$ & $+0.122$ & $0.280$ & $0.460$ \\
\hline
\end{tabular}
\end{table}

This action decoding analysis provide exploratory evidence that representations aligned with brain activity also carry TR-level information about participants' actions. We interpret them cautiously: they link brain-aligned features to behavior at the representational level but do not establish strong behavioral prediction, human-like policy use, or competent gameplay.

\section{Generalization across model families}
\label{app:vem_generalization}

\begin{figure}[t]
    \centering
    \includegraphics[width=0.7\linewidth]
      {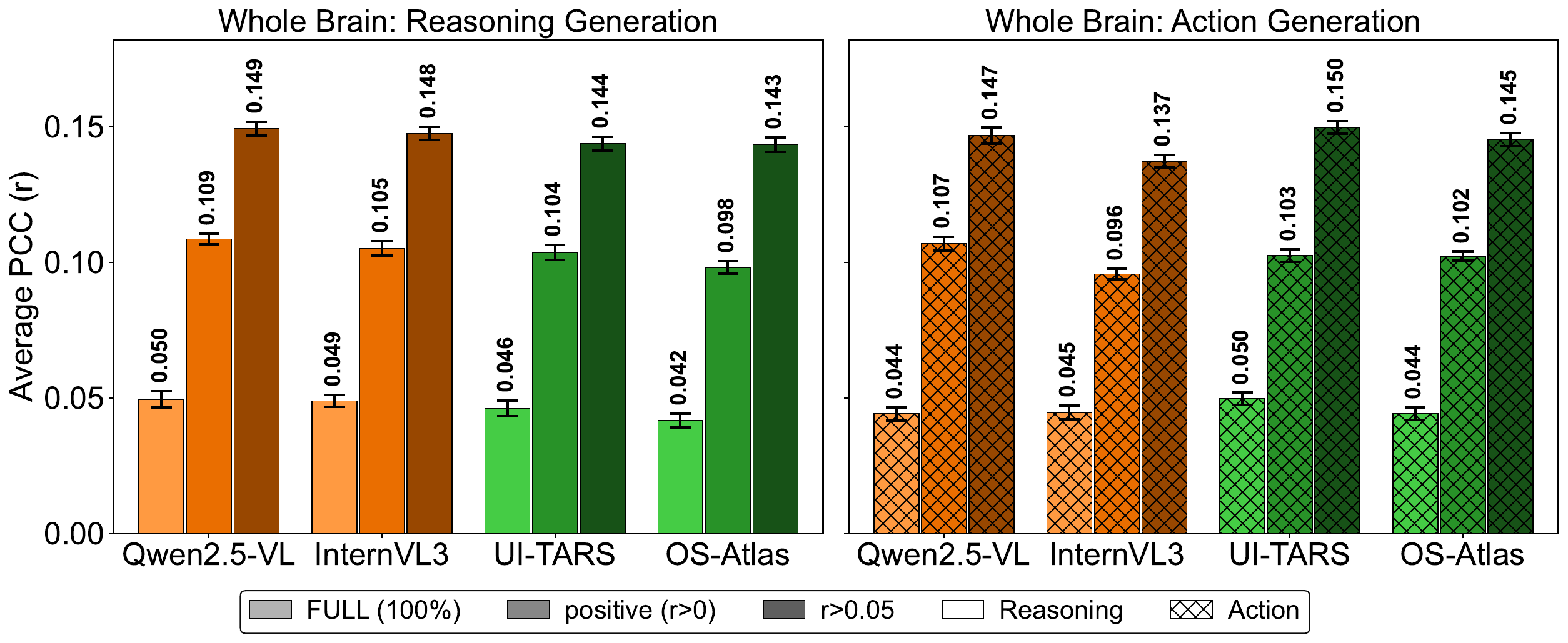}
      \vspace{-0.2cm}
    \caption{
    \textbf{Whole-brain alignment across two VLMs and two LAMs under reasoning and action prompts.} Average Pearson correlation across participants and voxels per model, under three voxel-selection criteria: all voxels (Full), positively predicted (r$>$0) and (r$>$0.05). Left: reasoning prompts. Right: action prompts (hatched). \textcolor{orange}{Orange}: VLMs; \textcolor{green}{Green}: LAMs. Darker shades denote the primary model in each family (Qwen2.5-VL, UI-TARS); lighter shades denote the secondary model (InternVL3, OS-Atlas). Error bars denote mean $\pm$ SEM across subjects.}
    \label{fig:vem_wholebrain_atari_allmodels}
\end{figure}

To verify that the prompted-alignment effect is not specific to a single VLM or LAM, we replicate the whole-brain analysis on a second model from 
each family: InternVL3-8B (VLM) and OS-Atlas-Pro-7B (LAM). Fig.~\ref{fig:vem_wholebrain_atari_allmodels} shows that all four models 
achieve substantial brain alignment under both reasoning and action prompts, with all four significantly exceeding the RL baselines (EMPA, 
DDQN). Within the VLM family, both Qwen2.5-VL (reasoning $r=0.050$, action $r=0.044$) and InternVL3 (reasoning $r=0.049$, action $r=0.045$) show a small reasoning-leaning pattern. Within 
the LAM family, both UI-TARS  
(reasoning $r=0.046$, action $r=0.050$) and OS-Atlas (reasoning $r=0.042$, action $r=0.044$) show a 
small action-leaning pattern. The within-model differences are small ($\le0.006$) and do not reach significance, so we treat this as a directional pattern rather than an established prompt preference;
For the remainder of the paper, we focus our ROI-level (RQ2) and 
variance-partitioning (RQ3) analyses on Qwen2.5-VL and UI-TARS as representative members of each family; corresponding results for 
InternVL3 and OS-Atlas are reported in App.~\ref{app:variancePartition}.

\section{Prompt Controls}
\label{app:promptcontrols}

\begin{figure}[!ht]
    \centering
    \includegraphics[width=0.9\linewidth]
      {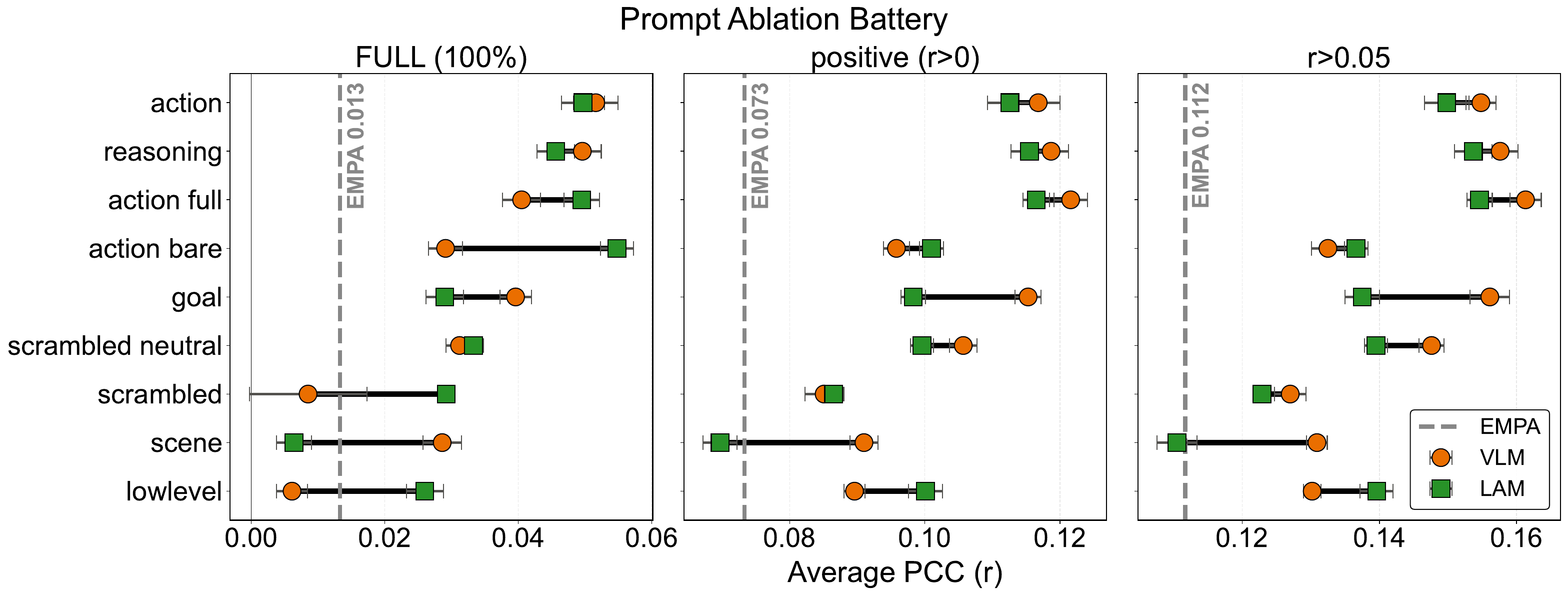}
      \vspace{-0.2cm}
    \caption{
    \textbf{Prompt-control battery.} Whole-brain alignment for the VLM (\textcolor{orange}{Orange} circles, Qwen2.5-VL) and LAM (\textcolor{green}{Green} squares, UI-TARS) under the two primary prompts and the seven control prompts of Table~\ref{tab:prompt_battery}, shown for all cortical vertices (FULL, primary metric), positively predicted vertices ($r>0$), and vertices with $r>0.05$. Dashed vertical line: EMPA under the same criterion. Error bars: mean $\pm$ SEM across $N=32$ subjects.}
    \label{fig:vem_wholebrain_atari_allmodels_prompt_controls}
\end{figure}

We next examine whether the observed prompt effects can be explained by text presence, visual description, goal information, or motor commitment alone (see Table~\ref{tab:prompt_battery}). In the primary unthresholded analysis, as shown in Fig.~\ref{fig:vem_wholebrain_atari_allmodels_prompt_controls}, (left), the original action and reasoning prompts produce consistently high brain alignment for both model families and exceed the EMPA baseline. The action-full condition also shows relatively high alignment ($\approx0.040$ and $0.049$), whereas action-bare is model-dependent, producing higher alignment for the LAM ($\approx0.055$) than for the VLM ($\approx0.029$). The goal condition yields intermediate and model-dependent performance ($\approx0.040$ for the VLM, $0.029$ for the LAM). In contrast, scrambled text, low-level colour descriptions, and scene descriptions generally produce weaker alignment than the original task-framed prompts ($r\approx0.006$--$0.029$), although the magnitude varies between the VLM and LAM. The same broad pattern is visible in the $r>0$ and $r>0.05$ sensitivity analyses. These results indicate that text presence or visual description alone is insufficient to reproduce the consistently stronger alignment elicited by task-instruction prompts. However, because the control effects are model-dependent, we do not interpret them as demonstrating that each prompt isolates a single action- or reasoning-specific semantic dimension.

\section{VLM and LAM: Prompt-driven alignment gains across ROIs}
\label{app:lamGainsScale}

\begin{figure}[!ht]
    \centering
    \includegraphics[width=0.7\linewidth]
      {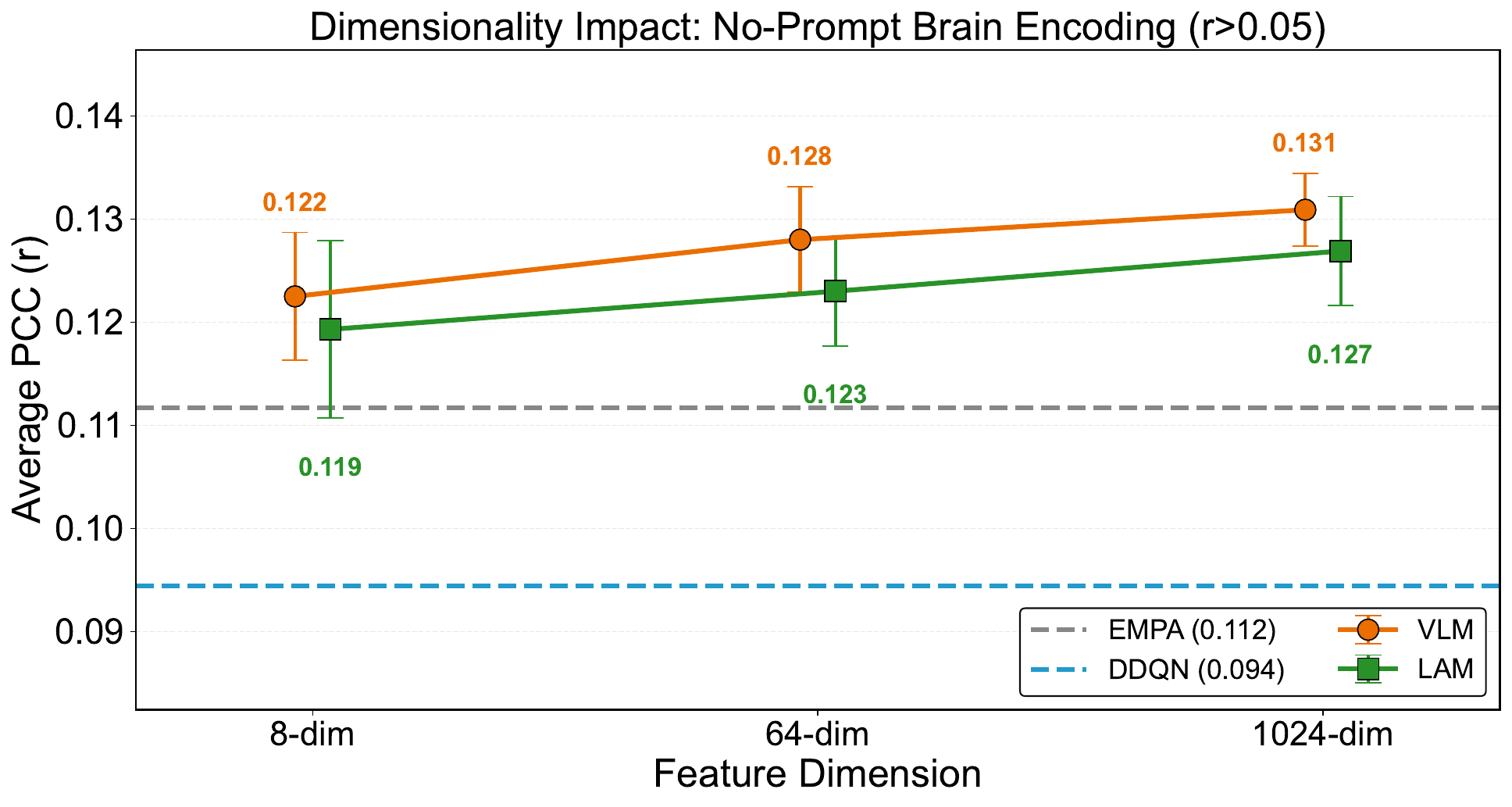}
      \vspace{-0.2cm}
    \caption{\textbf{Whole-brain voxel-wise encoding performance.} Effect of feature dimensionality under the \emph{no-prompt} condition, shown for voxels with $r>0.05$:: VLM (\textcolor{orange}{orange}) and LAM (\textcolor{green}{green}) features evaluated at three dimensionalities (8, 64, 1024), compared against the RL baselines EMPA (gray dashed, r=0.112) and DDQN (light blue dashed, r=0.094). * indicates significant pairwise gain from 8 to 64 dim (paired t-test, 
$p<0.05$); the 64$\to$1024 transition is non-significant, indicating saturation. At every dimensionality tested, both VLM and LAM significantly exceed both RL baselines. Error bars: mean $\pm$ SEM across subjects.}
    \label{fig:vem_wholebrain_atari_r005}
\end{figure}

As shown in Fig.~\ref{fig:vem_wholebrain_atari_r005}, both VLM and LAM representations outperform the classical baselines in whole-brain voxel-wise encoding without any prompt instruction (shown for voxels with $r>0.05$). Specifically, VLM (\emph{no-prompt}) achieves r=0.131 and LAM (\emph{no-prompt}) reaches r=0.127, compared to EMPA (r=0.112) and DDQN (r=0.094). The advantage is consistent across all three feature dimensionalities tested. Crucially, even when VLM and LAM features are 
reduced to 8 dimensions to match EMPA's feature count, they continue to 
significantly show higher brain alignment than both RL baselines (VLM r=0.122, LAM r=0.119), ruling out feature count as the source of the gain.
Performance saturates between 64-dim and 1024-dim, suggesting that the gain over RL baselines reflects what the features encode rather than feature dimensionality alone. Together, these results indicate that contemporary multimodal architectures learn richer latent representations of dynamic game environments than traditional task-optimized RL systems.



\begin{figure}[!ht]
    \centering
    \includegraphics[width=0.95\linewidth]
      {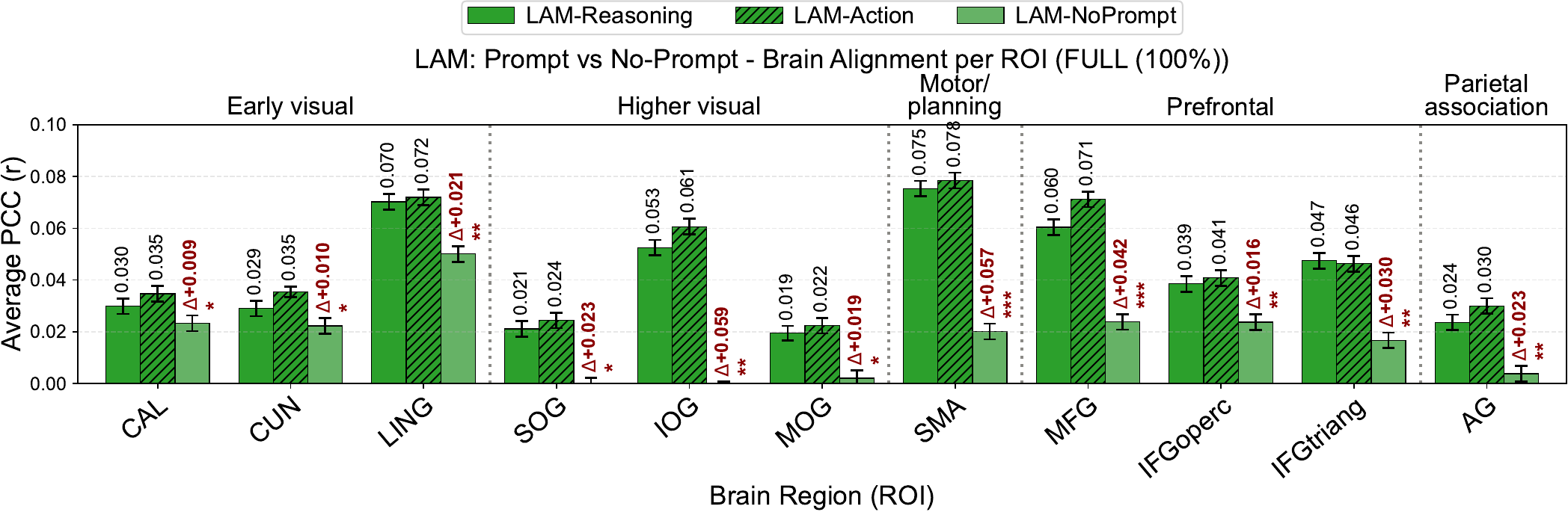}
    \includegraphics[width=0.95\linewidth]
      {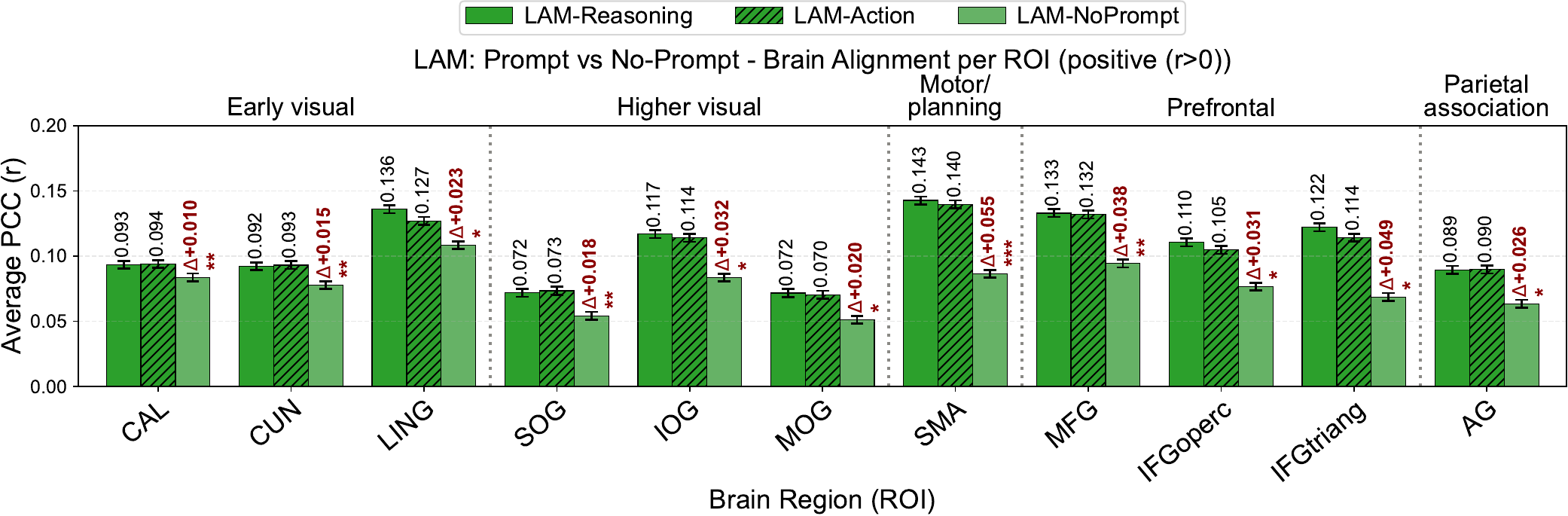}
      \includegraphics[width=0.95\linewidth]
      {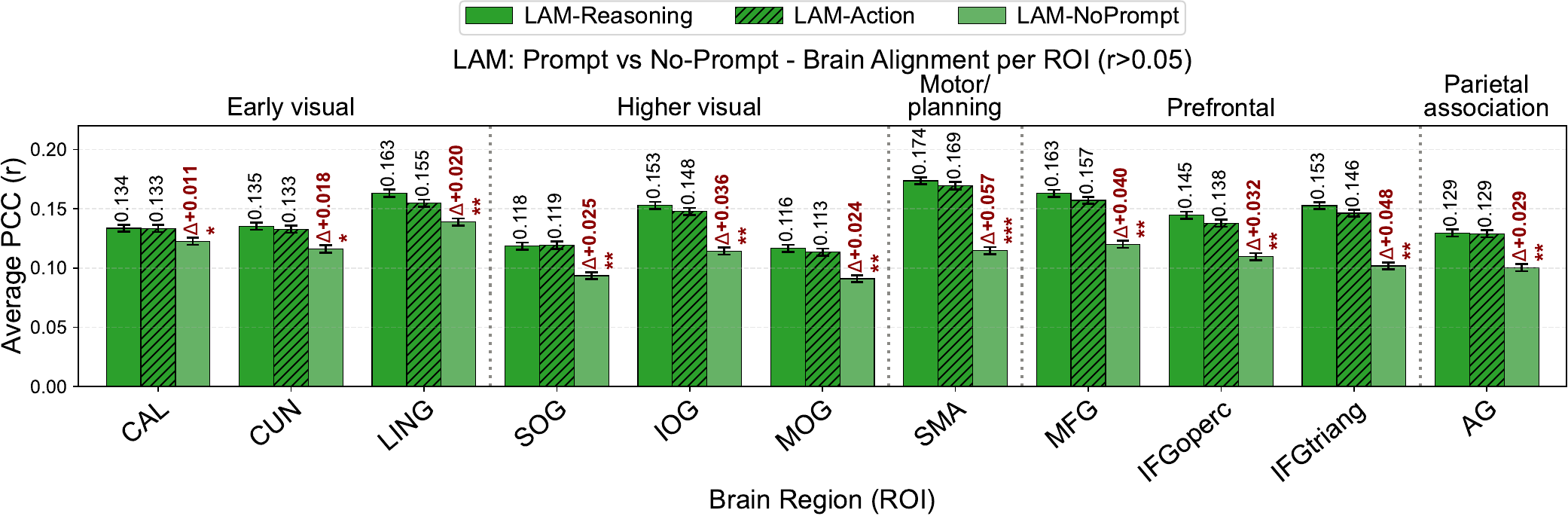}
    \caption{
    Brain alignment averaged across participants and voxels per ROI for prompted vs. no-prompt LAM representations. Average Pearson correlation between predicted and observed fMRI responses across 11 ROIs for LAM-Reasoning (solid), LAM-Action (hatched), and LAM-NoPrompt (light). Error bars denote mean $\pm$ SEM across subjects. Markers denote paired comparisons against no-prompt (*p $<$ 0.05, **p $<$ 0.01, ***p $<$ 0.001). Both prompted conditions show higher alignment than no-prompt across all ROIs.}
    \label{fig:vem_lam_rois_atari}
\end{figure}

We repeat the ROI analysis on LAMs under Action, Reasoning, and 
\emph{no-prompt} settings (Fig.~\ref{fig:vem_lam_rois_atari}). 
As with VLMs, prompting improves alignment in every ROI ($q<0.05$, FDR-corrected). Gains are largest in SMA (reasoning/action $r=0.075/0.078$ vs.\ \emph{no-prompt} $r=0.020$; $\Delta r=+0.057$, $q<0.001$) and MFG ($0.060/0.071$ vs.\ $0.024$; $\Delta r=+0.042$, $q<0.001$), followed by IOG and IFGtriang (both $\Delta r=+0.030$), SOG and AG (both $+0.023$), LING ($+0.021$), and MOG ($+0.019$), and smallest in IFGoperc ($+0.016$), CUN ($+0.010$), and CAL ($+0.009$). Within the LAM, action-prompt alignment is numerically equal to or higher than reasoning-prompt alignment in 10 of 11 ROIs, with IFGtriang the only exception ($0.047$ vs.\ $0.046$), whereas the VLM shows a mixed pattern (Fig.~\ref{fig:vem_vlm_rois_atari}). These within-model action--reasoning differences are small and not individually significant; we examine them with variance partitioning in the main text (Sec.~\ref{sec:results}, RQ3).The same pattern holds when restricting to vertices with $r>0$ or $r>0.05$ (Fig.~\ref{fig:vem_lam_rois_atari}, middle and bottom panels), where all 11 ROIs again show significant prompt gains.



\begin{figure}[!ht]
    \centering
    \includegraphics[width=0.95\linewidth]
      {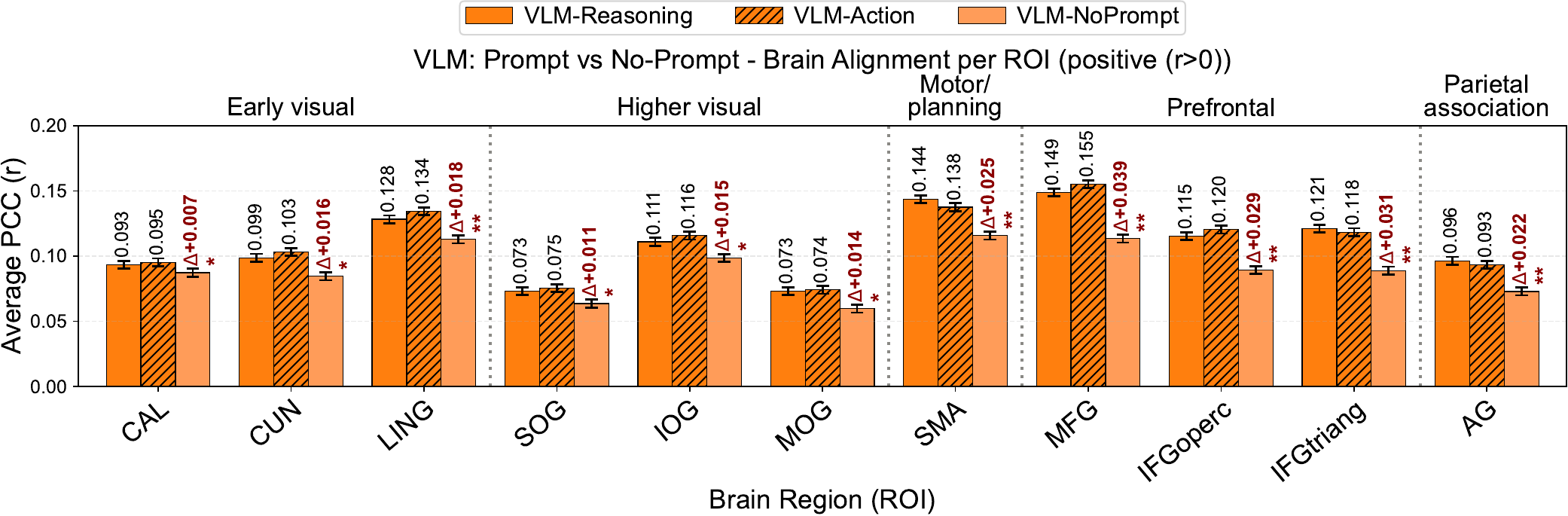}
      \includegraphics[width=0.95\linewidth]
      {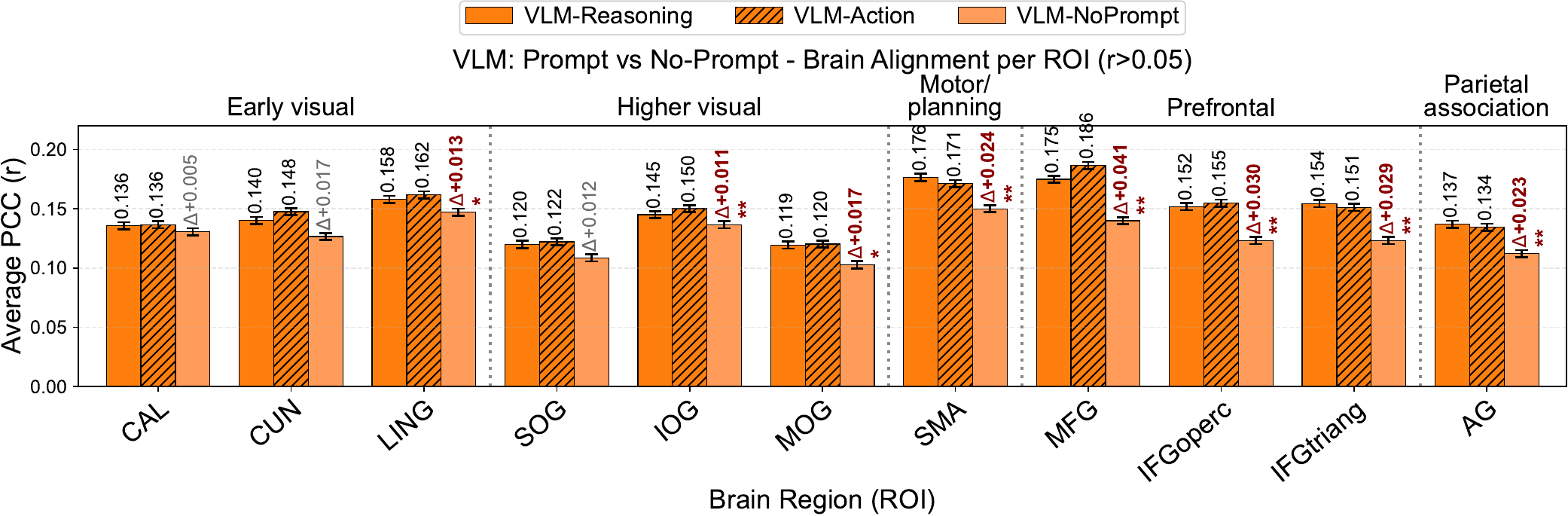}
    \caption{
    Brain alignment averaged across participants and voxels per ROI for prompted vs. no-prompt VLM representations. Average Pearson correlation between predicted and observed fMRI responses across 11 ROIs for VLM-Reasoning (solid), VLM-Action (hatched), and VLM-NoPrompt (light). Error bars denote mean $\pm$ SEM across subjects. Markers denote paired comparisons against no-prompt (*p $<$ 0.05, **p $<$ 0.01, ***p $<$ 0.001). Both prompted conditions show higher alignment than no-prompt across all ROIs.}
    \label{fig:vem_vlm_rois_atari_thresholded}
\end{figure}

\begin{figure}[!ht]
    \centering
    \includegraphics[width=0.49\linewidth]
      {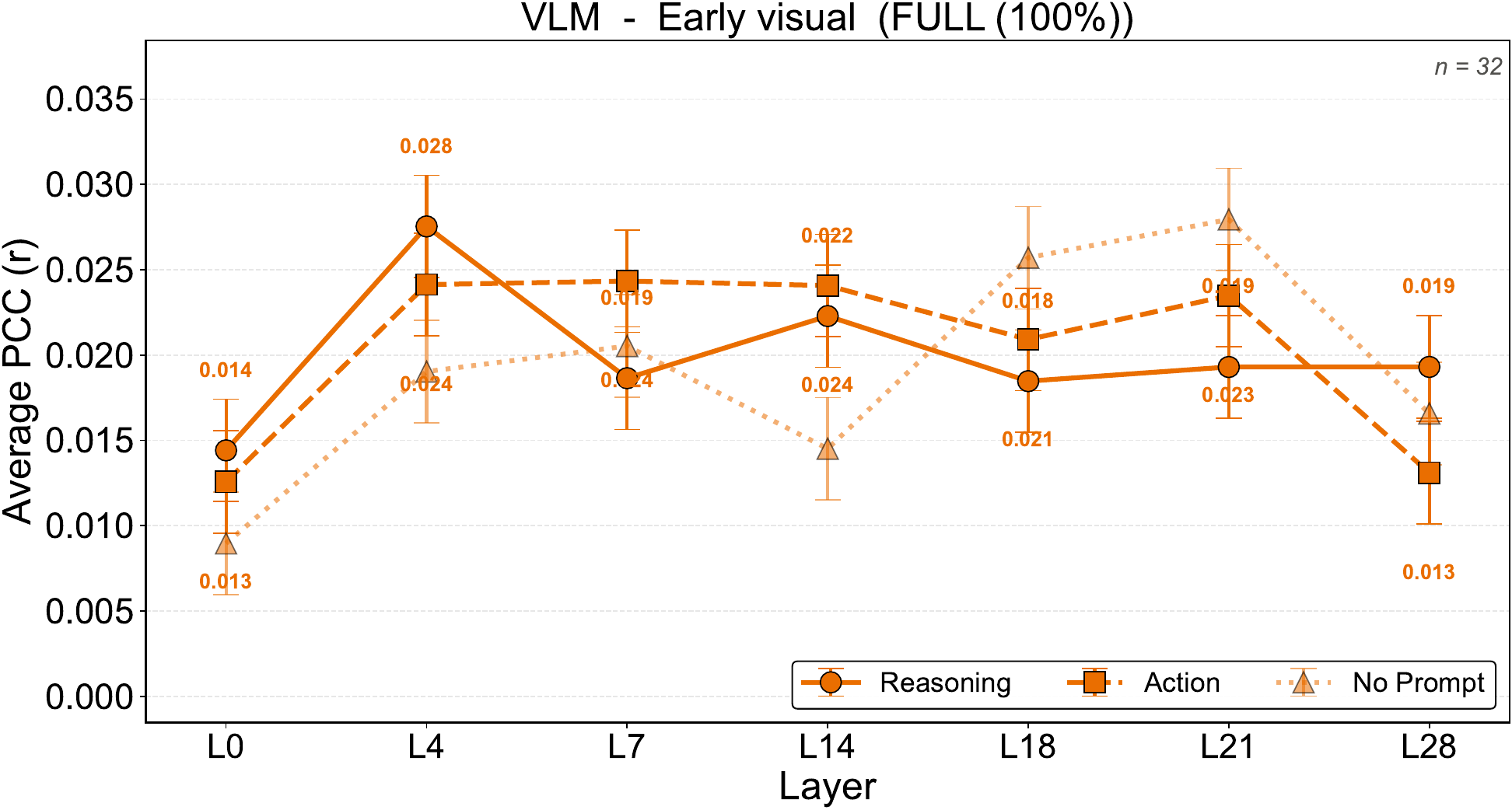}
      \includegraphics[width=0.49\linewidth]
      {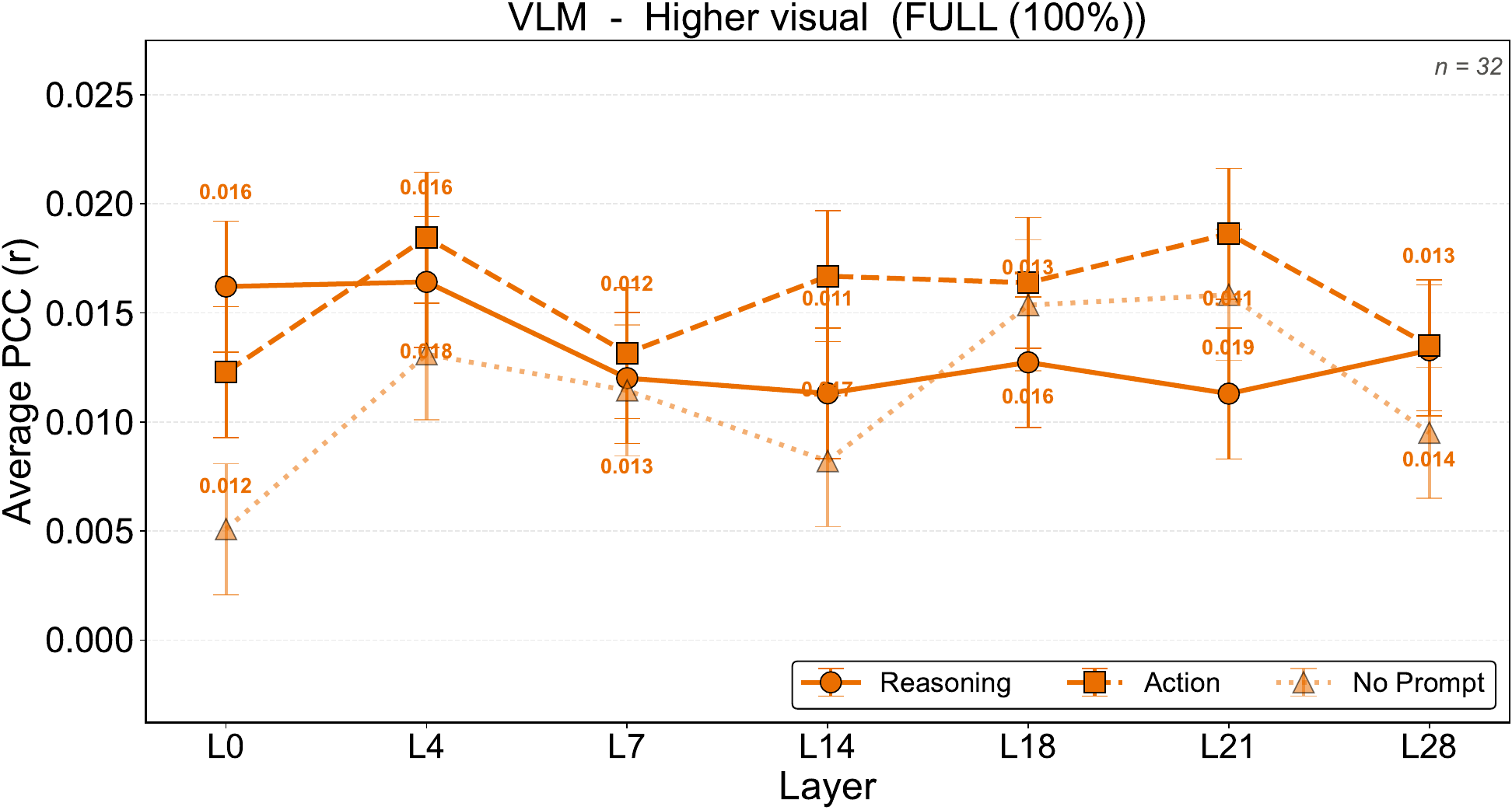}
      \includegraphics[width=0.49\linewidth]
      {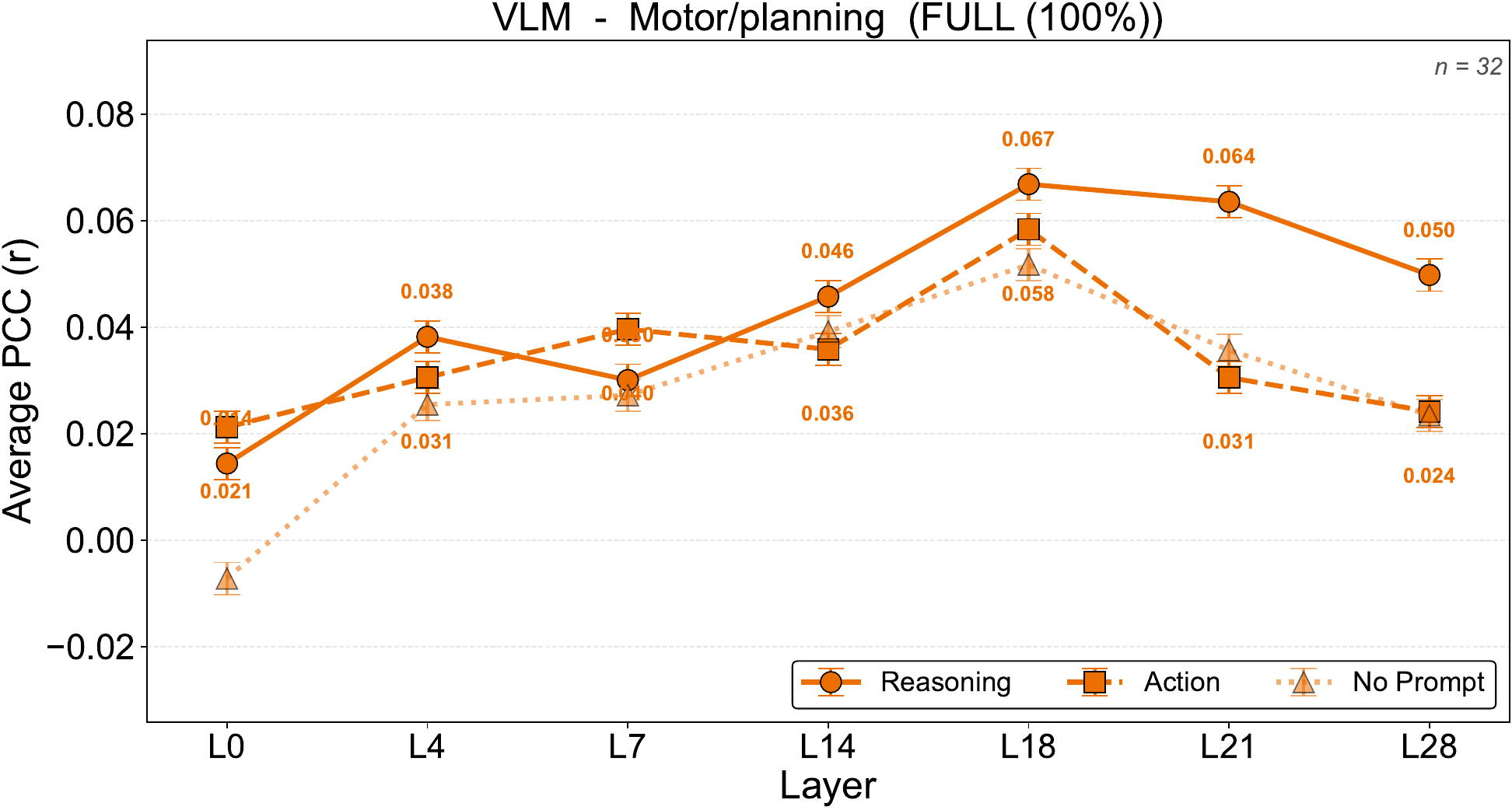}
      \includegraphics[width=0.49\linewidth]
      {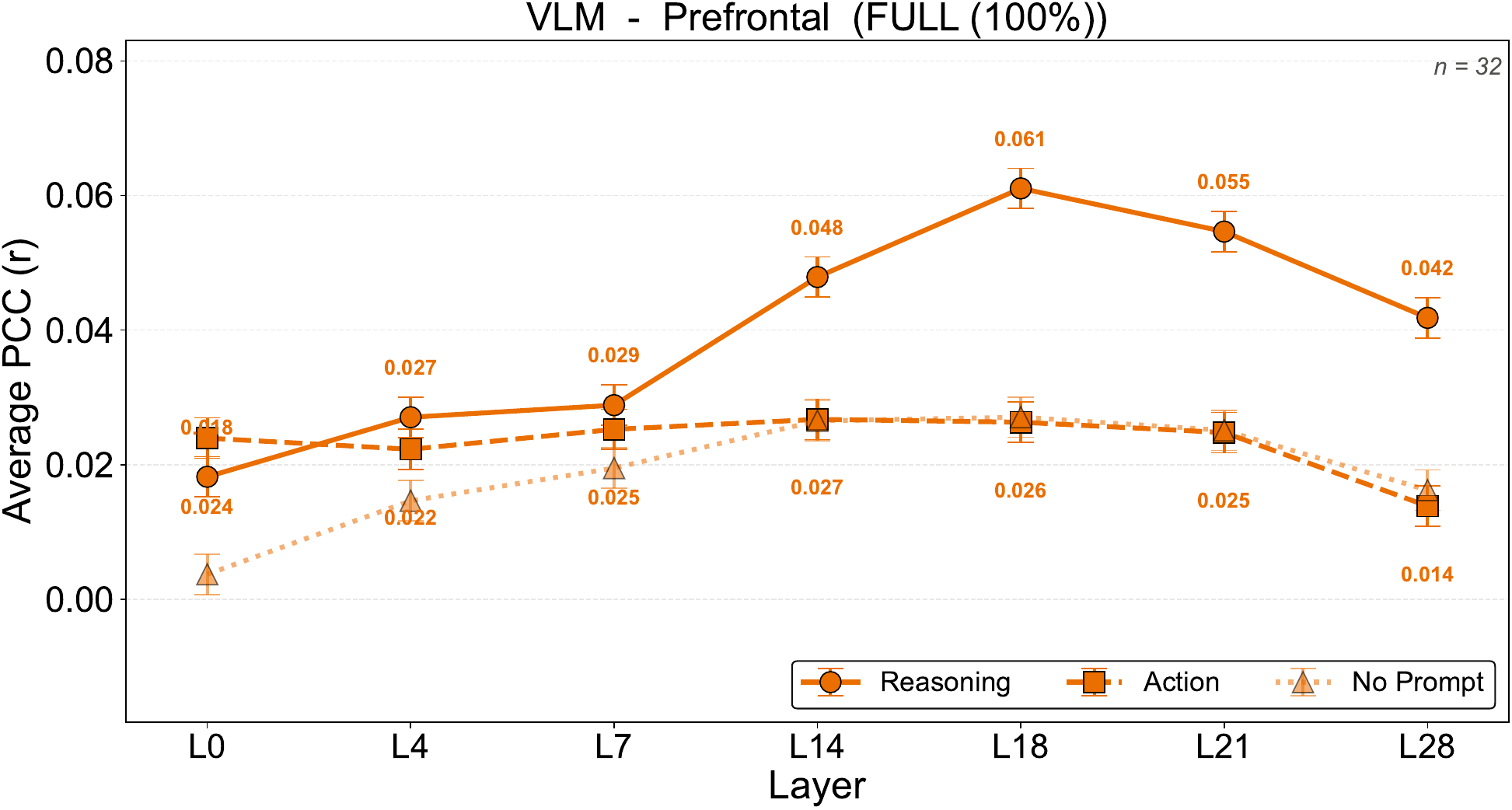}
      \includegraphics[width=0.49\linewidth]
      {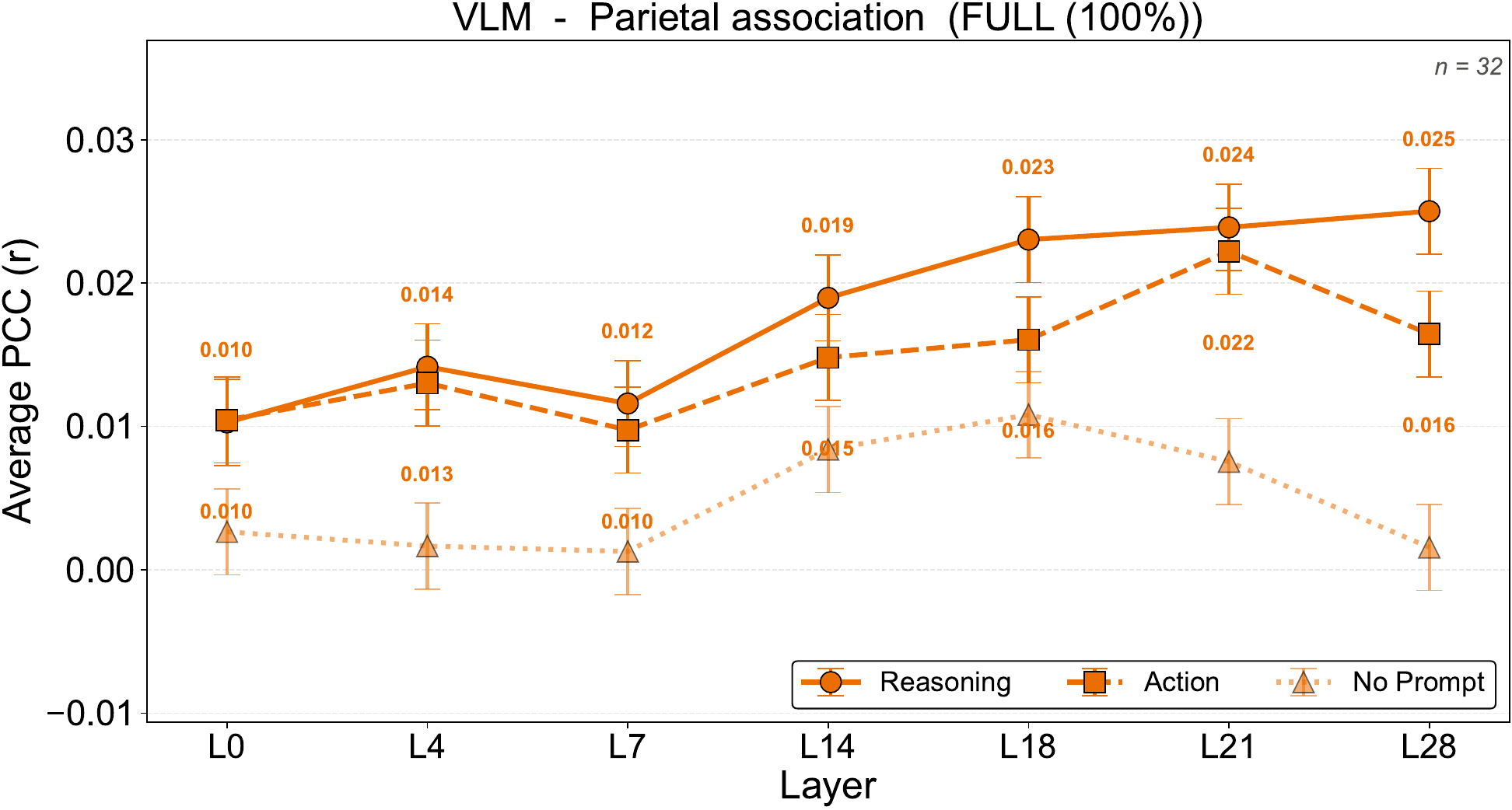}
     \vspace{-0.3cm}
    \caption{\textbf{Layer-wise brain alignment of VLM (Qwen2.5-VL) representations, all vertices (FULL).} Average Pearson correlation between predicted and observed fMRI responses at sampled layers (L0--L28) for reasoning (solid circles), action (dashed squares), and \emph{no-prompt} (dotted triangles) representations, shown separately for five ROI groups: early visual (CAL, CUN, LING), higher visual (SOG, IOG, MOG), motor/planning (SMA), prefrontal (MFG, IFGoperc, IFGtriang), and parietal association (AG). Error bars: mean $\pm$ SEM across $N=32$ participants. Visual groups peak at early layers, whereas motor/planning, prefrontal, and parietal groups peak at intermediate-to-deep layers (Table~\ref{tab:roi_peak_layers})
    }
    \label{fig:vem_vlm_rois_atari_laye_hierarchy_full}
\end{figure}

\begin{figure}[!ht]
    \centering
    \includegraphics[width=0.49\linewidth]
      {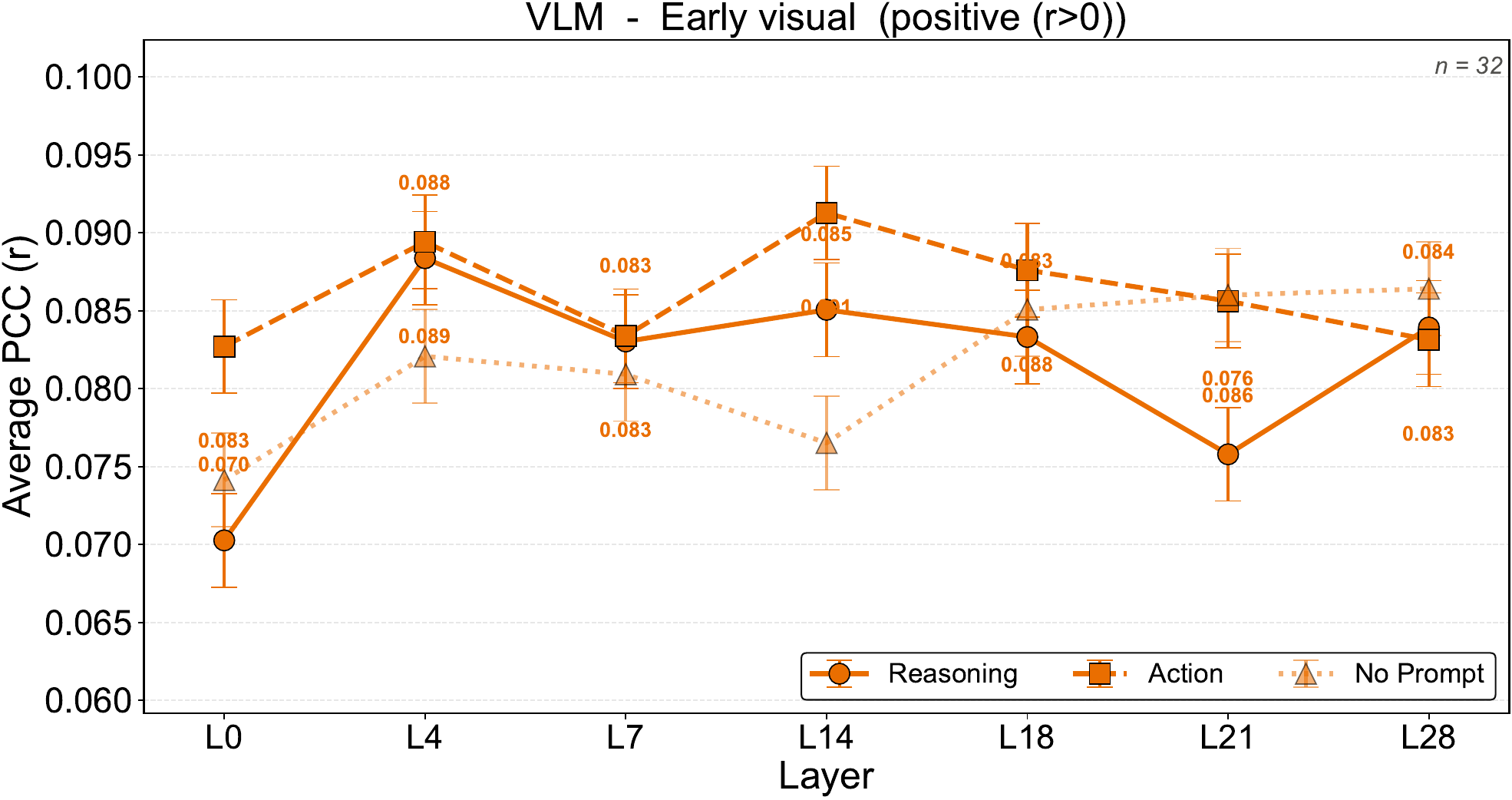}
      \includegraphics[width=0.49\linewidth]
      {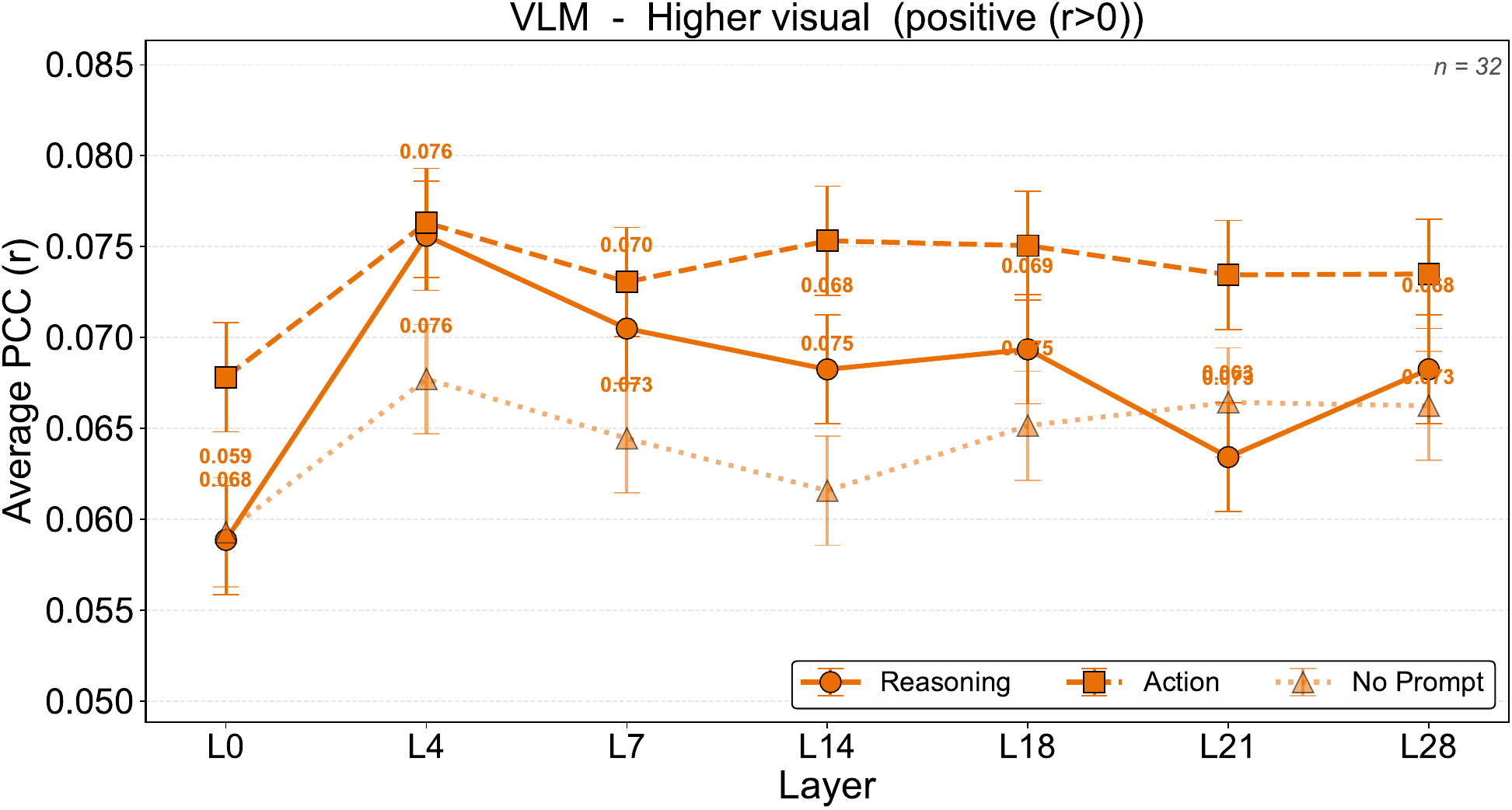}
      \includegraphics[width=0.49\linewidth]
      {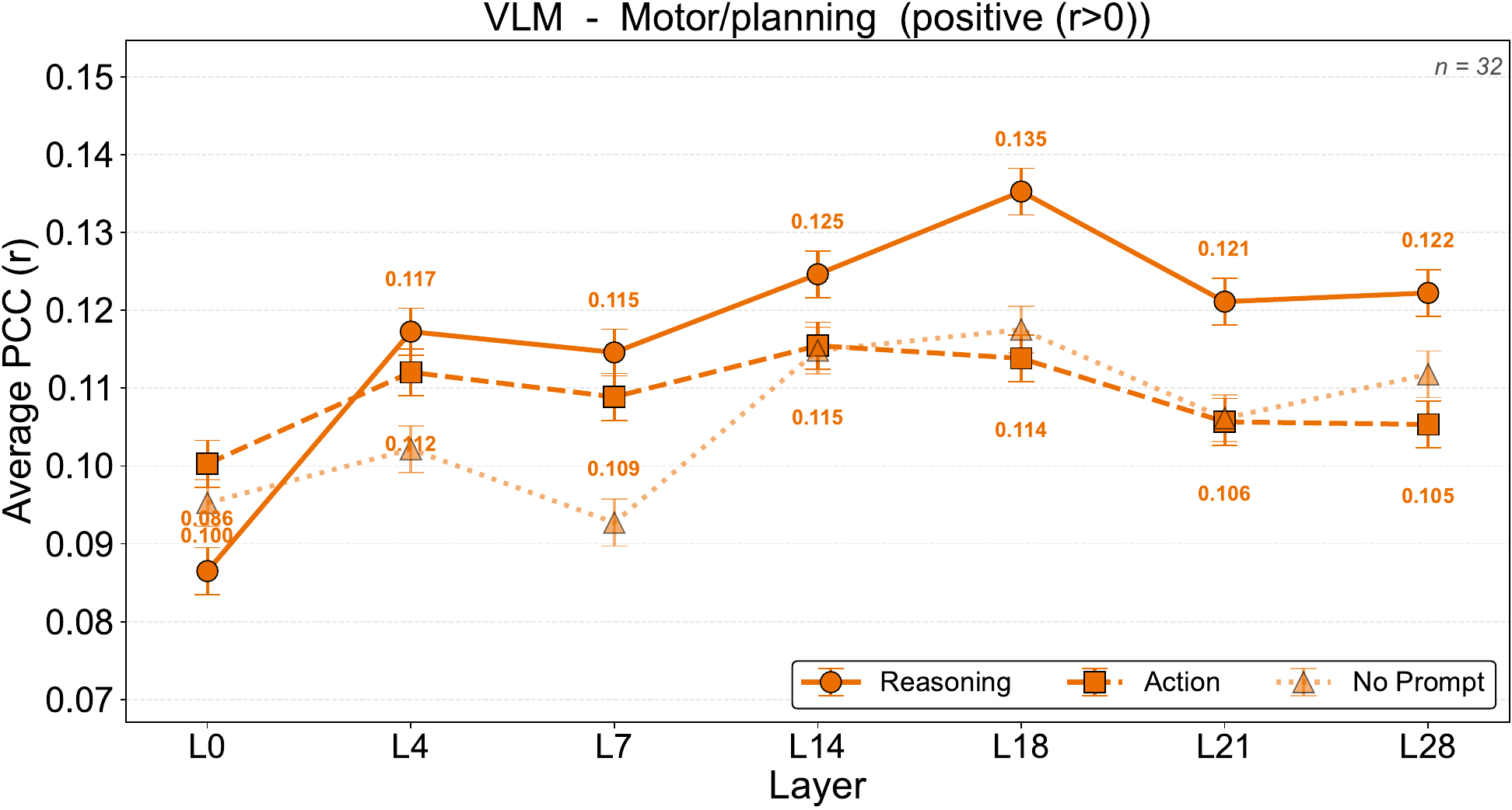}
      \includegraphics[width=0.49\linewidth]
      {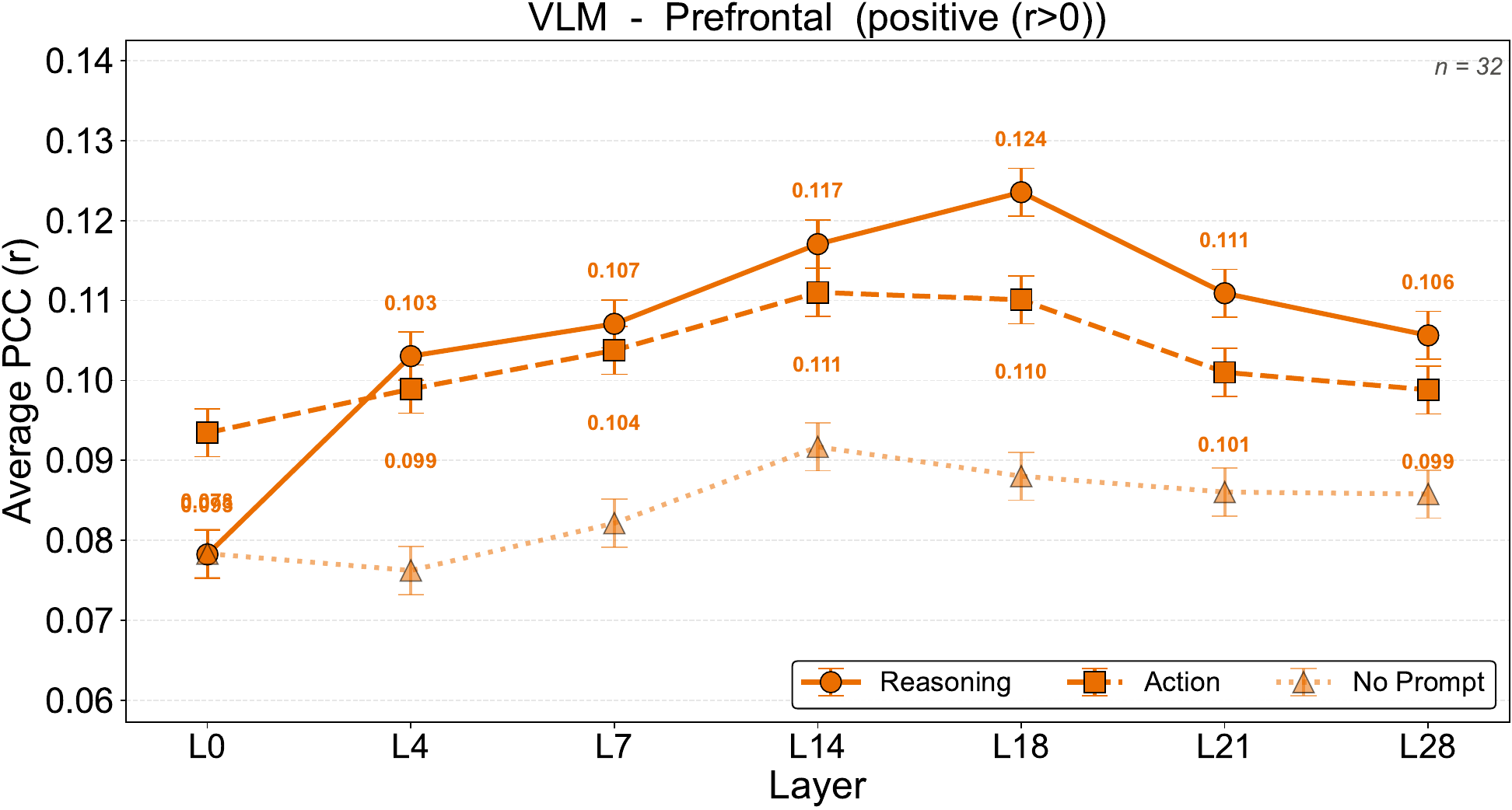}
      \includegraphics[width=0.49\linewidth]
      {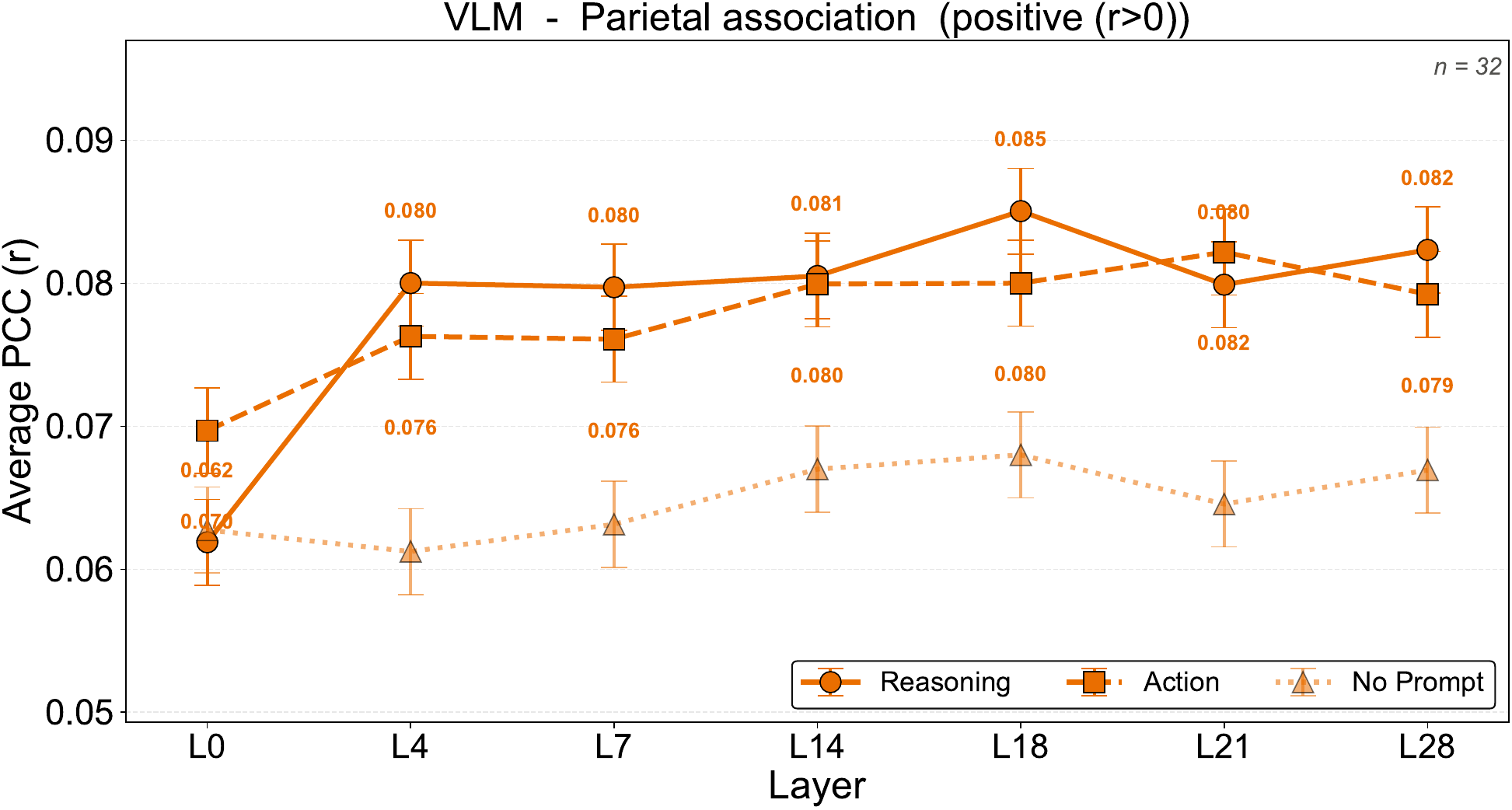}
     \vspace{-0.3cm}
    \caption{\textbf{Layer-wise brain alignment of VLM (Qwen2.5-VL) representations, positively predicted vertices ($r>0$).} Same as Fig.~\ref{fig:vem_vlm_rois_atari_laye_hierarchy_full}, restricted to vertices with $r>0$. Error bars: mean $\pm$ SEM across $N=32$ participants.
    }
    \label{fig:vem_vlm_rois_atari_laye_hierarchy_positive}
\end{figure}

\begin{figure}[!ht]
    \centering
    \includegraphics[width=0.49\linewidth]
      {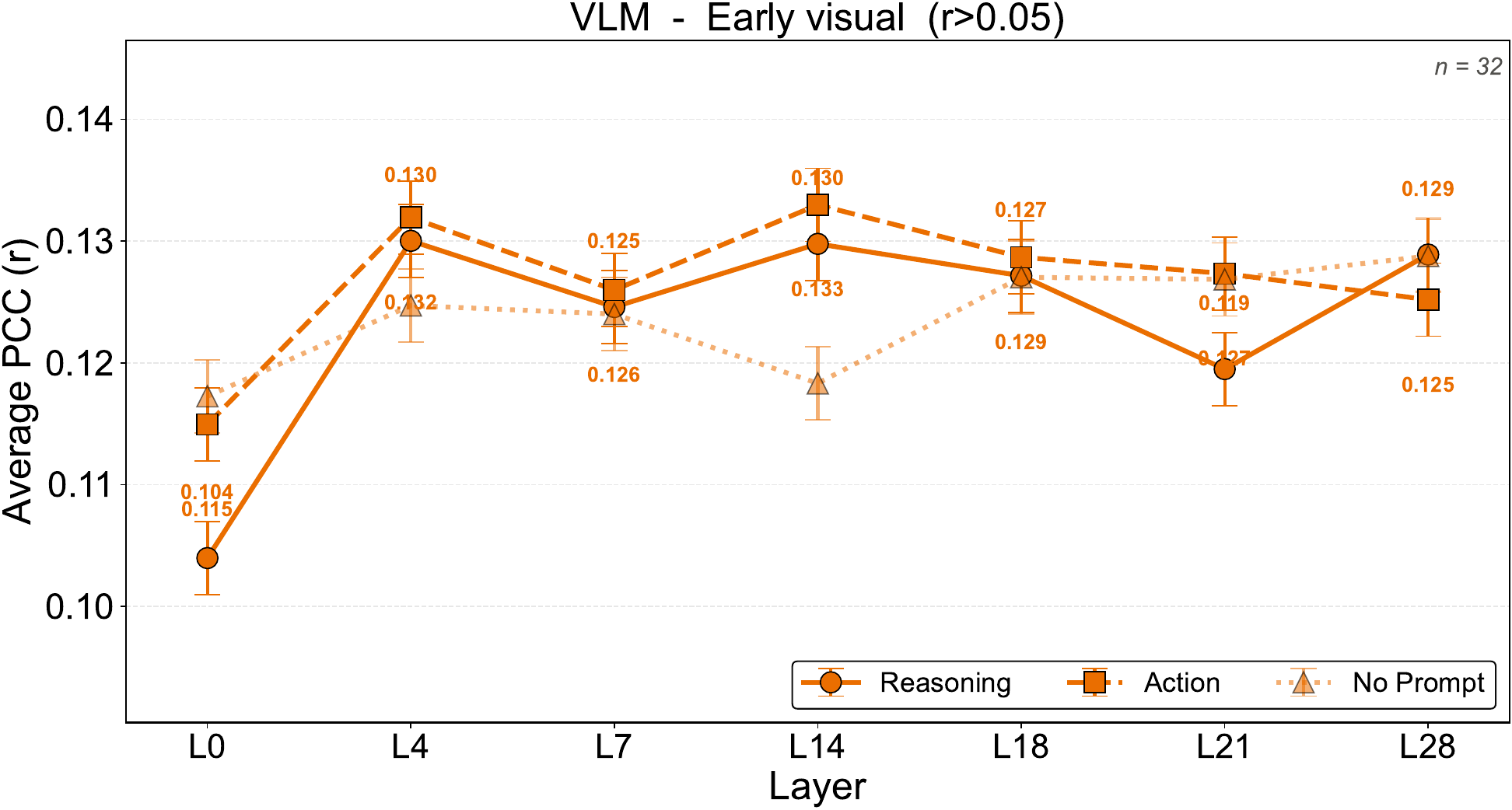}
      \includegraphics[width=0.49\linewidth]
      {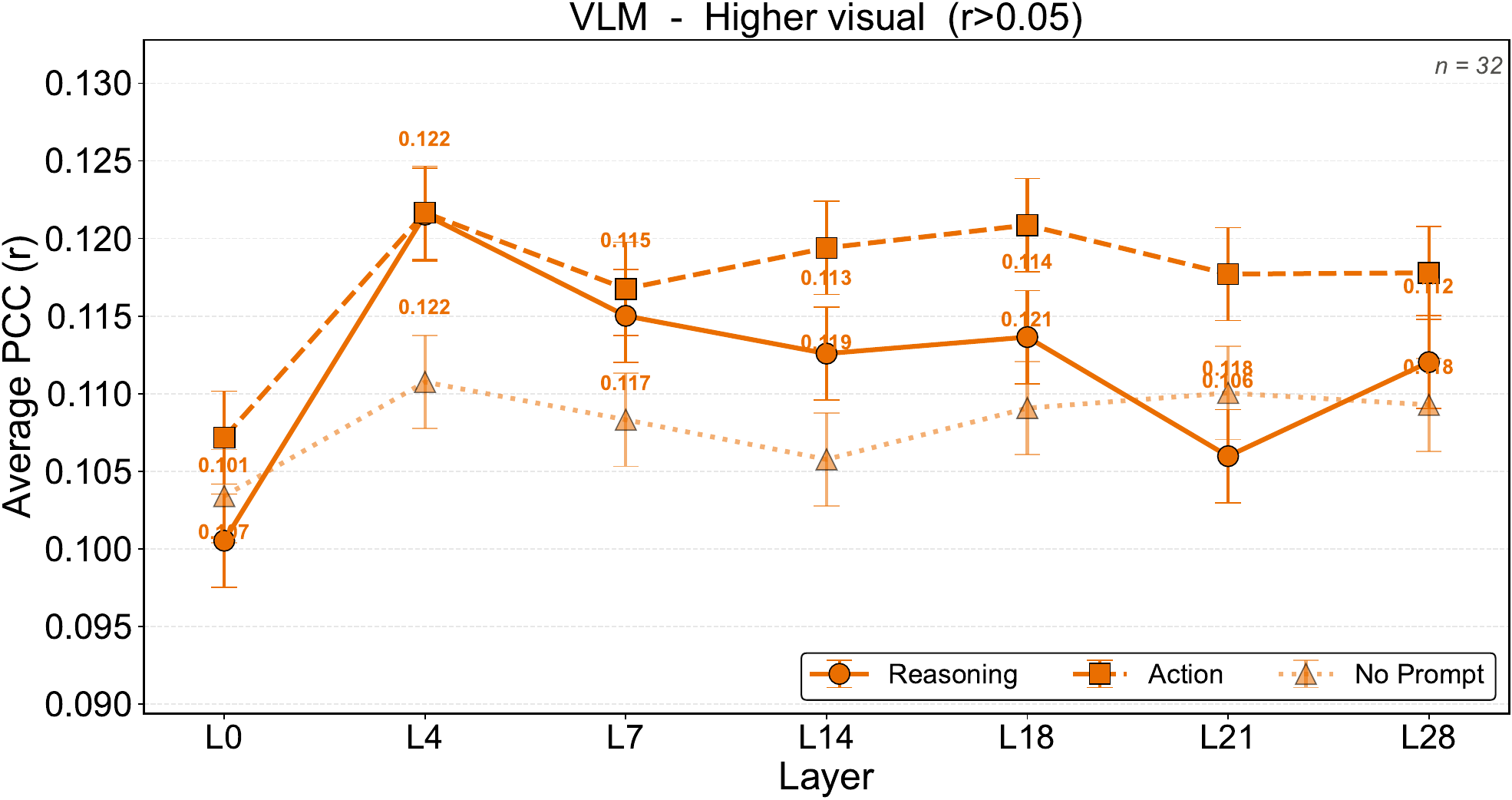}
      \includegraphics[width=0.49\linewidth]
      {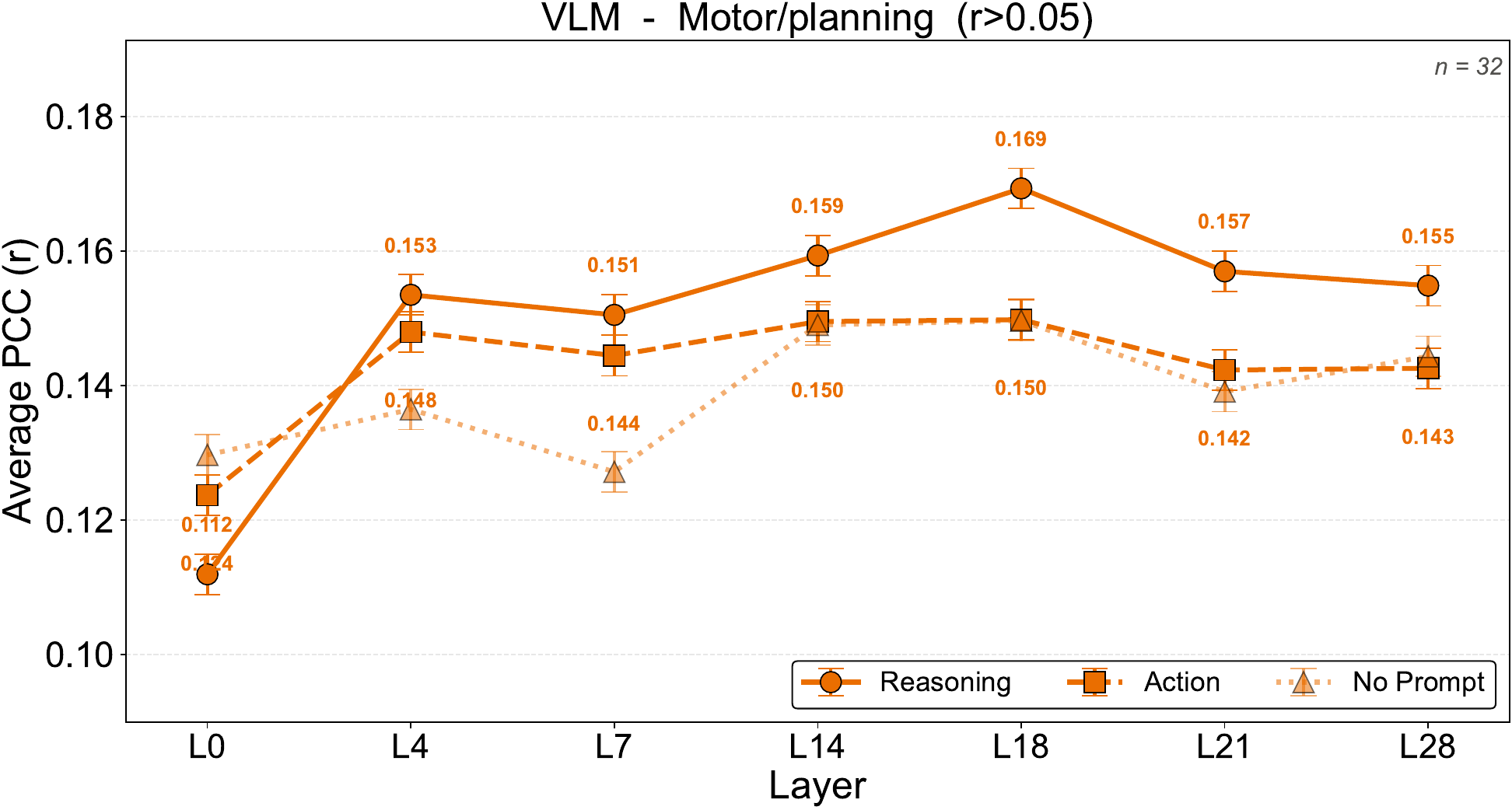}
      \includegraphics[width=0.49\linewidth]
      {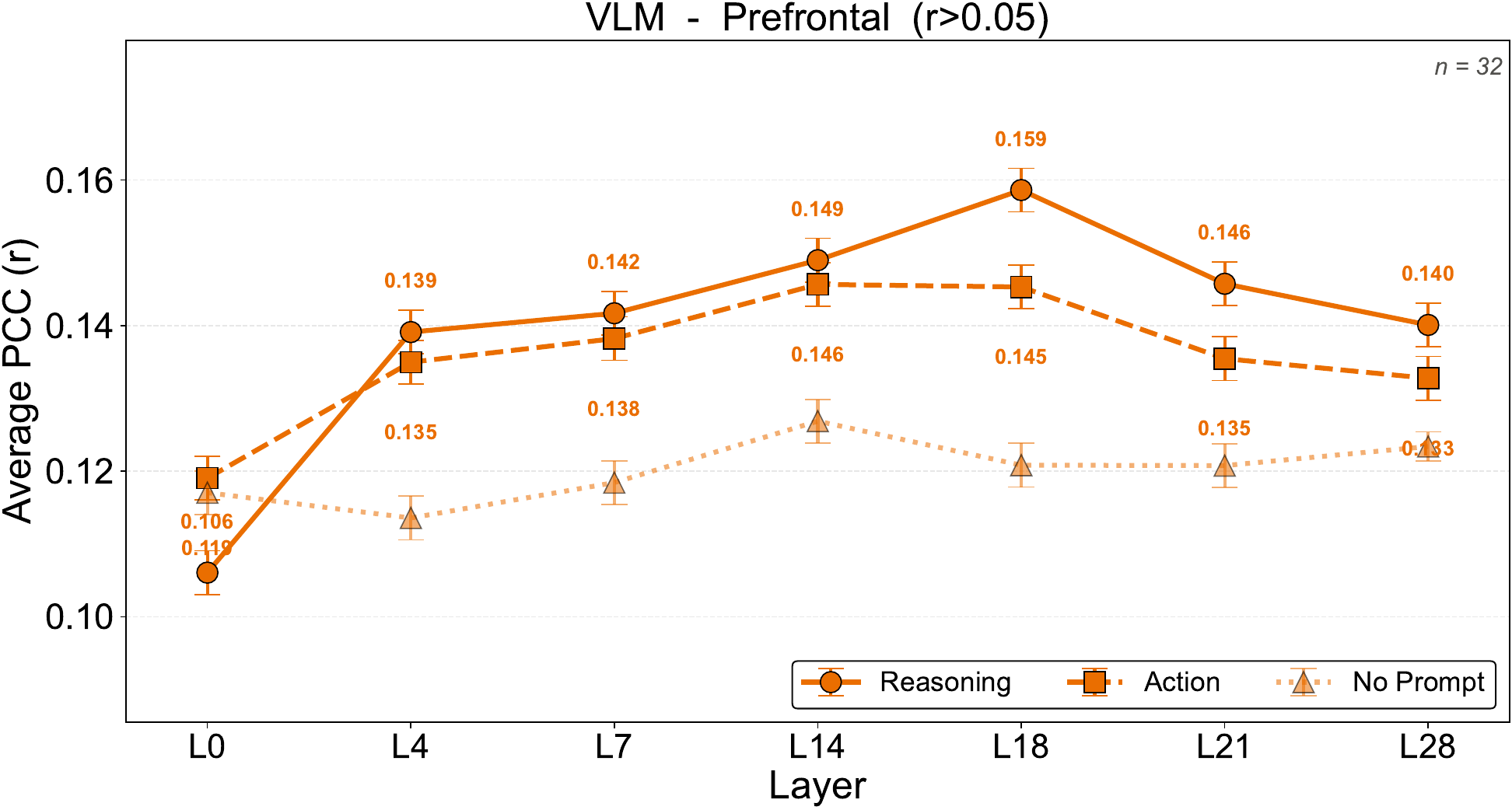}
      \includegraphics[width=0.49\linewidth]
      {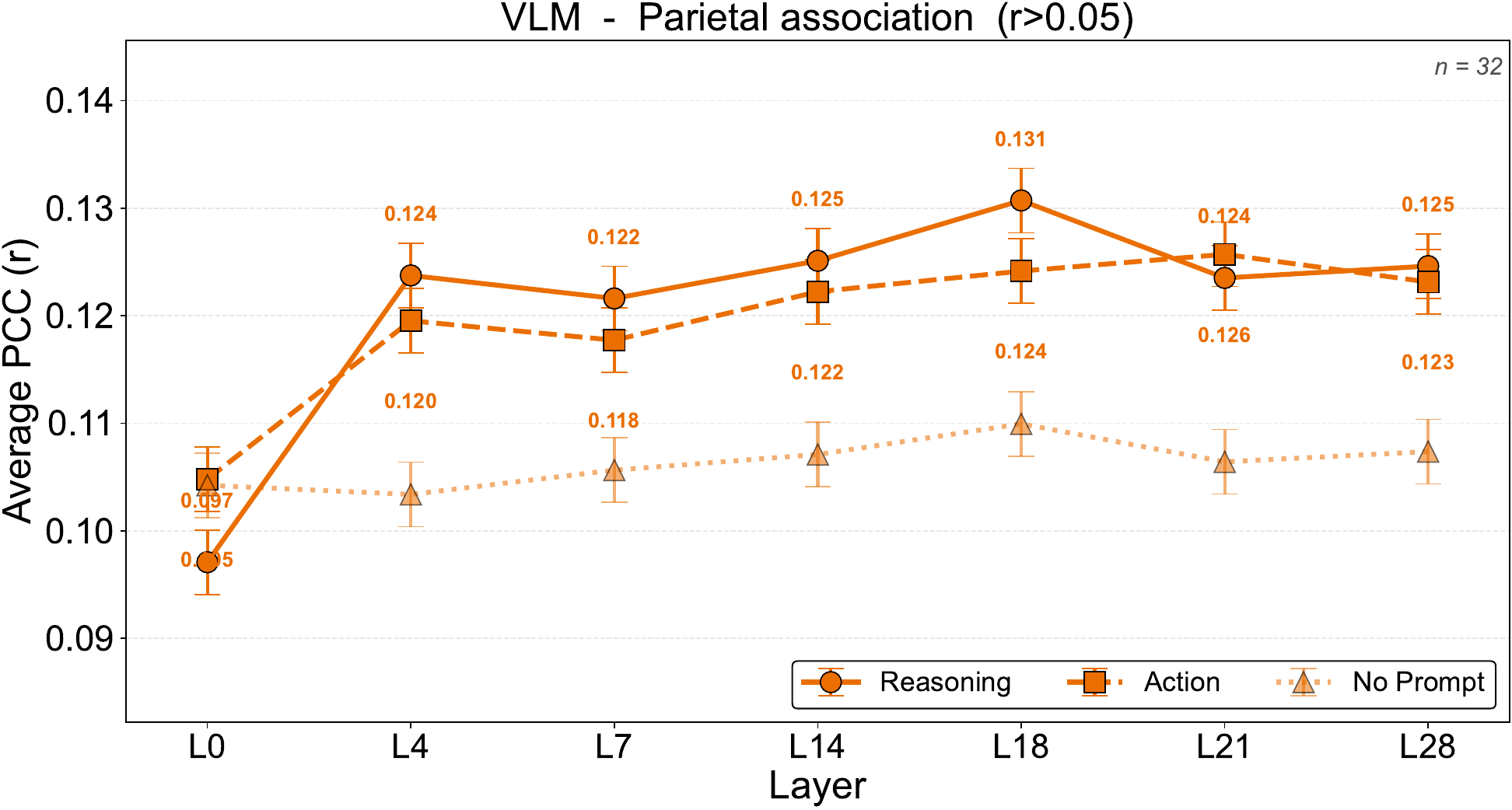}
     \vspace{-0.3cm}
    \caption{\textbf{Layer-wise brain alignment of VLM (Qwen2.5-VL) representations, vertices with $r>0.05$.} Same as Fig.~\ref{fig:vem_vlm_rois_atari_laye_hierarchy_full}, restricted to vertices with $r>0.05$. Error bars: mean $\pm$ SEM across $N=32$ participants.
    }
    \label{fig:vem_vlm_rois_atari_laye_hierarchy_r005}
\end{figure}

\begin{figure}[!ht]
    \centering
    \includegraphics[width=0.9\linewidth]
      {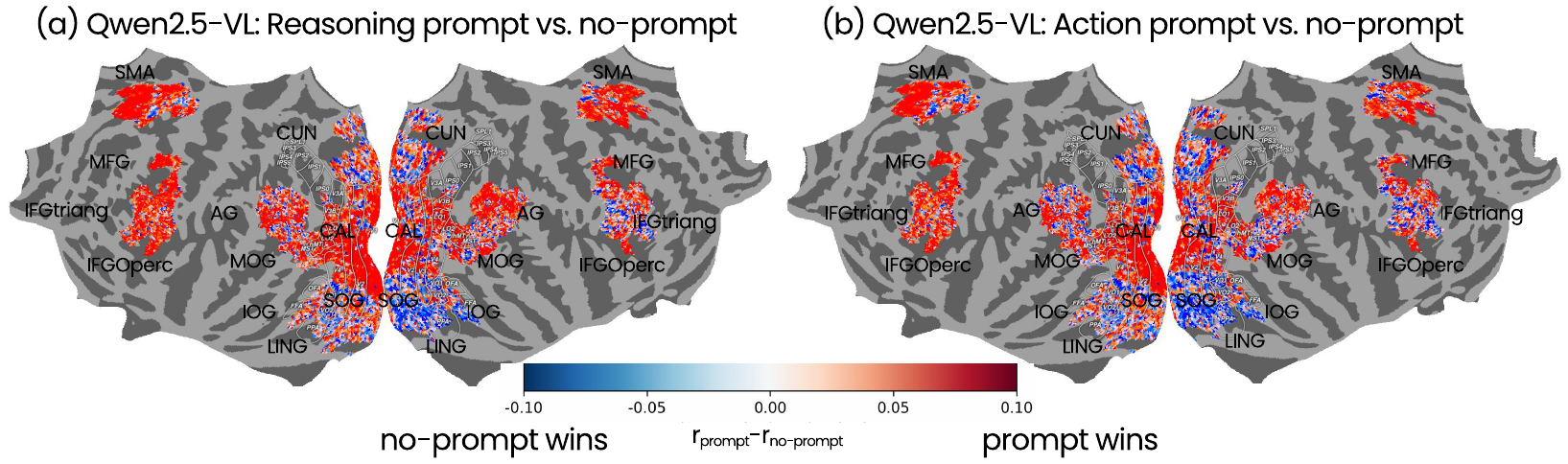}
    \caption{\textbf{Voxel-wise difference in brain alignment between prompted and \emph{no-prompt} VLM representations.} Group-averaged across subjects cortical flatmap (both hemispheres, `fsaverage') showing per-voxel differences in Pearson correlation, $r_{\text{prompted}}$-$r_{\text{no-prompt}}$, for Qwen2.5-VL under (a) reasoning prompts and (b) action prompts. Red voxels indicate higher alignment under the prompted condition; blue voxels indicate higher alignment under no-prompt.
    }
    \label{fig:vlm_prompt_noprompt_flatmap_groupmean_atari}
\end{figure}

\begin{figure}[!ht]
    \centering
    \includegraphics[width=0.49\linewidth]
      {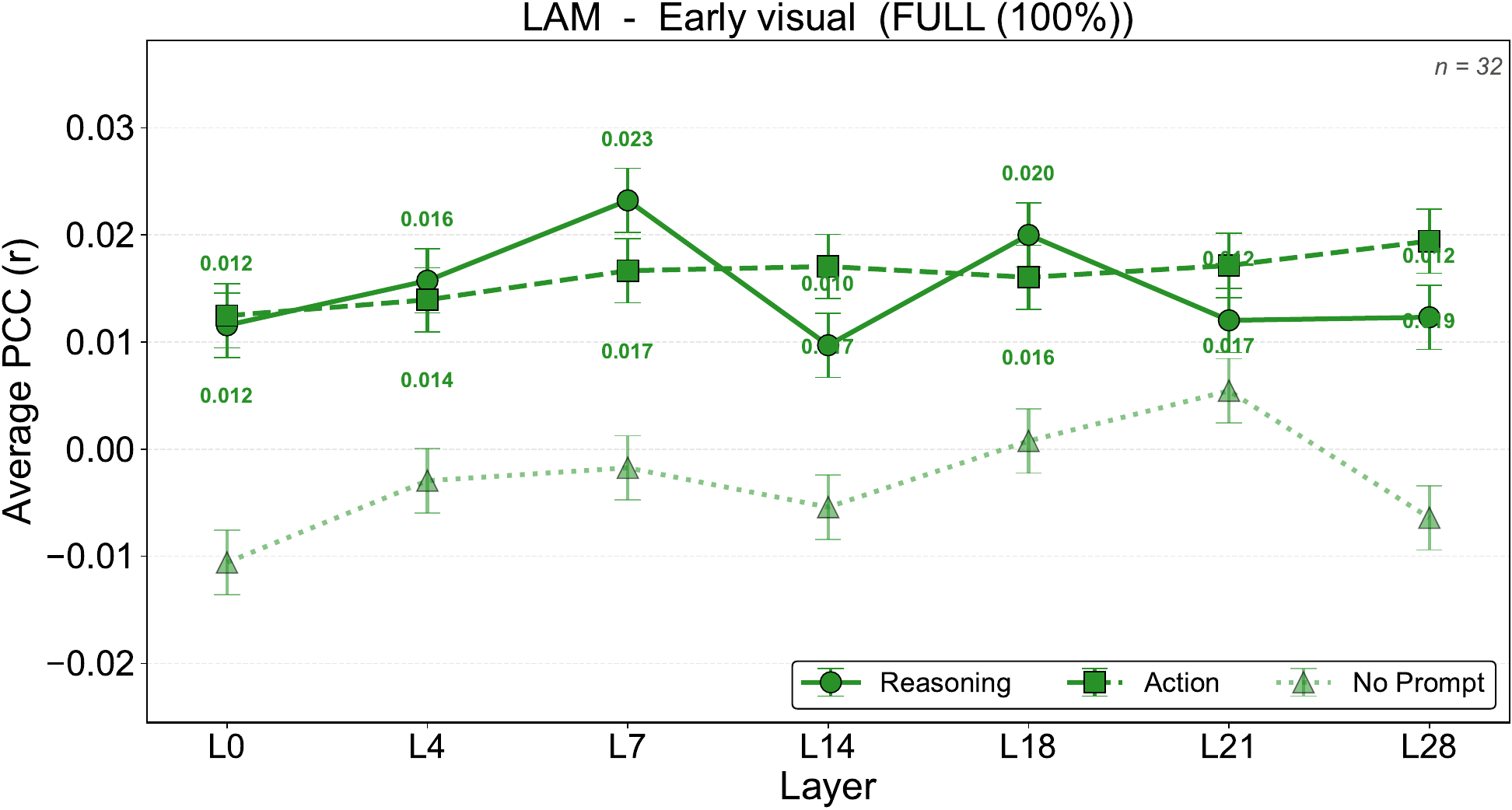}
      \includegraphics[width=0.49\linewidth]
      {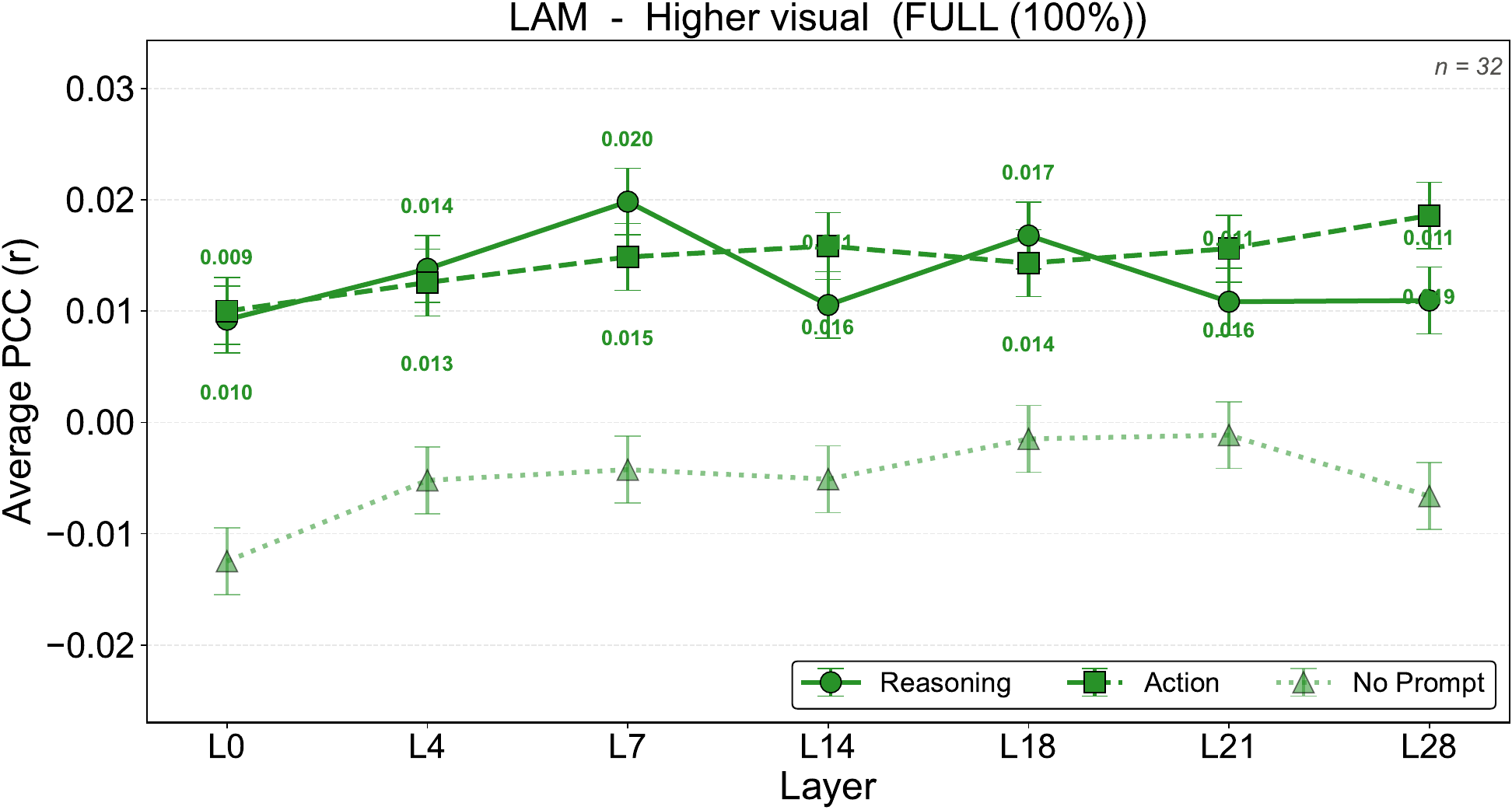}
      \includegraphics[width=0.49\linewidth]
      {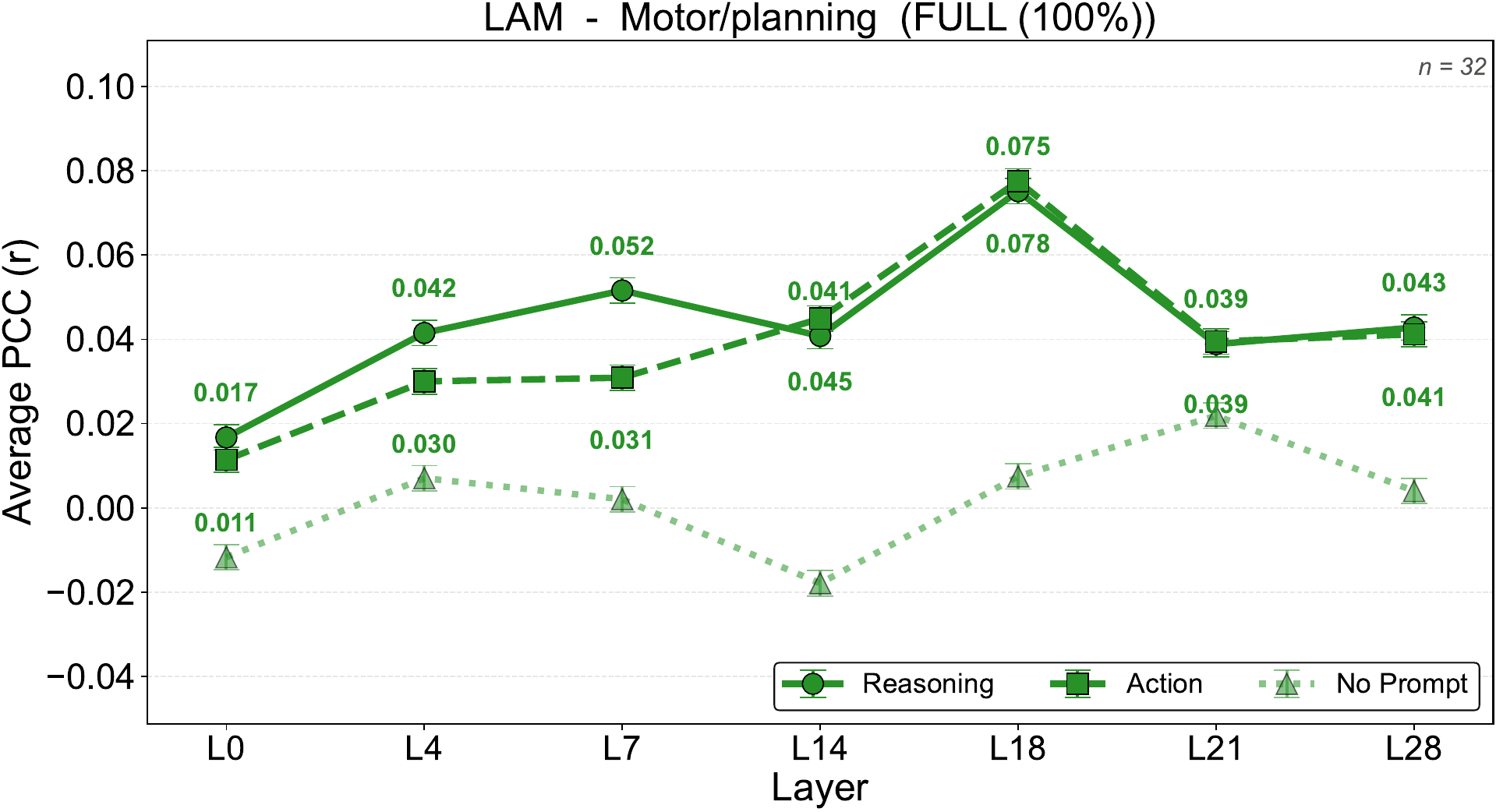}
      \includegraphics[width=0.49\linewidth]
      {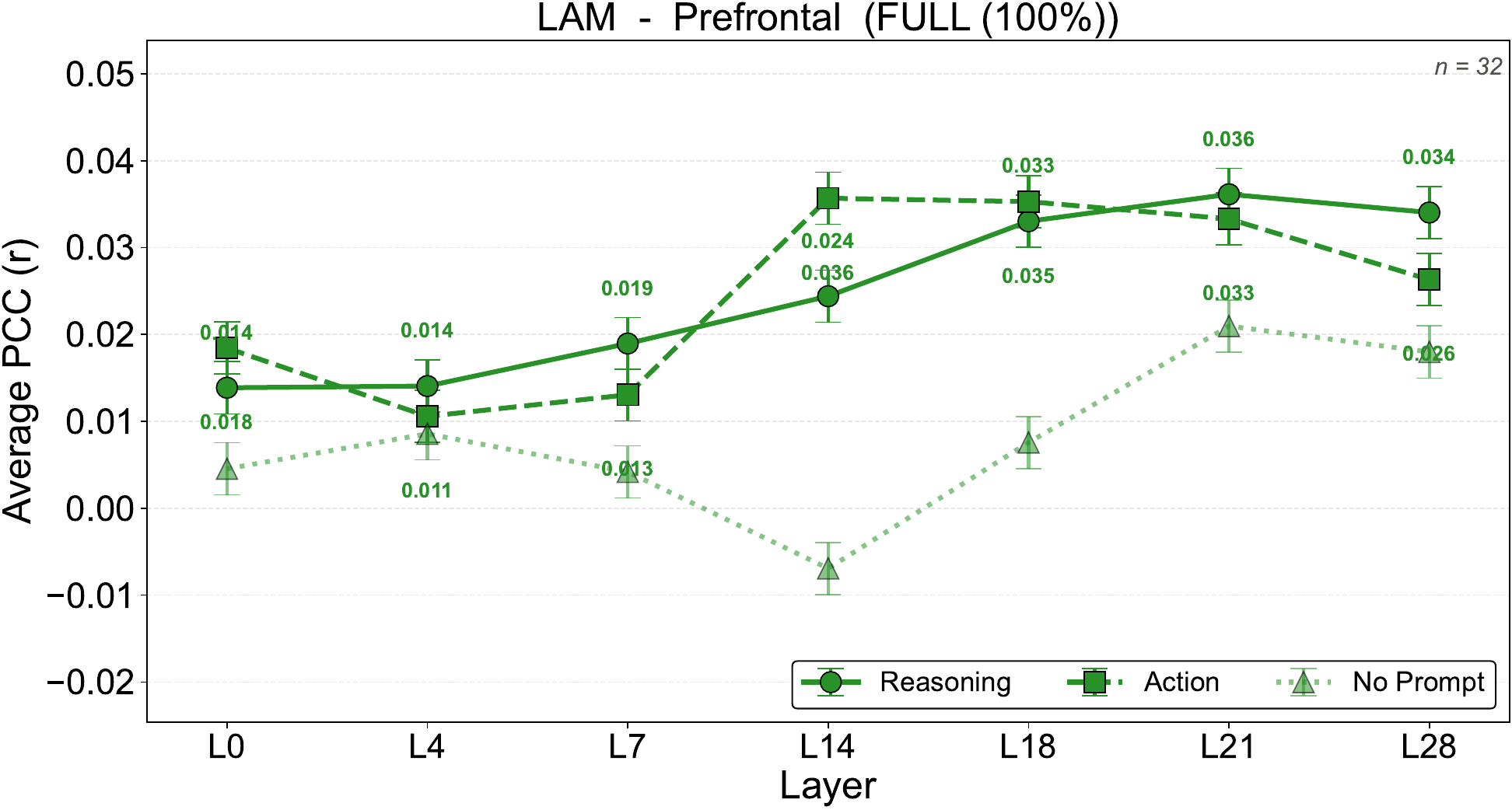}
      \includegraphics[width=0.49\linewidth]
      {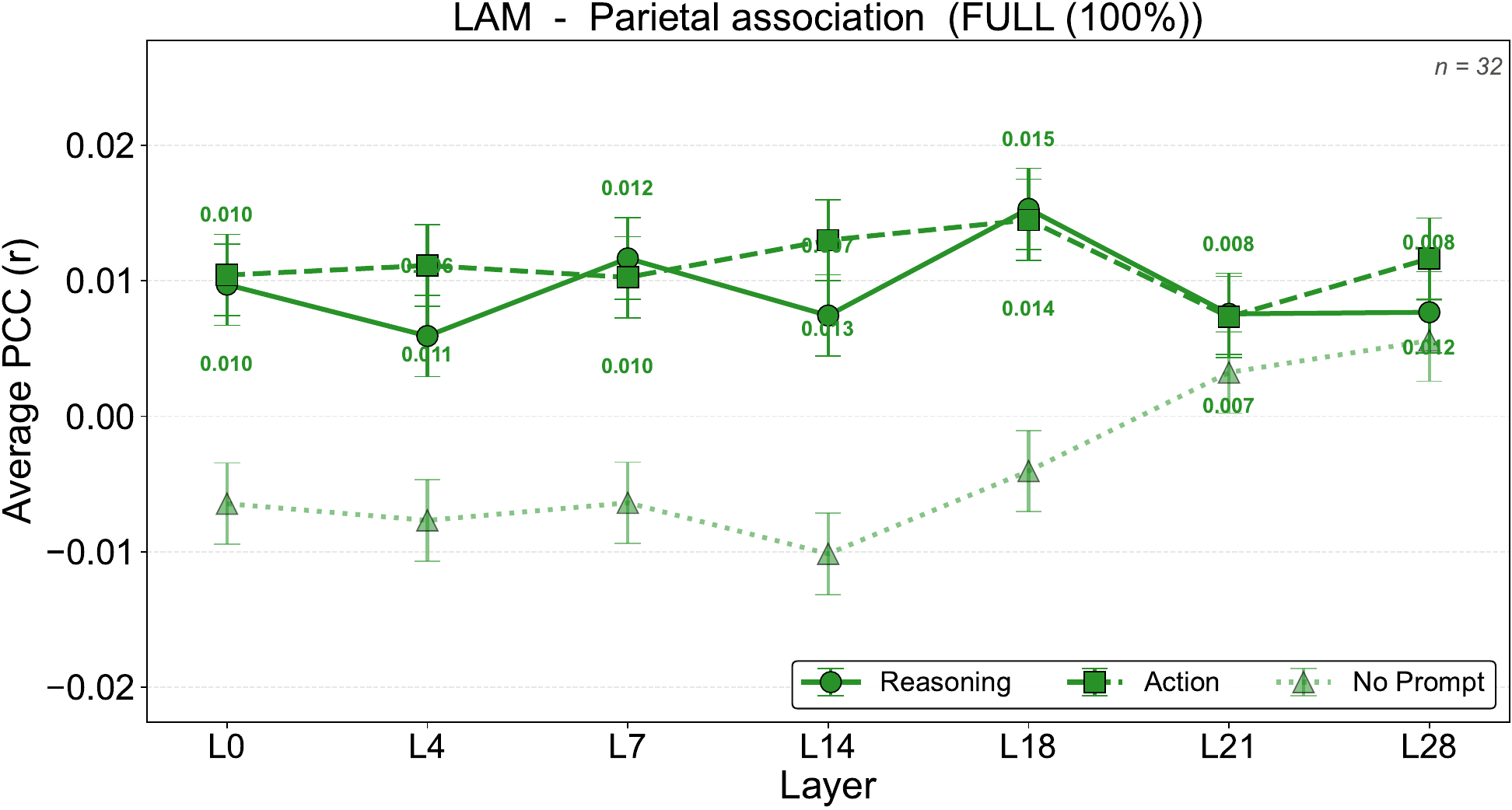}
     \vspace{-0.3cm}
    \caption{\textbf{Layer-wise brain alignment of LAM (UI-TARS) representations, all vertices (FULL).} Average Pearson correlation between predicted and observed fMRI responses at sampled layers (L0--L28) for reasoning (solid circles), action (dashed squares), and \emph{no-prompt} (dotted triangles) representations, shown for the same five ROI groups as Fig.~\ref{fig:vem_lam_rois_atari}. Error bars: mean $\pm$ SEM across $N=32$ participants. Prompted representations exceed \emph{no-prompt} representations at all layers in every ROI group, and higher-order groups peak at intermediate-to-deep layers (Table~\ref{tab:roi_peak_layers}).
    }
    \label{fig:vem_lam_rois_atari_laye_hierarchy_full}
\end{figure}

\begin{figure}[!ht]
    \centering
    \includegraphics[width=0.49\linewidth]
      {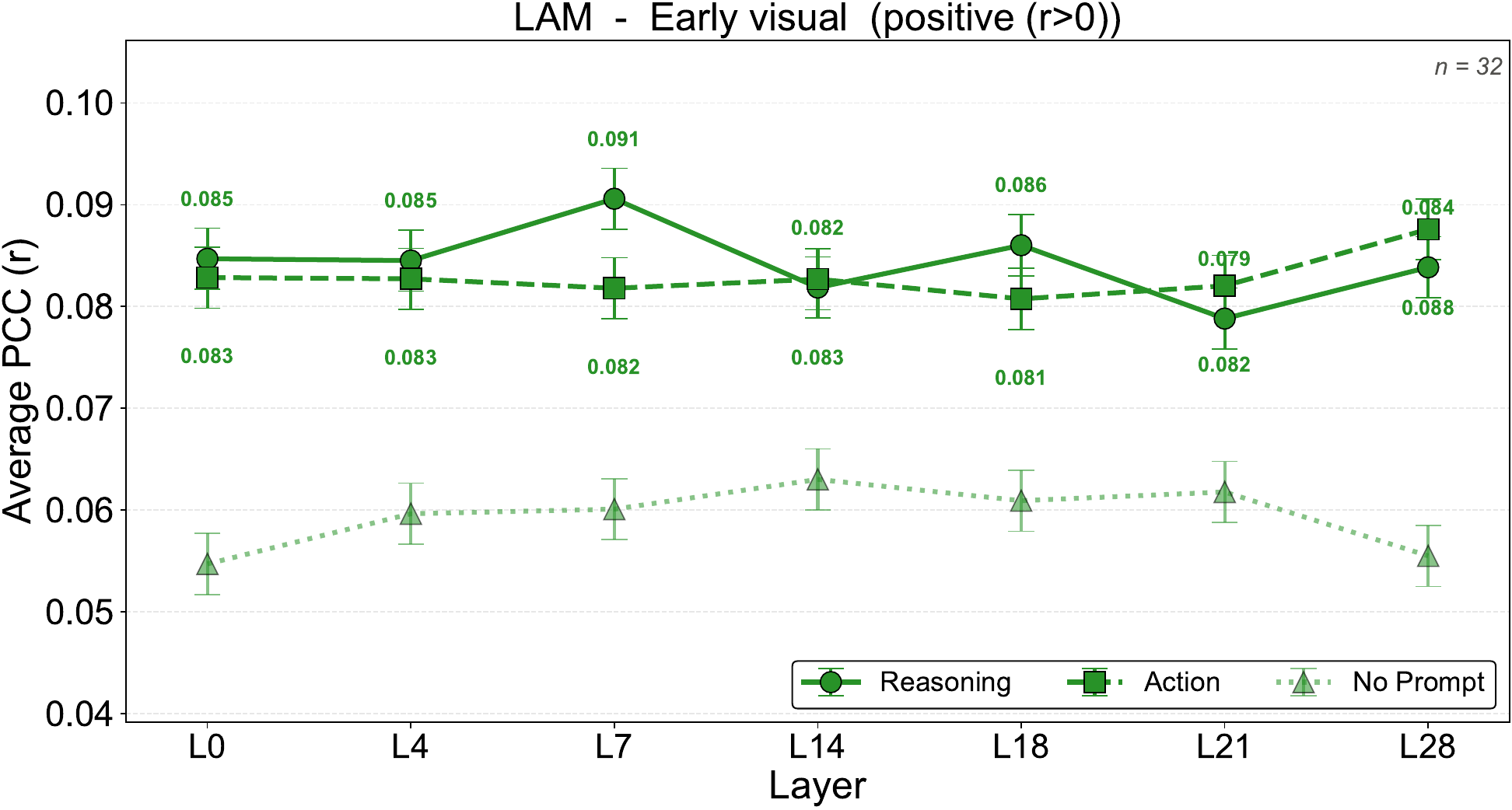}
      \includegraphics[width=0.49\linewidth]
      {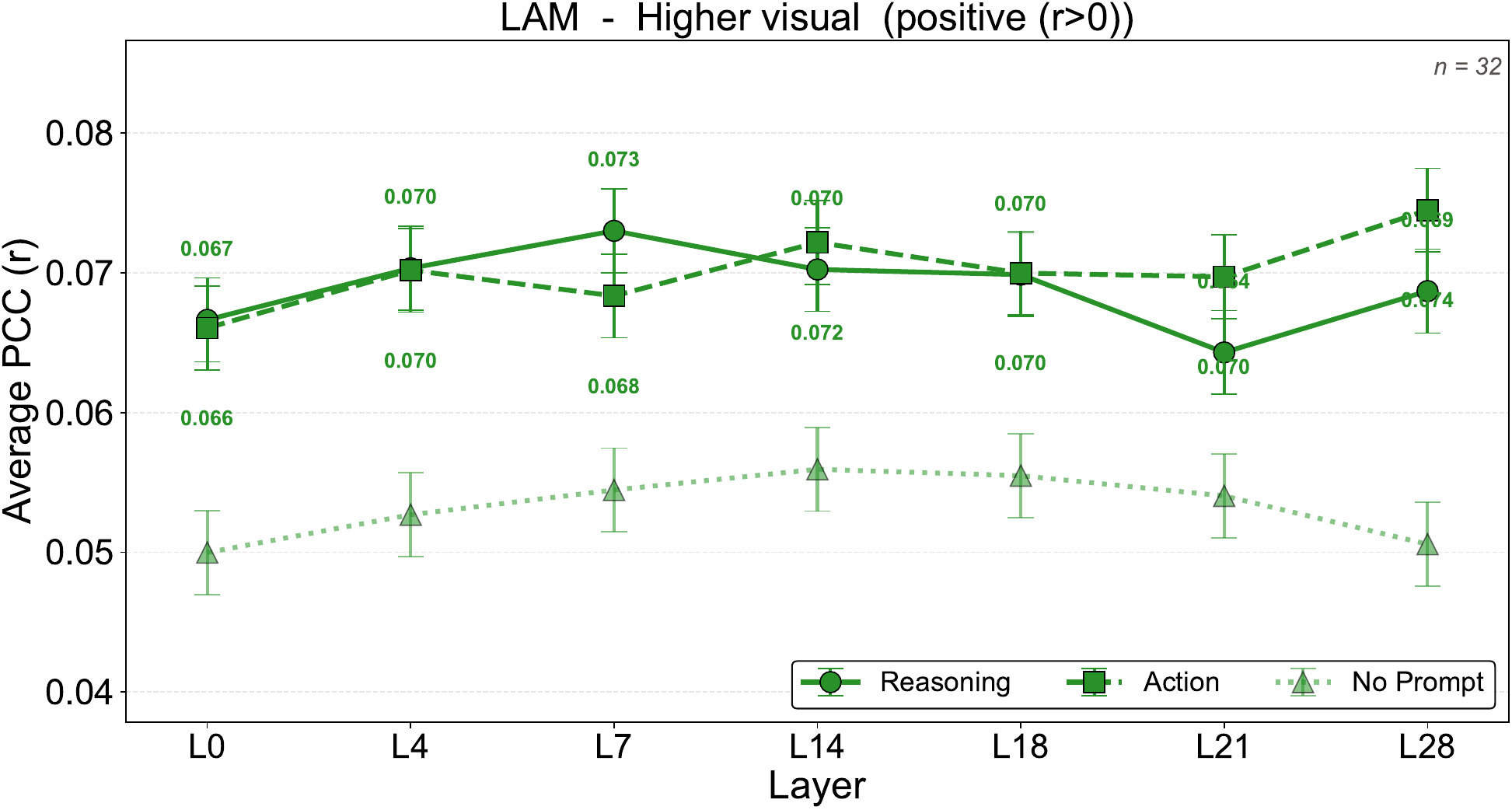}
      \includegraphics[width=0.49\linewidth]
      {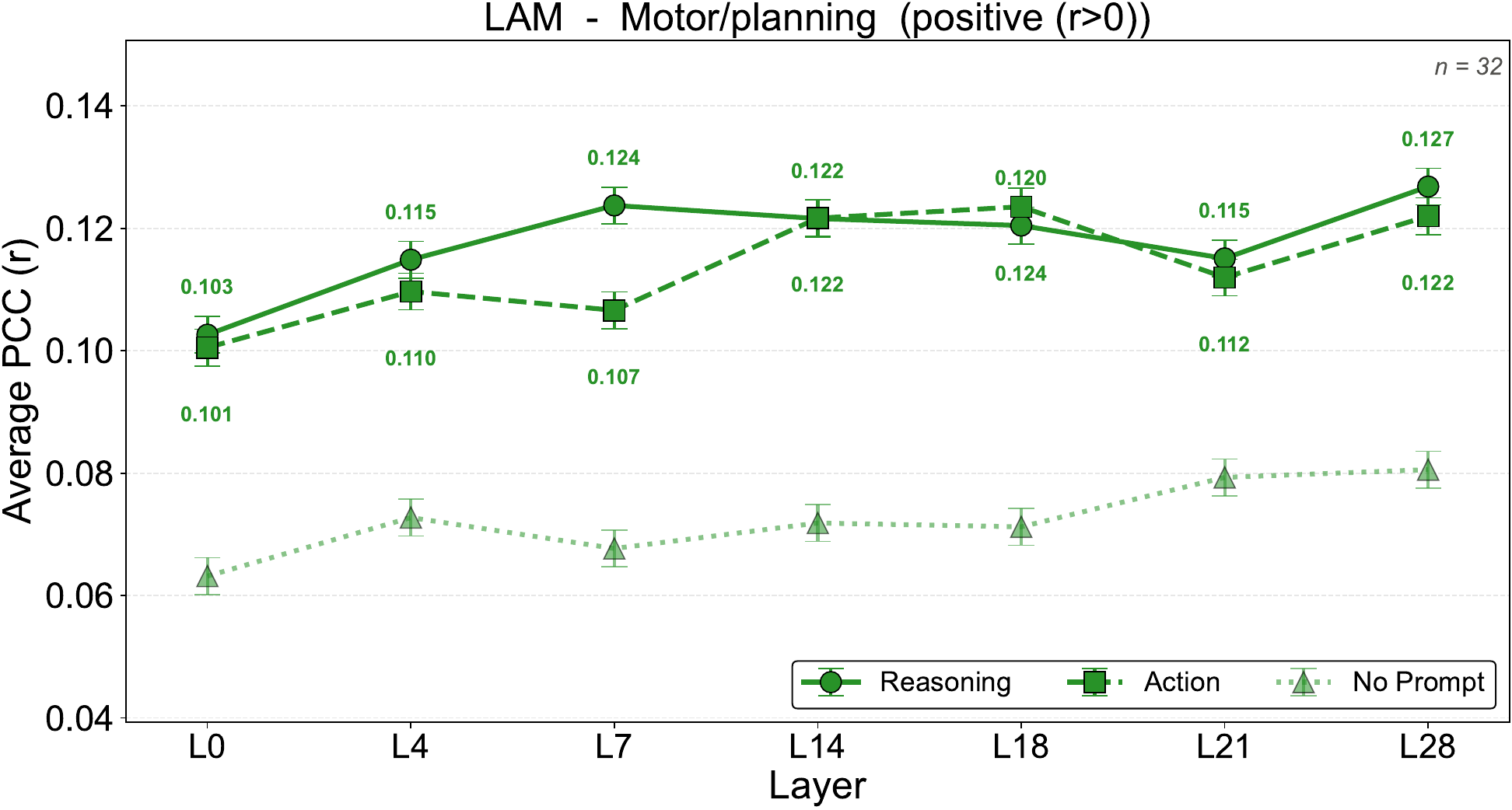}
      \includegraphics[width=0.49\linewidth]
      {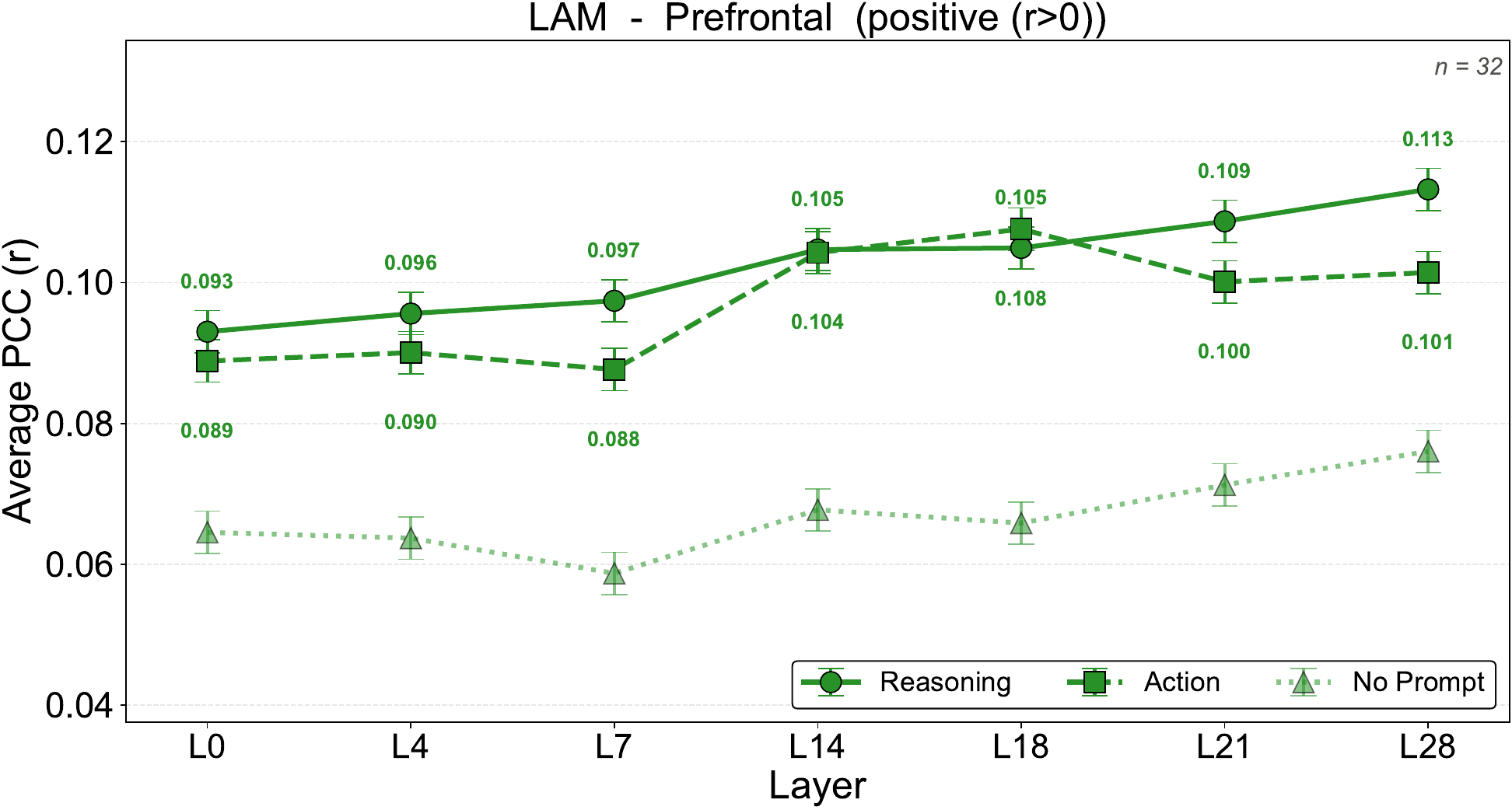}
      \includegraphics[width=0.49\linewidth]
      {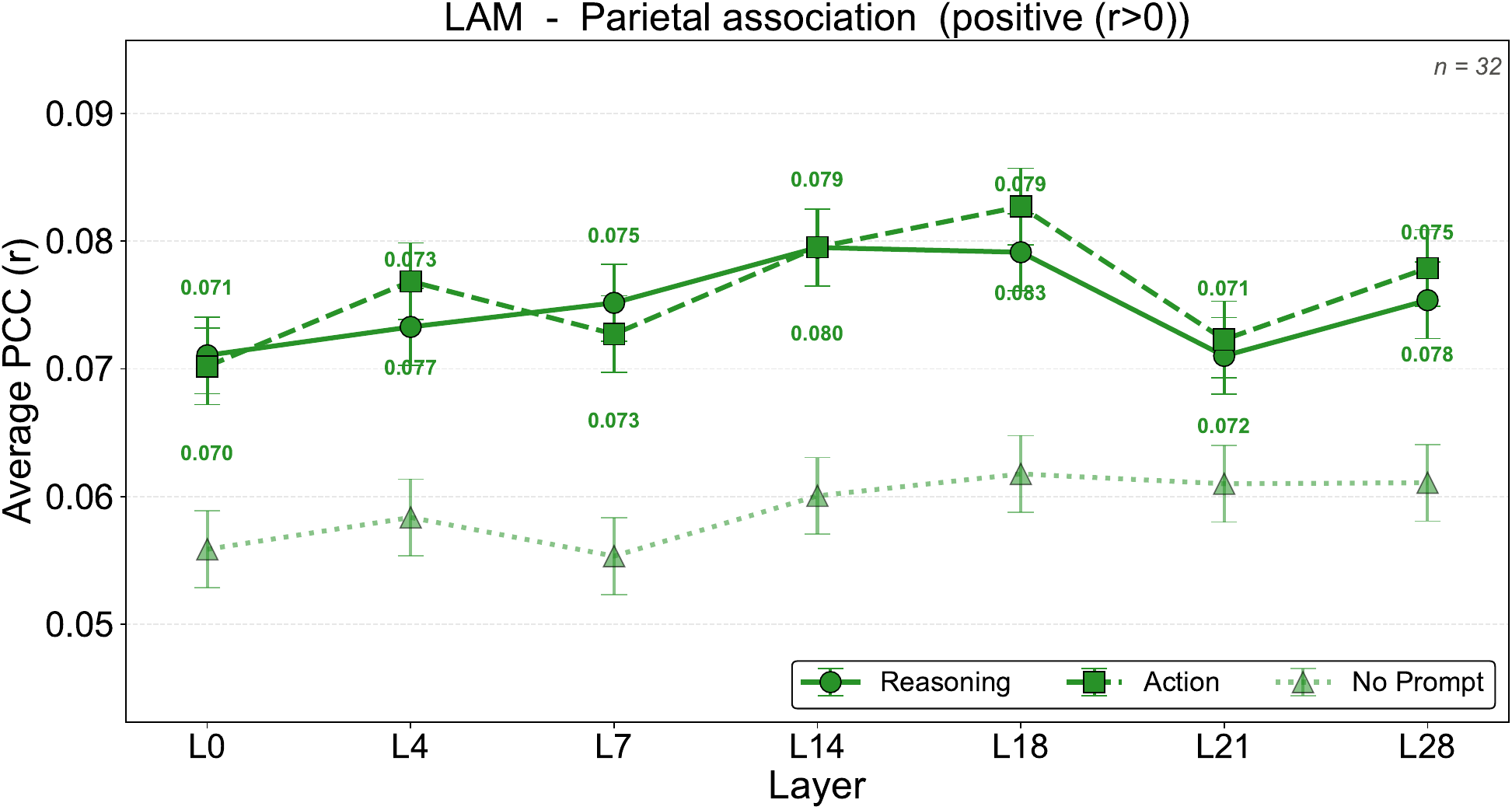}
     \vspace{-0.3cm}
    \caption{\textbf{Layer-wise brain alignment of LAM (UI-TARS) representations, positively predicted vertices ($r>0$).} Same as Fig.~\ref{fig:vem_lam_rois_atari_laye_hierarchy_full}, restricted to voxels with $r>0$. Error bars: mean $\pm$ SEM across $N=32$ participants.
    }
    \label{fig:vem_lam_rois_atari_laye_hierarchy_positive}
\end{figure}

\begin{figure}[!ht]
    \centering
    \includegraphics[width=0.49\linewidth]
      {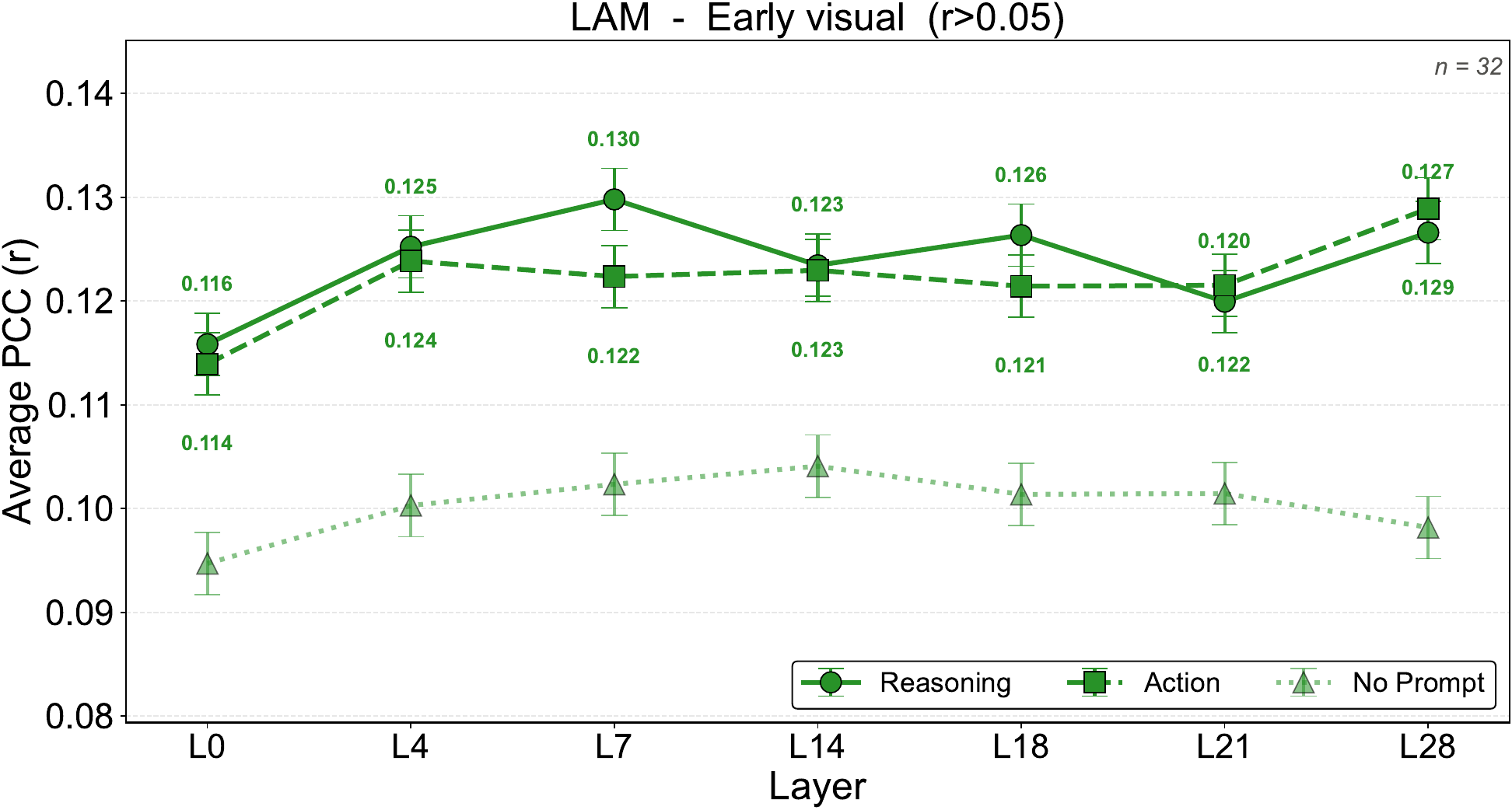}
      \includegraphics[width=0.49\linewidth]
      {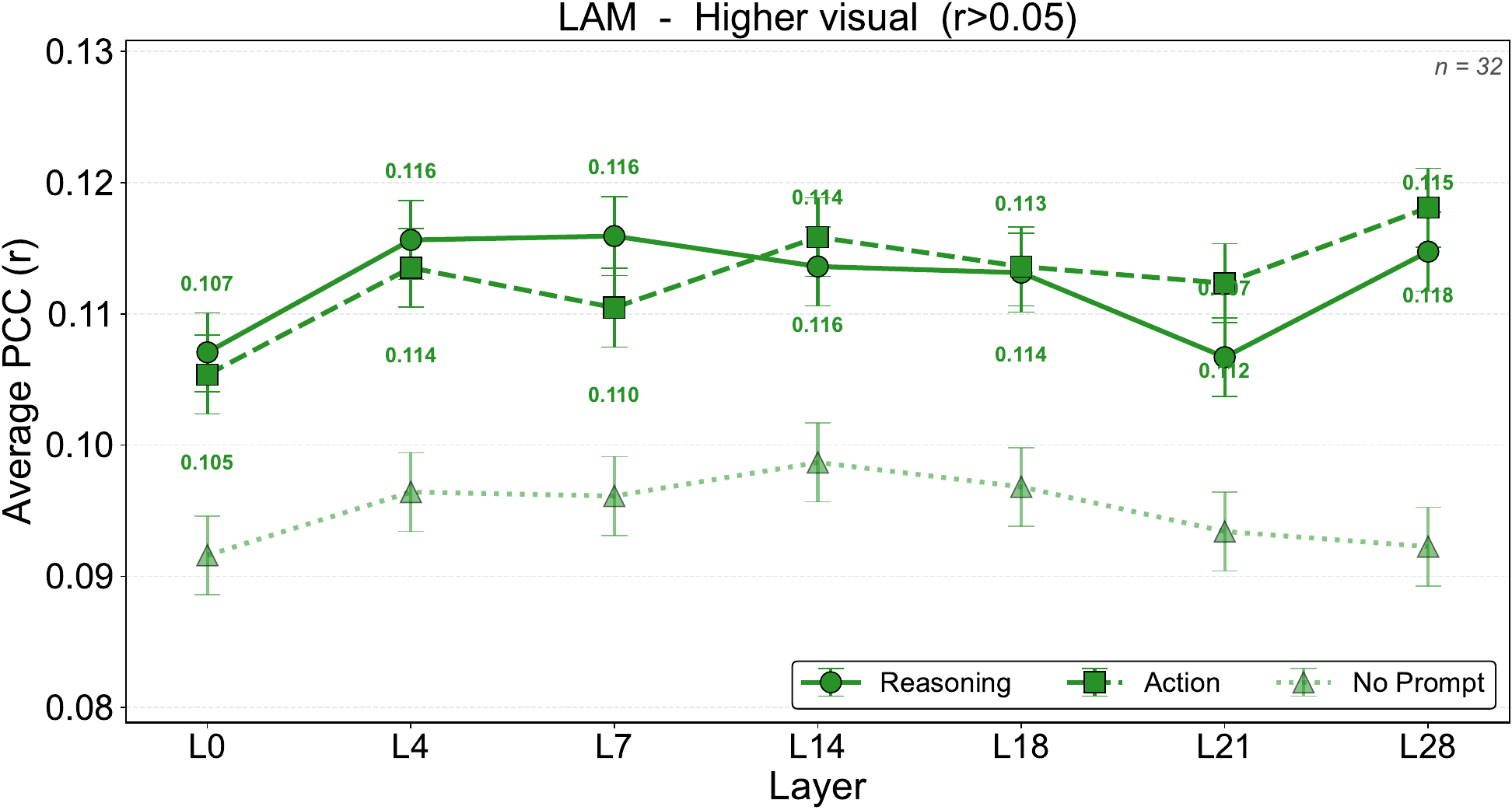}
      \includegraphics[width=0.49\linewidth]
      {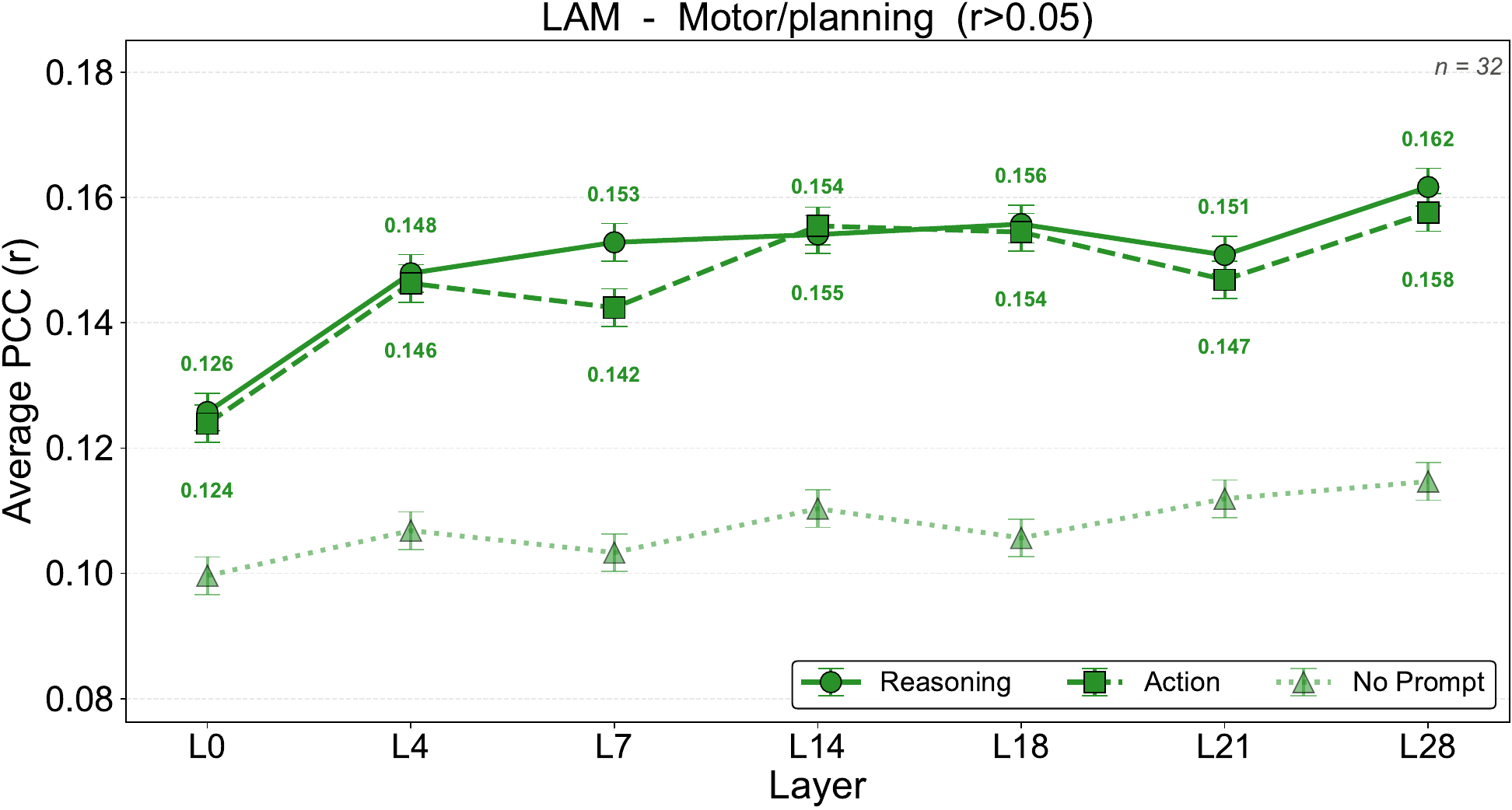}
      \includegraphics[width=0.49\linewidth]
      {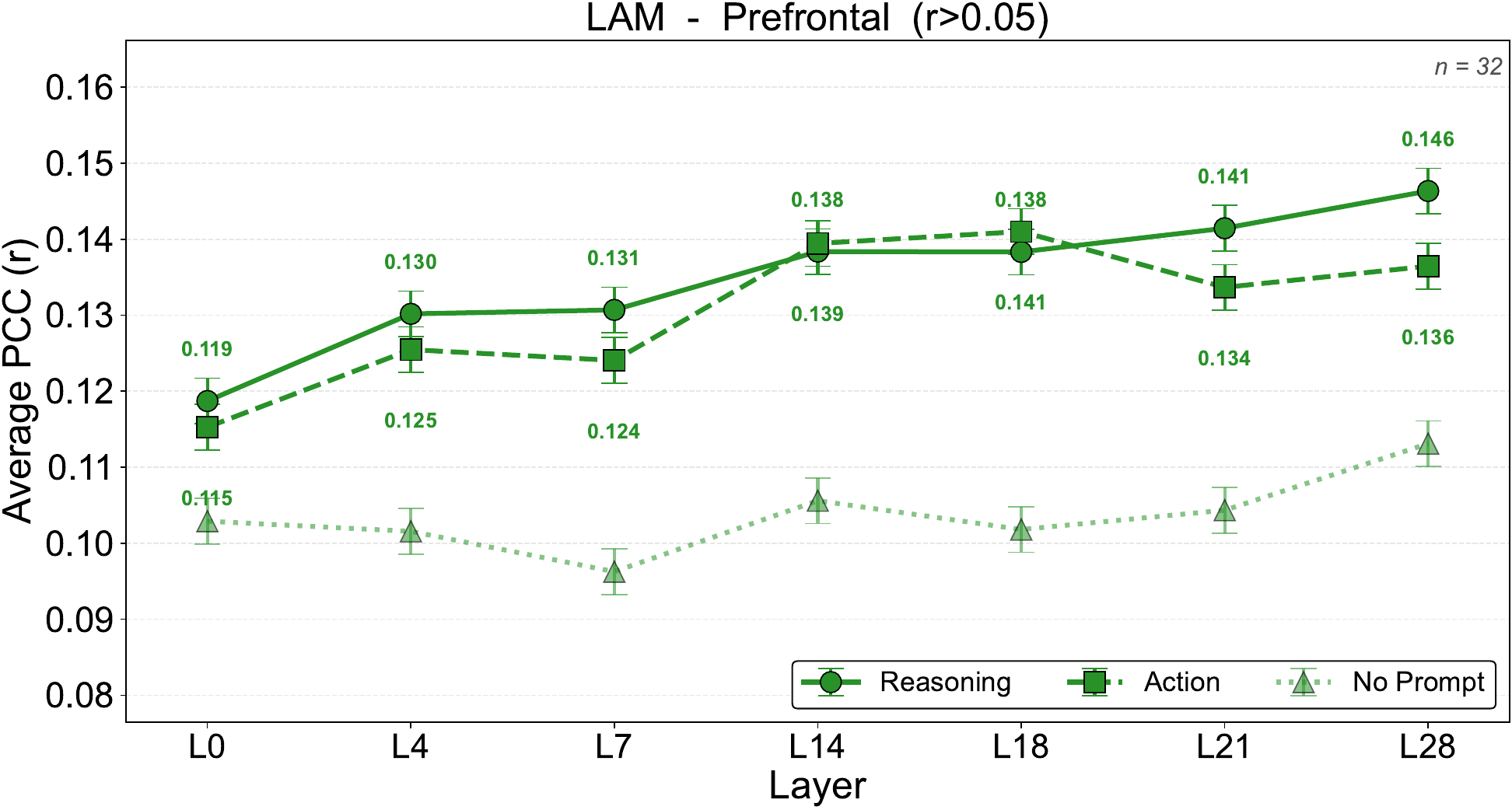}
      \includegraphics[width=0.49\linewidth]
      {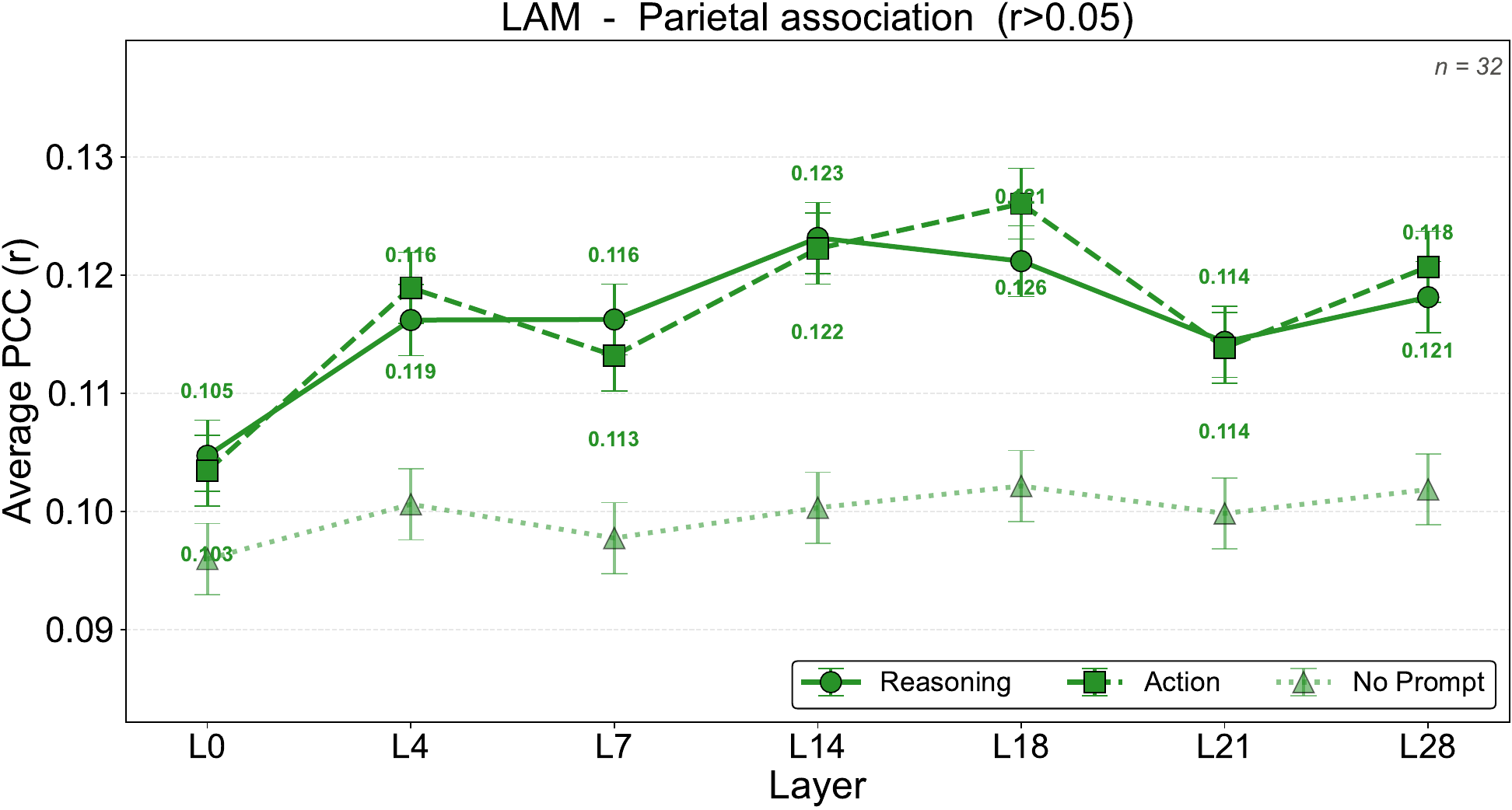}
     \vspace{-0.3cm}
    \caption{\textbf{Layer-wise brain alignment of LAM (UI-TARS) representations, vertices with $r>0.05$.} Same as Fig.~\ref{fig:vem_lam_rois_atari_laye_hierarchy_full}, restricted to vertices with $r>0.05$. Error bars: mean $\pm$ SEM across $N=32$ participants.
    }
    \label{fig:vem_lam_rois_atari_laye_hierarchy_r005}
\end{figure}

\begin{table*}[!ht]
\centering
\caption{\textbf{Peak model layers across cortical ROI groups for the reasoning condition.}
Each entry reports the peak layer and its group-mean held-out Pearson correlation in parentheses, using all vertices and $N=32$ participants. Ties indicate equal values at three-decimal precision. Complete layer profiles are provided in Appendix Fig.~\ref{fig:vem_vlm_rois_atari_laye_hierarchy_full} and~\ref{fig:vem_lam_rois_atari_laye_hierarchy_full}.}
\label{tab:roi_peak_layers}
\scriptsize
\setlength{\tabcolsep}{7pt}
\begin{tabular}{|l|c|c|c|c|c|}
\hline
\textbf{Model} &
\textbf{Early visual} &
\textbf{Higher visual} &
\textbf{Motor/planning} &
\textbf{Prefrontal} &
\textbf{Parietal association} \\
\hline
VLM & L4 (0.028) & L0/L4 (0.016) & L18 (0.067) & L18 (0.061) & L28 (0.025) \\
LAM & L7 (0.023) & L7 (0.020) & L7/L18 (0.052) & L21 (0.036) & L18 (0.015) \\
\hline
\end{tabular}
\end{table*}

\noindent\textbf{Layer-wise regional preferences.}
We next examine which model layers provide the highest unthresholded encoding performance within each cortical group (Table~\ref{tab:roi_peak_layers}). For the VLM reasoning condition, the early- and higher-visual groups peak at L4 and L0/L4, respectively, whereas motor/planning, prefrontal, and parietal regions peak at L18, L18, and L28. For the LAM, the early- and higher-visual groups both peak at L7, while motor/planning peaks jointly at L7/L18, prefrontal at L21, and parietal cortex at L18. Thus, higher-order association regions generally attain their peak alignment at deeper layers than visual regions. Because these peaks are descriptive and were selected from the evaluated layers, we do not interpret them as evidence of a statistically established monotonic hierarchy. Complete layer profiles are provided in Appendix Figs.~\ref{fig:vem_vlm_rois_atari_laye_hierarchy_full} (VLM) and~\ref{fig:vem_lam_rois_atari_laye_hierarchy_full} (LAM).

\section{Variance Partitioning Results}
\label{app:variancePartition}

\begin{figure}[!ht]
    \centering
    \includegraphics[width=\linewidth]
      {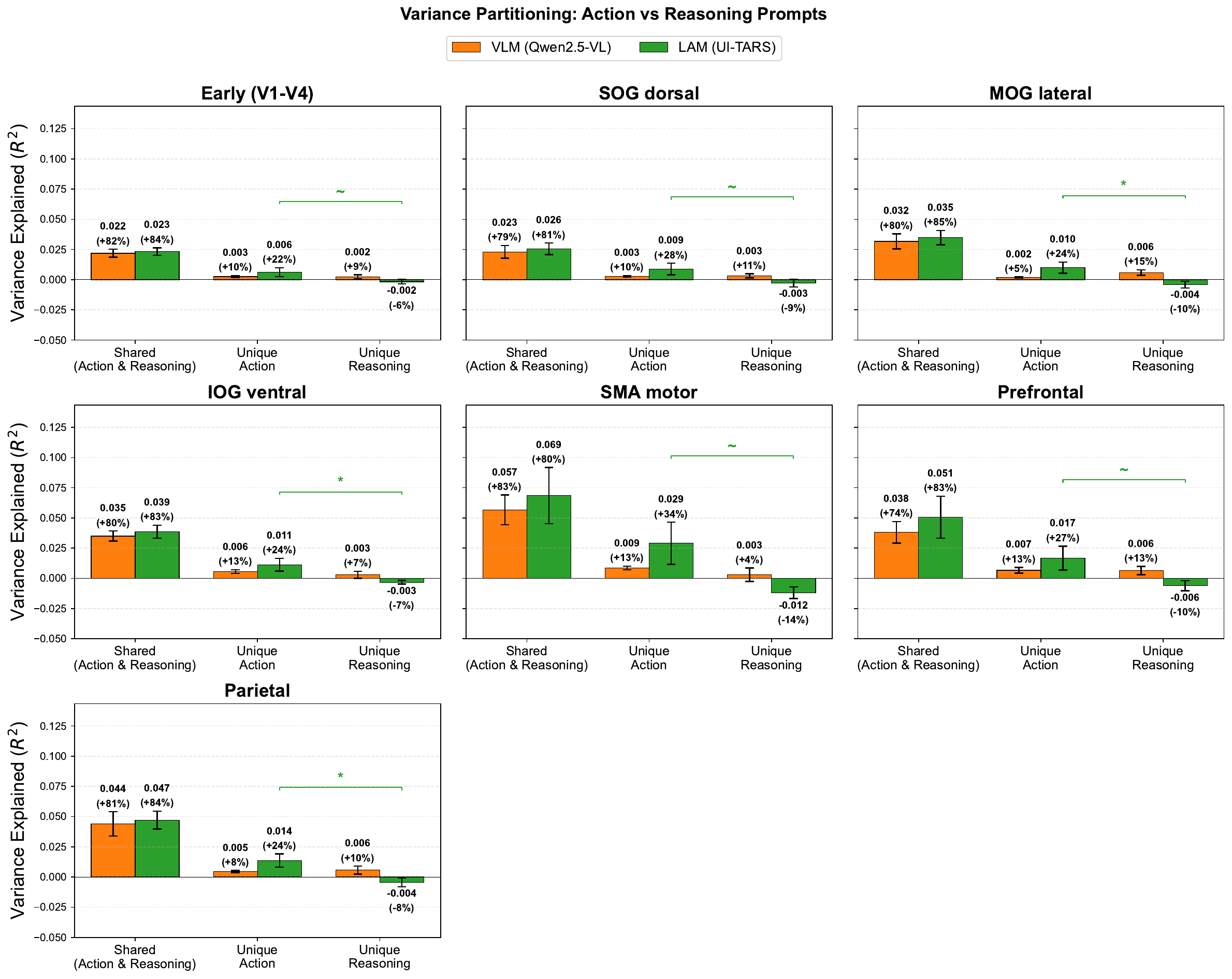}
    \caption{
    \textbf{Variance partitioning of action vs.\ reasoning prompts.} $R^2$ decomposed into shared variance and prompt-unique components for VLM (Qwen2.5-VL, orange) and LAM (UI-TARS-7B-DPO, green), averaged across participants. Numbers above bars show absolute $R^2$ and 
(in parentheses) percentage of joint explained variance. VLM unique variance is balanced across prompts; LAM is action-dominant, with negative unique-reasoning variance across ROIs. Brackets: VLM-vs-LAM comparison on unique reasoning, * $p<0.05$, $\sim$ marginal 
($p<0.10$). Error bars: mean $\pm$ SEM across subjects.}
\label{fig:variance_partition_remain_rois_atari}
\end{figure}

Supplementary Fig.~\ref{fig:variance_partition_remain_rois_atari} extends 
the variance partitioning from Fig.~\ref{fig:variance_partition_wholebrain_atari} 
to the remaining seven ROIs and ROI groups (early visual, SOG, MOG, IOG, SMA, prefrontal, and parietal). Three patterns hold consistently across regions.

\textbf{The reasoning-redundancy asymmetry varies across cortical regions.} 
The LAM's unique-reasoning variance is negative in all seven ROIs ($-6\%$ to $-14\%$ of joint explained variance), whereas the VLM's remains positive throughout ($4\%$--$15\%$). The asymmetry is largest in frontal-motor cortex (SMA: $-14\%$; prefrontal: $-10\%$) and MOG ($-10\%$), intermediate in SOG ($-9\%$), parietal ($-8\%$), and IOG ($-7\%$), and smallest in early visual cortex ($-6\%$). The VLM--LAM difference in unique-reasoning variance is significant in MOG, IOG, and parietal cortex ($p<0.05$) and marginal in the remaining ROIs ($p<0.10$). The magnitude therefore varies across regions but does not follow a strictly monotonic cortical ordering.

\textbf{LAM unique-action variance is consistently $2$--$5\times$ larger than VLM's} ($22\%$--$34\%$ vs.\ $5\%$--$13\%$ across ROIs), following 
the same hierarchical ordering: SMA ($34\%$ vs.\ $13\%$) and is lowest in early visual cortex ($22\%$ vs.\ $10\%$), so the largest action-unique and most negative reasoning-unique components both occur in SMA.
The action-uniqueness ordering thus mirrors the reasoning-redundancy ordering, both peaking in motor-planning cortex.

\textbf{Shared variance dominates in both models} ($74\%$--$85\%$ of joint explained variance across all ROIs), indicating that most prompt-driven brain prediction reflects representations the two prompts 
engage in common. The VLM/LAM dissociation therefore does not lie in this shared bulk but in how each model allocates the residual 
prompt-unique variance: VLM splits it roughly evenly between action and reasoning, while LAM concentrates it on action, with reasoning becoming actively redundant in higher-order cortex.

\subsection{Generalization across VLM–LAM pairs for variance partitioning: OS-Atlas-Pro vs InternVL3}
\label{app:cross_family_vp}
\begin{figure}[!ht]
    \centering
    \includegraphics[width=\linewidth]{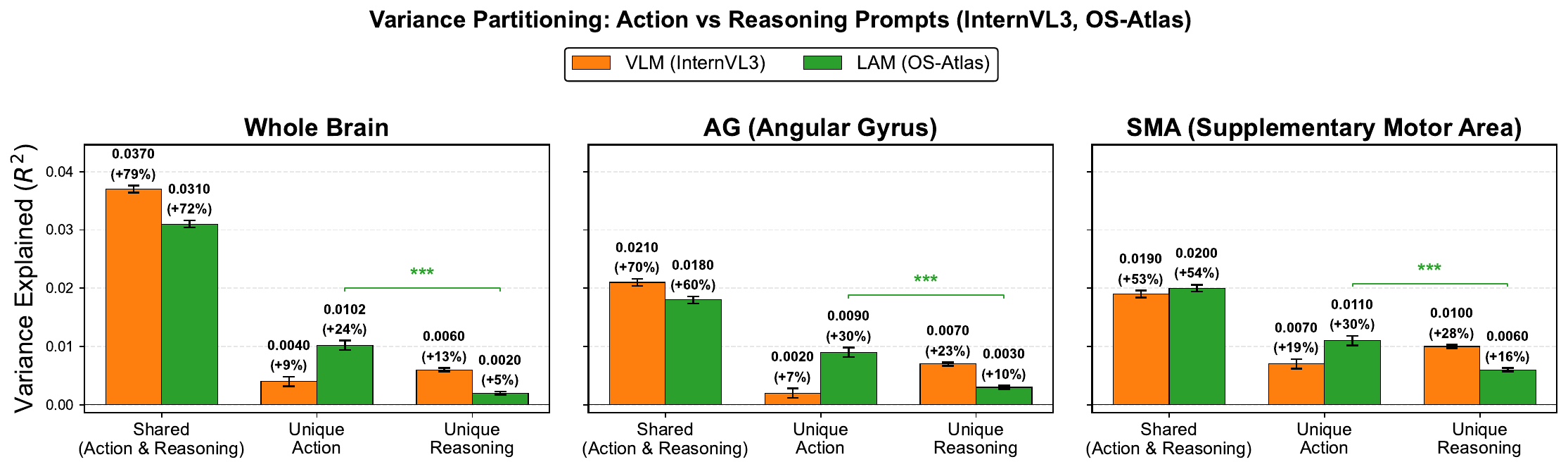}
    \caption{
    \textbf{Variance partitioning of action vs.\ reasoning prompts for the second VLM--LAM pair.} $R^2$ decomposed into shared variance and prompt-unique components for the VLM (InternVL3, orange) and LAM (OS-Atlas-Pro, green) at the whole-brain level and in AG and SMA, averaged across participants. Numbers above bars show absolute $R^2$ and (in parentheses) percentage of joint explained variance. InternVL3 is approximately prompt-symmetric to reasoning-leaning, whereas OS-Atlas-Pro is action-dominant. Brackets and significance markers follow.}
    \label{fig:internvl_osworld_variance_partition_wholebrain_atari}
\end{figure}

To test whether the action-dominance vs prompt-symmetric dissociation observed within the Qwen-VL family (Section~\ref{sec:rq3}) generalizes to other architectures, we ran variance partitioning, comparing OS-Atlas-Pro-7B (Qwen2-VL-based LAM) and InternVL3-8B (Qwen2.5-VL-based VLM, cross-family). Because OS-Atlas and InternVL3 differ in architecture family, this comparison cannot isolate fine-tuning effects from architectural effects. We therefore treat it as supporting evidence of generalization across VLM–LAM pairs rather than a controlled within-family contrast. Figure~\ref{fig:internvl_osworld_variance_partition_wholebrain_atari} shows the explained variance across whole-brain, AG, and SMA. The qualitative dissociation is preserved across all three regions:
\begin{itemize}
\item \textbf{Whole-brain:} OS-Atlas shows strong action-dominance ($u_{\text{Action}}$ = 0.0102, $u_{\text{Reasoning}}$ = 0.002); InternVL3 is approximately prompt-symmetric ($u_{\text{Action}}$ = 0.004, $u_{\text{Reasoning}}$ = 0.006).
\item \textbf{AG:} OS-Atlas action-asymmetric (
0.009 vs. 0.003); InternVL3 Reasoning-leaning (
0.002 vs. 0.007).
\item \textbf{SMA:} OS-Atlas action-asymmetric (
0.011 vs. 0.006); InternVL3 Reasoning-leaning (0.007 vs. 0.010).
\end{itemize}

The shared component is comparable between the two models in all three regions (0.031-0.037 at whole-brain, 0.018 to 0.021 at AG and SMA), indicating both models capture similar overlapping information. The dissociation manifests in the unique-variance components: action-tuning preferentially carves Action-relevant structure into the LAM representation, while the VLM's representation distributes variance more evenly between prompt conditions or shifts toward Reasoning in higher-order cortex. The cross-family preservation of the dissociation supports the interpretation that the within-family Qwen2-VL pattern reflects a general property of action-tuned vs general-purpose multimodal foundation models, not a quirk specific to the Qwen2-VL $\rightarrow$ UI-TARS fine-tuning step.

\section{[Additional analysis]: Alignment using generated reasoning traces of thinking-mode models.}
\label{app:qwen35}
Beyond prompt-conditioned representations from VLMs and LAMs (RQ1-RQ3), we conduct a preliminary study to examine whether the \emph{generated} reasoning traces of thinking-mode models align with brain activity. Recent vision-language models such as Qwen3.5~\citep{qwen35blog}, positioned as ``native multimodal agents,'' produce explicit reasoning sequences between \texttt{<think>} and \texttt{</think>} tags before generating a final answer. We exploit this thinking-mode capability to extract two distinct embeddings for each gameplay frame context: one from the reasoning span and one from the final answer (see the Qwen3.5 prompt template in App.~\ref{app:prompttemplates}.

Specifically, we run generation with thinking enabled and extract two readouts from the resulting sequence: (i) the last-token hidden state at the end of the reasoning trace (before \texttt{</think>}), referred to as the \emph{Reasoning last-token} readout; and (ii) the last-token hidden state at the end of the full generation (after \texttt{</think>}), referred to as the \emph{Final-answer last-token readout}.
To enable a consistent internal comparison, we apply the same generation-based feature extraction procedure used for Qwen2.5-VL and UI-TARS. Note that this comparison places Qwen3.5's generated reasoning content alongside VLM/LAM's prompt-conditioned encodings, related but distinct measurements.

Across subjects, we observe a substantial asymmetry in Qwen3.5. The final-answer readout (r=$0.031\pm0.005$) shows higher brain alignment than the Reasoning last-token readout (r=$0.012\pm0.005$), paired $t$-test $p=0.032$, Wilcoxon $p=0.043$. For reference, prompt-conditioned VLM and LAM achieve r$\approx$0.047 for both Action and Reasoning prompts, indicating that Qwen3.5's final answer output is comparably brain-aligned to prompt-conditioned multimodal models, while its reasoning trace is substantially weaker.
To further examine whether this gap in Qwen3.5 reasoning alignment reflects the content of the reasoning trace or merely the readout position, we compare mean-pooled readouts across the same spans. 
Mean-pooling the reasoning span raises alignment to $r=0.036\pm0.010$, numerically comparable to the final-answer last-token readout ($r=0.031\pm0.005$), suggesting that the observed reasoning--final-answer difference is sensitive to the readout method.
The persistence of this gap under matched pooling suggests that the asymmetry is not solely a readout-position artifact. Why explicit reasoning traces show weaker brain alignment than the model's eventual action commitment - whether due to the content of the reasoning trace, residual positional effects, or trace-length variability - remains an open question. Together, these observations document a substantial reasoning-action asymmetry in thinking-mode brain alignment and motivate further investigation of how reasoning-trace structure interacts with brain alignment in such models.

\section{Limitations}
\label{app:limitations}

Our study has several limitations that scope its claims and motivate future work.
One possible limitation our study lies in matched-architecture VLM-vs-LAM comparison is restricted to the Qwen-VL family (Qwen2.5-VL; UI-TARS-7B-DPO and OS-Atlas-Pro-7B, both Qwen2-VL-based); InternVL3-8B is included only as a cross-family VLM, since no comparable InternVL-based LAM is publicly available. Our LAMs are also GUI/desktop-trained rather than physical-embodiment or game-specific. Cross-family LAM evaluation and comparison with embodied models (e.g., OpenVLA-class) are left for future work.
Second limitation lies in our extension to thinking-mode foundation models is preliminary in three respects: (i) we test only a single model (Qwen3.5), and the observed asymmetry between reasoning-trace and final-answer representations may not generalize to other thinking-mode models such as DeepSeek-R1 or Qwen3 variants.  (ii) the prompt and extraction setup for Qwen3.5 differ from those used for VLM/LAM extraction: Qwen3.5 was given only the Reasoning prompt, with both readouts (reasoning-span last-token and final-answer last-token) drawn from the same generation; (iii) reasoning-trace length varies across TRs, introducing residual confounds that our matched mean-pooling controls only partially. Future work will focus on systematic evaluation across multiple thinking-mode models with controlled trace-length and matched prompts.
Lastly, our study compares the internal representations of the foundation models with the brain activity during naturalistic gameplay, characterizing how each system encodes the state of the game, the perceptual content and the structure relevant to the decision. Our behavioral analyses (App.~\ref{app:behavior}) are limited to offline model--human action agreement and linear decoding of participants' actions; we do not relate model representations to finer-grained behavioral measures such as button-press latencies or trial-by-trial reaction times, and we do not evaluate the models in closed-loop gameplay. Such analyses could test more directly whether brain-aligned representations also predict moment-to-moment human gameplay decisions, providing a complementary axis of model--human comparison. We consider this a natural extension of the current study.

\end{document}